\documentclass[preprint,12pt, 3p, sort&compress]{elsarticle}
\usepackage{fullpage}
\usepackage[T1]{fontenc}
\usepackage{amssymb,amsmath,amsthm,amsfonts,amsbsy,mathrsfs}
\usepackage{bbold}
\allowdisplaybreaks
\usepackage{graphicx,color}

\usepackage{caption}
\usepackage{subcaption}
\usepackage{multirow}
\usepackage{float}
\usepackage[colorlinks   = true, urlcolor     = cyan, linkcolor    = blue, citecolor   = red]{hyperref}
\usepackage{bm}
\usepackage{pict2e}
\usepackage{epsfig}
\usepackage{epstopdf}
\usepackage{comment}
\usepackage{url}
\usepackage[ruled,vlined]{algorithm2e}

\newcommand{\REV}[1]{{\color{black}#1}}

\usepackage{tabularx,multirow,booktabs}
\newcolumntype{L}{>{\raggedright\arraybackslash}X}
\newcolumntype{C}{>{\centering\arraybackslash}X}
\newcolumntype{R}{>{\raggedleft\arraybackslash}X}

\journal{Journal of Computational Physics}

\newcommand{\rz}{\mathbb{R}}

\newcommand{\cz}{\mathbb{C}}

\newcommand{\nz}{\mathbb{N}}

\newcommand{\bfp}{{\bf p}}

\newcommand{\bfx}{{\bf x}}

\newcommand{\bfQ}{{\bf Q}}
\newcommand{\bfR}{{\bf R}}

\newcommand{\beq}{\begin{equation}}
\newcommand{\eeq}{\end{equation}}
\newcommand{\beqs}{\begin{eqnarray}}
\newcommand{\eeqs}{\end{eqnarray}}
\newcommand{\beql}{\begin{equation} \label}

\newcommand{\half}{\frac{1}{2}}

\newcommand{\calB}{{\cal B}}

\newcommand{\calD}{{\cal D}}

\newcommand{\calG}{{\cal G}}

\newcommand{\calI}{{\cal I}}

\newcommand{\calK}{{\cal K}}
\newcommand{\calL}{{\cal L}}
\newcommand{\calM}{{\cal M}}

\newcommand{\calO}{{\cal O}}

\newcommand{\calU}{{\cal U}}
\newcommand{\calV}{{\cal V}}

\newcommand{\hamil}{\mathfrak{H}}

\newcommand{\innprod}[3]{\langle#1,#2\rangle_{#3}}

\newcommand{\Lpspc}[3]{\textsf{L}^{#1}_{#2}(#3)}

\let\oldFootnote\footnote
\newcommand\nextToken\relax

\renewcommand\footnote[1]{%
    \oldFootnote{#1}\futurelet\nextToken\isFootnote}

\newcommand\isFootnote{%
    \ifx\footnote\nextToken\textsuperscript{,}\fi}

\begin{document}


\begin{frontmatter}
\title{Solution of the Schr\"{o}dinger equation for quasi-one-dimensional materials using helical waves}

\author[shivang]{Shivang Agarwal}\corref{asb_ucla}

\address[shivang]{Department of Electrical and Computer Engineering, University of California, Los Angeles, CA 90095, U.S.A}

\author[asb]{Amartya S. Banerjee}\corref{asb_ucla}
\ead{asbanerjee@ucla.edu}

\address[asb]{Department of Materials Science and Engineering, University of California, Los Angeles, CA 90095, U.S.A}

\begin{abstract}
We formulate and implement a spectral method for solving the Schr\"odinger equation, as it applies to quasi-one-dimensional materials and structures.  This allows for computation of the electronic structure of important technological materials such as nanotubes (of arbitrary chirality), nanowires, nanoribbons, chiral nanoassemblies, nanosprings and nanocoils, in an accurate, efficient and systematic manner. Our work is motivated by the observation that one of the most successful methods for carrying out electronic structure calculations of bulk/crystalline systems --- the plane-wave method --- is a spectral method based on eigenfunction expansion. Our scheme avoids computationally onerous approximations involving periodic supercells often employed in conventional plane-wave calculations of quasi-one-dimensional materials, and also overcomes several limitations of other discretization strategies, e.g.,  those based on finite differences and atomic orbitals. The basis functions in our method  --- called  \textit{helical waves} (or \textit{twisted waves}) --- are eigenfunctions of the Laplacian with symmetry adapted boundary conditions, and are expressible in terms of plane waves and Bessel functions in \textit{helical coordinates}. 

We describe the setup of fast transforms to carry out discretization of the governing equations using our basis set, and the use of matrix-free iterative diagonalization to obtain the electronic eigenstates. Miscellaneous computational details, including the choice of eigensolvers, use of a preconditioning scheme, evaluation of oscillatory radial integrals and the imposition of a kinetic energy cutoff are discussed. We have implemented these strategies into a computational package called HelicES (Helical Electronic Structure). We demonstrate the utility of our method in carrying out systematic electronic structure calculations of various quasi-one-dimensional materials through numerous examples involving nanotubes, nanoribbons and nanowires.  We also explore the convergence properties of our method, and assess its accuracy \REV{and computational efficiency} by comparison against reference finite difference, transfer matrix method and plane-wave results. We anticipate that our method will find applications in computational nanomechanics and multiscale modeling, for carrying out transport calculations of interest to the field of semiconductor devices, and for the discovery of novel chiral phases of matter that are of relevance to the burgeoning quantum hardware industry. 


\end{abstract}


\end{frontmatter}

\section{Introduction}
\label{sec:introduction}
Low dimensional materials have been intensely investigated in the past few decades due to their remarkable electronic, optical, transport and mechanical characteristics \cite{bhushan2007springer,cao2004nanostructures}. The properties of these materials often provide sharp contrasts with the bulk phase, and have led to various technological applications, including e.g., new kinds of sensors, actuators and energy harvesting devices \citep{kang2006carbon, chopra2003selective, baughman1999carbon, kong2014carbon, sun2017energy, fan2016flexible}. Quasi-one-dimensional materials --- which include nanotubes, nanoribbons, nanowires, nanocoils, as well as miscellaneous structures of biological origin \citep{xu2022chiral, James_OS} --- are particularly interesting in this regard. This is due to the unique electronic properties that emerge as a result of the availability of a single extended spatial dimension in these structures \citep{giamarchi2003quantum, bockrath1999luttinger, egger2010helical, zaitsev2000luttinger}, the possibility that they are associated with ferromagnetism, ferroelectricity, and superconductivity \cite{SuperconductivityIn1D, qin2017superconductivity, krusin2004room, liu2015giant}, and the fact that the behavior of these materials may be readily modulated via imposition of mechanical  deformation modes such as torsion and/or stretching. \citep{Dumitrica_Tight_Binding2, yu2022density, ding2002analytical}. Quasi-one-dimensional materials have also been investigated as hardware components for computing platforms --- both conventional \citep{tans1998room, li2004carbon} and quantum \citep{aiello2022chirality}. The applications of such materials in the latter case are connected to anomalous transport (the Chiral Induced Spin Selectivity effect \citep{naaman2015ciss}) and exotic electronic states \citep{zhang2019next} that can be observed in such systems.

Given the importance of quasi-one-dimensional materials, it is highly desirable to have available computational methods that can efficiently characterize the unique electronic properties of these systems. However, conventional electronic structure calculation methods --- based e.g. on plane-waves \citep{Martin_ES, Hutter_abinitio_MD}  --- are generally inadequate in handling them. This is a result of the non-periodic symmetries in the atomic arrangements of such materials. As a result of these symmetries, the single particle Schr\"odinger equation associated with the electronic structure problem exhibits special invariances \citep{My_PhD_Thesis, banerjee2021ab}, which plane-waves, being intrinsically periodic, are unable to handle. For example, ground state plane-wave calculations of a twisted nanoribbon (see Fig.~\ref{fig:twisted_nanoribbon}) will usually involve making the system artificially periodic along the direction of the twist axis --- thus resulting in a periodic supercell containing a very large number
of atoms, as well as the inclusion of a substantial amount of vacuum padding in the directions orthogonal to the twist axis, so as to minimize interactions between periodic images. Together, these conditions can make such calculations extremely challenging even on high performance computing platforms, if not altogether impractical. There have been a few attempts to treat quasi-one-dimensional materials using Linear Combination of Atomic Orbitals (LCAO) based techniques \citep{d2009single, dovesi2017crystal17, Mintmire_White1, CNT_1, CRYSTAL, CNT_4}. However, such methods suffer from basis incompleteness and superposition errors \citep{balabin2010communications, gutowski1993critical, simon1996does}, which can make it difficult to obtain systematically convergent and improvable results.

In view of these limitations of conventional methods, a series of recent contributions has explored the use of real space techniques to study quasi-one-dimensional materials and their natural deformation modes \citep{ghosh2019symmetry, banerjee2021ab, yu2022density, sharma2021real}. Specifically, this line of work incorporates the helical interaction potentials present in such systems using \textit{helical Bloch waves} and employs higher order finite differences to discretize the single particle Schr\"odinger equation in \textit{helical coordinates}. While this technique shows systematic convergence, and has enabled the exploration of various fascinating electromechanical properties, it also has a number of significant drawbacks. First, due to the curvilinearity of helical coordinates, the discretized Hamiltonian appearing in these calculations is necessarily non-Hermitian \citep{gygi1995real, banerjee2016cyclic}. This complicates the process of numerical diagonalization and makes many of the standard iterative eigensolvers \citep{Saad_large_eigenvalue_book} unusable. Second, the discretized equations have a coordinate singularity along the system axis which restricts the use of the methods to tubular structures and prevents important nanomaterials such as nanowires and nanoribbons from being studied. The presence of the singularity also tends to ill condition the discretized Hamiltonian, which further restricts the applicability of the method to systems in which the atoms lie far enough away from the system axis (e.g. larger diameter nanotubes). Finally, while the finite difference approach does allow for the simulation of materials with twist (intrinsic or applied), the sparsity pattern of the discretized Hamiltonian worsens upon inclusion of twist, making simulations of such systems significantly more burdensome.

In this work we formulate and implement a novel computational technique that remedies all of the above issues and allows one to carry out systematic numerical solutions of the Schr\"odinger equation, as it applies to quasi-one-dimensional materials and structures. The technique presented here can be thought of as an analog of the classical plane-wave method, and is similar in spirit to the spectral scheme for clusters presented in \citep{Banerjee2015spectral}. Like the classical plane-wave method, a single parameter (the kinetic energy cutoff) dictates the overall quality of solution of our numerical scheme. We present a derivation of the basis functions of our method --- called \textit{helical waves} (or \textit{twisted waves}) --- as eigenfunctions of the Laplacian under suitable boundary conditions. We describe how helical waves may be used to discretize the symmetry adapted Schr\"odinger equation for quasi-one-dimensional materials, and how matrix-free iterative  techniques can be used for diagonalization. A key feature of our technique is the handling of convolution sums through the use of fast basis transforms, and we describe in detail how these transforms are formulated and implemented. We also discuss various other computational aspects, including the choice of eigensolvers and preconditioners, and the handling of oscillatory radial integrals that appear in our method. We have implemented these techniques into a MATLAB \citep{MATLAB:2019} package called HelicES (\textbf{Helic}al \textbf{E}lectronic \textbf{S}tructure), which we use for carrying out demonstrative electronic structure calculations of various quasi-one-dimensional materials. We also present results related to the convergence, \REV{computational efficiency} and accuracy properties of our method, while using finite difference, transfer matrix and plane-wave methods for reference data.

We remark that our technique has connections with methods presented in earlier work concerning electronic structure calculations in cylindrical geometries \citep{solid_state_finite, d2010linear,hussain2018calculation,d2012cylindrical,makaev2008linearized}, but is more general in that the use of helical waves automatically allows both chiral (i.e., twisted) and achiral (i.e., untwisted) structures to be naturally handled.  Additionally, some of these earlier studies have employed the strategy of setting up of the discretized Hamiltonian explicitly and then using direct diagonalization techniques, which scales in a significantly worse way (both in memory and computational time) compared to the transform based matrix-free strategies adopted by us. We also note in passing that the basis functions presented here appear to be scalar versions of twisted wave fields explored recently in the x-ray crystallography \citep{justel2016bragg, friesecke2016twisted} and elastodynamics \citep{chaplain2022elastic_1, chaplain2022elastic_2} literature. 

The rest of this paper is organized as follows. In Section \ref{sec:formulation}, we specify the class of systems of interest to this work, formalize the relevant computational problem, and describe our discretization strategy. Numerical techniques and algorithms are presented in Section \ref{sec:MATLAB}, following which we present  results in Section \ref{sec:results}. We conclude in Section \ref{sec:Conclusions} and also discuss the future outlook of the work. Miscellaneous derivations and computational details are presented in the Appendices.
\section{Formulation}
\label{sec:formulation}
In what follows, $\textbf{e\textsubscript{X}}$, $\textbf{e\textsubscript{Y}}$, $\textbf{e\textsubscript{Z}}$ will denote the standard orthonormal basis of $\mathbb{R}^3$. Position vectors will be typically denoted using boldface lower case letters (e.g., $\bfp)$ and rotation matrices using boldface uppercase (e.g., ${\bfQ}$).The atomic unit system of $m_e = 1, \hbar = 1, \frac{1}{4 \pi \epsilon_0} = 1$ will be used throughout the paper, unless otherwise mentioned. Cartesian and cylindrical coordinates will be typically denoted as $(x,y,z)$ and $(r,\vartheta,z)$ respectively. The $\times$ sign will be reserved for denoting dimensions of matrices (e.g. using $M\times N$ to denote the dimensions of a matrix with $M$ rows and $N$ columns), while $*$ will be used to explicitly denote multiplication by or in between scalars, vectors and  matrices.
\subsection{Description of Physical System and Computational Problem}
\label{sec:system}
\begin{figure}[htb]
    \centering
    \begin{subfigure}[b]{0.46\textwidth}
    \centering
    \includegraphics[trim={12cm 2cm 11cm 1cm}, clip, width=0.75\textwidth]{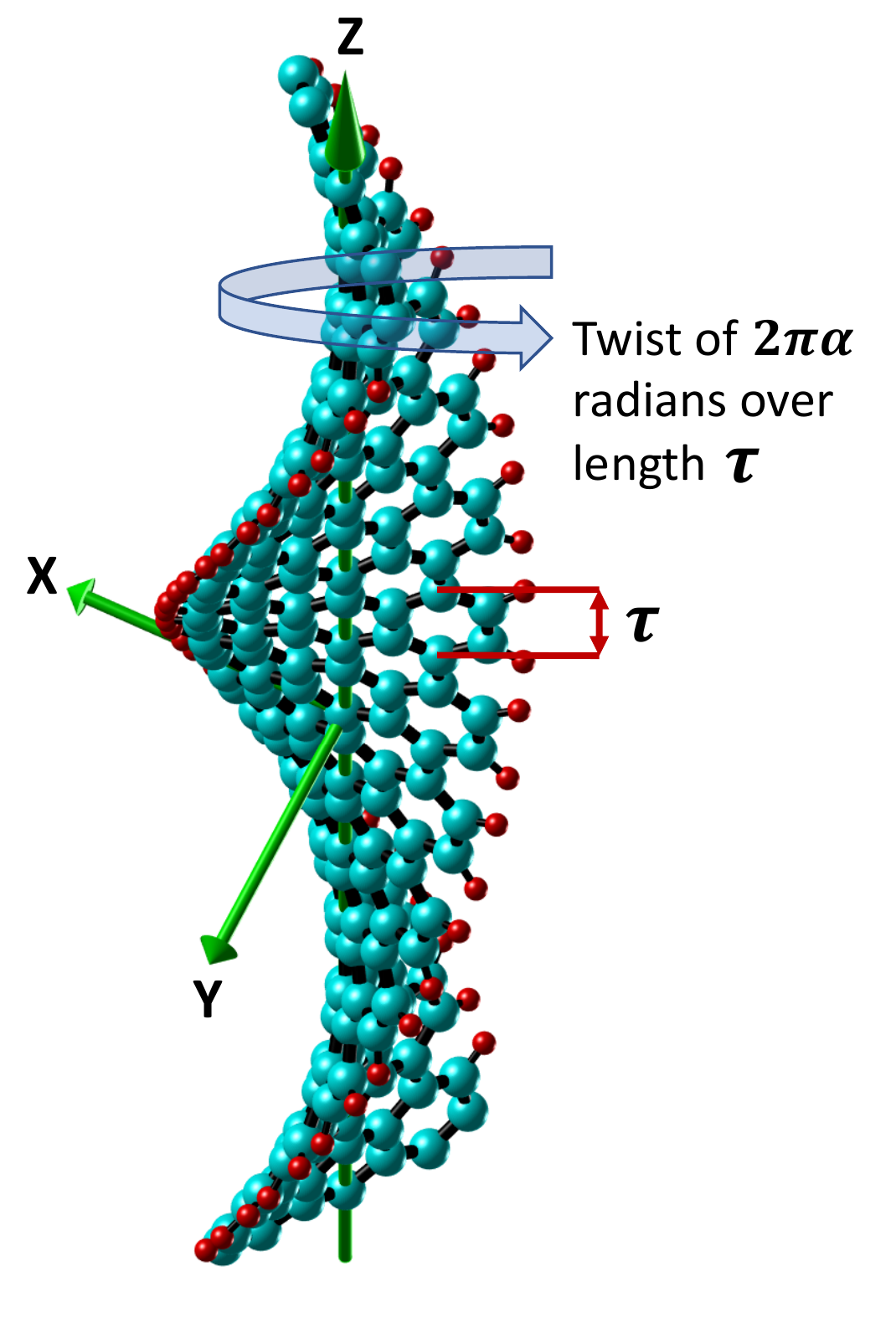}
    \vspace{0.8cm}
    \caption{A twisted nanoribbon}
    \label{fig:twisted_nanoribbon}
    \end{subfigure}
    $\quad$
    \begin{subfigure}[b]{0.35\textwidth}
    \centering
    \includegraphics[trim={15cm 2cm 12cm 1cm}, clip, width=0.75\textwidth]{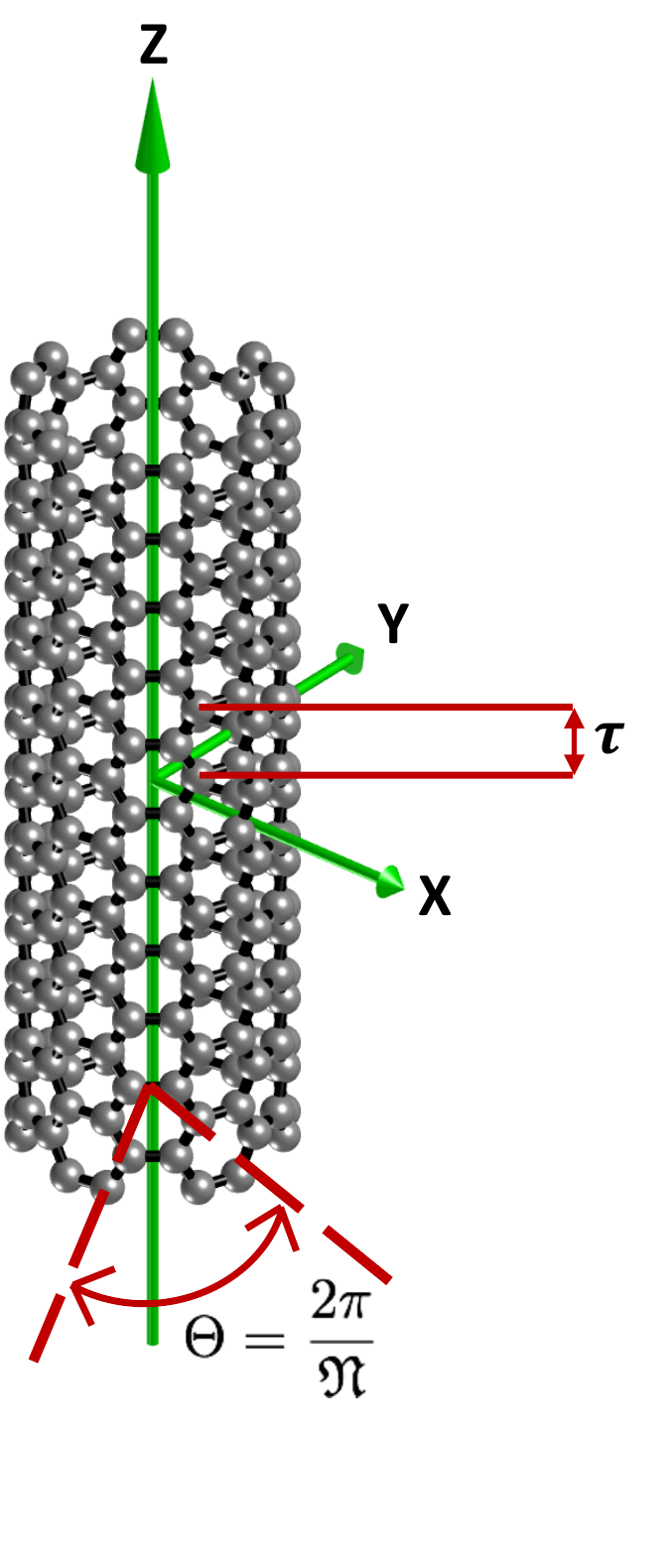}
    \caption{An armchair nanotube}
    \label{fig:CNT}
    \end{subfigure}
    \caption{Examples of the type of nanostructures that can be investigated using the computational framework presented in this work. Helical and cyclic symmetry parameters associated with the geometries of the structures are shown.}
 \label{fig:geometry}   
\end{figure}

We consider a quasi-one-dimensional nanostructure of infinite  extent aligned along $\textbf{e\textsubscript{Z}}$ (see Fig.~\ref{fig:geometry}). We assume the structure to be of limited extent along $\textbf{e\textsubscript{X}}$ and $\textbf{e\textsubscript{Y}}$. Let the atoms of the structure have coordinates:
\begin{align}
\mathcal{S} = \{ \mathbf{p}_1, \mathbf{p}_2, \mathbf{p}_3,\dots:\mathbf{p}_{i} \in \mathbb{R}^3\}\,.
\label{Eqn:S_coord}
\end{align}
Quasi-one-dimensional structures in their undeformed states, or while being subjected to natural deformation modes such as extension, compression or torsion, can often be described using helical (i.e., screw transformation) and cyclic symmetries \citep{banerjee2021ab, yu2022density, My_PhD_Thesis, James_OS}. Accordingly, we may identify a finite subset of atoms of the structure with coordinates:
\begin{align}
\mathcal{P} = \{ \mathbf{r}_1, \mathbf{r}_2, \mathbf{r}_3,\dots,\mathbf{r}_{M}:\mathbf{r}_{i} \in \mathbb{R}^3\}\,,
\label{Eqn:P_coord}
\end{align}
and a corresponding set of symmetry operations:
\begin{align}
\mathcal{G} = \Big\{ \Upsilon_{\zeta,\mu}=\big(\bfR_{(2\pi \zeta\alpha+\mu\Theta)}|\,\zeta\tau \textbf{e\textsubscript{Z}}):\zeta \in \mathbb{Z},\mu=0,1,\dots,\mathfrak{N} - 1\Big\}\,,
\label{Eqn:symm_opera}
\end{align}
such that:
\begin{align}
\mathcal{S} = \underset{\mu=0,1,\dots,\mathfrak{N}-1}{\bigcup_{\zeta \in \mathbb{Z}}}\!\bigcup^{M}_{i=1}\bfR_{(2\pi \zeta\alpha+\mu\Theta)}\mathbf{r}_{i}+\zeta\tau \textbf{e\textsubscript{Z}}\,.
\label{Eqn:sim_atoms}
\end{align}
Here, the $\Upsilon_{\zeta,\mu}$ are symmetry operations of the structure --- specifically, each $\Upsilon_{\zeta,\mu}$  is an isometry whose action on an arbitrary point $\mathbf{x} \in \mathbb{R}^3$ (denoted as $\Upsilon_{\zeta,\mu} \circ \bfx$) is to rotate it by the angle $2\pi \zeta\alpha+\mu\Theta$ about $\textbf{e\textsubscript{Z}}$, while simultaneously translating it by $\mu\tau$ about the same axis. The natural number $\mathfrak{N}$ is related to cyclic symmetries in the nanostructure about the axis $\textbf{e\textsubscript{Z}}$, with $\Theta=2\pi/\mathfrak{N}$ denoting the cyclic symmetry angle. The quantity $\tau$ is the pitch of the screw transformation part of  $\Upsilon_{\zeta,\mu}$, the parameter $\alpha$ takes values $0 \leq \alpha < 1$, and $\beta = 2\pi\alpha/\tau$ captures the rate of twist (imposed or intrinsic) in the structure. The case $\alpha = 0$ usually represents achiral or untwisted structures (see Fig.~\ref{fig:geometry}) .

The electronic properties of a quasi-one-dimensional material under study can be investigated by calculating the spectrum of the single particle Schr\"{o}dinger operator:
\begin{align}
\hamil = -\half \Delta + V(\bfx)\,,
\end{align} 
associated with the system. Determination of the spectrum in an efficient manner, especially for realistic quasi-one-dimensional nanomaterials serves as the primary computational problem of interest in this work. Here, $V(\bfx)$ represents the ``effective potential'' as perceived by the electrons. The potential can be computed through self-consistent means (for example, as part of Density Functional Theory calculations \citep{banerjee2021ab, yu2022density}), or through the use of empirical pseudopotentials \citep{walter1970wave, fong1970energy}, as done here. Due to the presence of global structural symmetries, the potential is expected to obey:
\begin{align}
\label{eq:Potential_Symmetry}
V(\bfx) = V(\Upsilon_{\zeta,\mu} \circ \bfx)\,, \forall \Upsilon_{\zeta,\mu} \in \calG\,.
\end{align}
As a consequence of the quasi-one-dimensional nature of the system, and the above symmetry conditions, the eigenstates of the Hamiltonian can be characterized in terms of \emph{Helical Bloch waves} \citep{My_PhD_Thesis, banerjee2021ab}. Specifically, solutions of the Schr\"odinger equation:
\begin{align}
\big(-\half \Delta + V(\bfx)\big)\psi = \lambda\,\psi\,,
\label{eq:Schrodinger_eq}
\end{align}
can be labeled using band indices $j\in \nz$, and symmetry adapted quantum numbers $\eta \in \left[-\half,\half \right )$, $\nu \in \{0,1,2,\ldots, \mathfrak{N}-1\}$. Moreover, these solutions obey the following condition for any symmetry operation $\Upsilon_{\zeta,\mu} \in \calG$:
\begin{align}
\label{eq:Bloch_theorem_1}
\psi_j(\Upsilon_{\zeta,\mu} \circ \bfx;\eta,\nu) = e^{-2\pi i \big({\zeta \eta + \frac{\mu\nu}{ \mathfrak{N}} }\big)} \psi_j(\bfx;\eta,\nu)\,.
\end{align}
The above relation can be used to reduce the computational problem of determining the eigenstates of the Schr\"odinger operator over all of space, to a \emph{fundamental domain} or \emph{symmetry-adapted unit cell}. 

\begin{figure}[htb]
    \centering
    \includegraphics[width=0.8\textwidth]{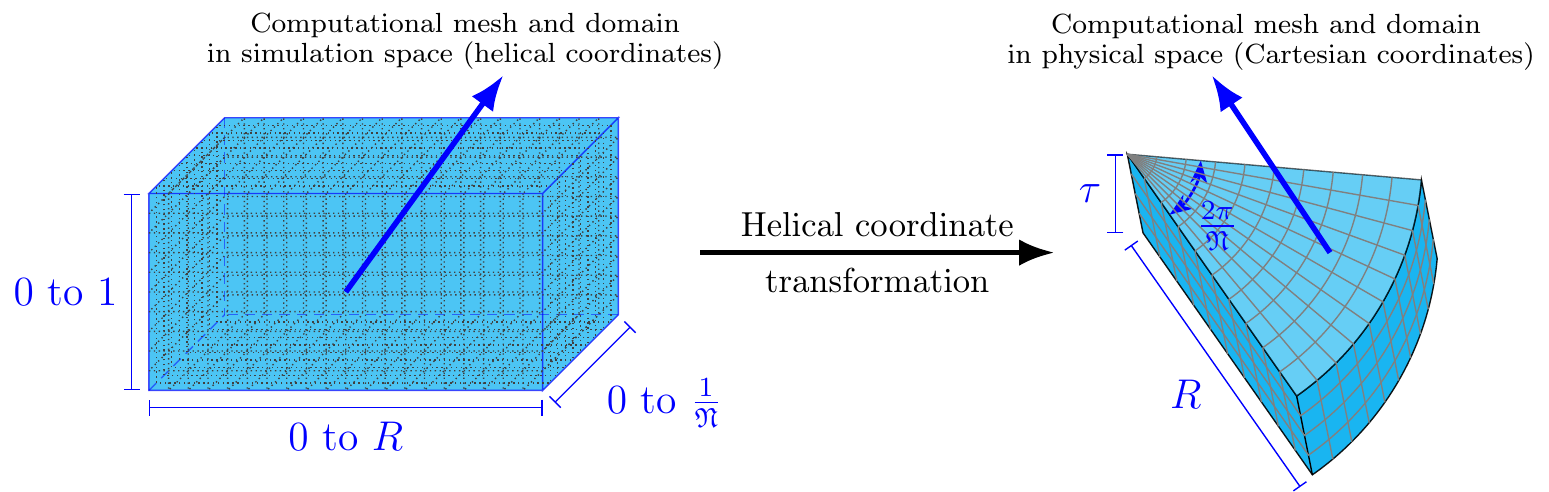}
    \caption{The computational mesh represented in simulation space using helical coordinates (\emph{left}), and physical space using Cartesian coordinates (\emph{right}). The slanted walls of the fundamental domain $\calD$ in physical space (\emph{right}) arise due to possibly arbitrary values of twist associated with the system.}
    \label{fig:coordinate_conversion}
\end{figure}

Since the structures considered here have limited spatial extent in the $\textbf{e\textsubscript{X}}-\textbf{e\textsubscript{Y}}$ plane, so does the computational unit cell. We denote the maximum radial coordinate of the points in the computational domain as $R$. Then, this region of space (see Fig.~\ref{fig:coordinate_conversion}) can be parametrized in cylindrical coordinates as:
\begin{align}
\calD =  \big\{(r,\vartheta,z):0 \leq r\leq R, \frac{2\pi\alpha z}{\tau} \leq \vartheta \leq \frac{2\pi\alpha z}{\tau} + \Theta, 0 \leq z \leq \tau\big\}\,.
\label{eq:fundamental_domain}
\end{align}
Due to the decay of the wavefunctions in the radial direction \citep{wavefunc_decay1, wavefunc_decay2}, it is often appropriate to enforce Dirichlet boundary conditions on the surface $r = R$, as done here. In practice, the value of $R$ can be chosen so as to ensure a sufficient amount of vacuum exists between the structure under study and this lateral boundary surface \citep{banerjee2021ab, ghosh2019symmetry}.
\subsection{The Helical Coordinate System and Transformation of Schr\"{o}dinger's Equation}
\label{sec:helical_coordinate_system}
\begin{figure}[htb]
    \centering
    \includegraphics[trim={2cm 2cm 3cm 1cm}, clip, width=0.4\textwidth]{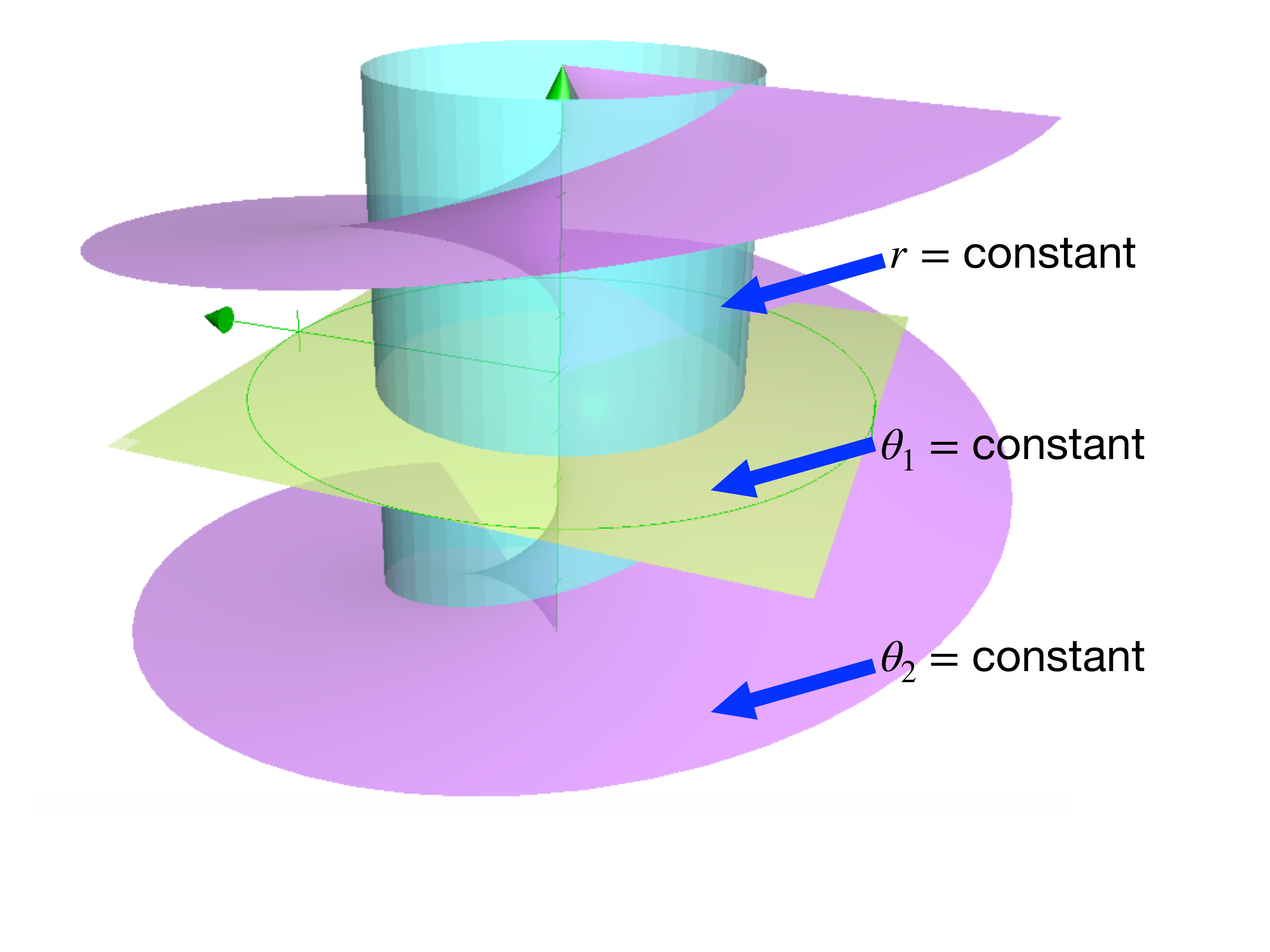}
    \caption{The helical coordinate system represented as constant surfaces of the parameters $r,\theta_1,\theta_2$ (the twist parameter $\alpha$ is nonzero here).}
    \label{fig:helical_coordinates}
\end{figure}

For computational purposes, it is useful to utilize a coordinate system that describes the computational domain $\calD$, and the quasi-one-dimensional system's symmetries more naturally. To this end, we employ \emph{helical coordinates} \citep{waldron1958helical, hochberg1997representing, My_PhD_Thesis} in this work (Fig.~ \ref{fig:helical_coordinates}). For a point $\bfp \in \rz^3$ with Cartesian coordinates $\left(x_{\mathbf{p}},y_{\mathbf{p}},z_{\mathbf{p}}\right)$, cylindrical coordinates $\left(r_{\mathbf{p}},\vartheta_{\mathbf{p}},z_{\mathbf{p}}\right)$, and helical coordinates $\left(\theta_{1\;\mathbf{p}},\theta_{2\;\mathbf{p}}, r_{\mathbf{p}}\right)$, the following relations hold:
\begin{align}
    \label{eq:coordinates_conversion}
   \begin{split}
        r_{\mathbf{p}} &= \sqrt{x_{\mathbf{p}}^2 + y_{\mathbf{p}}^2}\, , \, \theta_{1\;\mathbf{p}} = \frac{z_{\mathbf{p}}}{\tau}\,, \\
        \theta_{2\;\mathbf{p}} &= \frac{1}{2\pi}\arctan 2\left(y_{\mathbf{p}},x_{\mathbf{p}}\right) - \alpha\frac{z_{\mathbf{p}}}{\tau} = \frac{\vartheta_{\mathbf{p}}}{2\pi} - \alpha\frac{z_{\mathbf{p}}}{\tau} \,.
    \end{split}
\end{align}
Regardless of the amount of twist or cyclic symmetries present in the system, the fundamental domain $\calD$ (eq.~\ref{eq:fundamental_domain}) can be conveniently expressed as a cuboid in helical coordinates, i.e., 
\begin{align}
\calD =  \big\{(\theta_1, \theta_2, r):0 \leq \theta_1 \leq 1,  0 \leq \theta_2 \leq  \frac{1}{\mathfrak{N}}, 0 \leq r\leq R\big\}\,.
\label{eq:fundamental_domain_helical}
\end{align}
Thus, it is easier to setup a computational mesh over the fundamental domain using helical coordinates. Moreover, the action of the symmetry operations $\Upsilon_{\zeta,\mu} \in \calG$ is to simply result in translations of the helical coordinates: if $\bfp \in \rz^3$ has helical coordinates $\left(\theta_{1\;\mathbf{p}},\theta_{2\;\mathbf{p}}, r_{\mathbf{p}}\right)$, then $ \Upsilon_{\zeta,\mu} \circ \bfp$ has helical coordinates $\left(\theta_{1\;\mathbf{p}}+\zeta,\theta_{2\;\mathbf{p}}+\frac{\mu}{\mathfrak{N}}, r_{\mathbf{p}}\right)$.
In particular, this implies that a function that is group invariant may be represented over the computational domain by means of periodic boundary conditions along the $\theta_1$ and $\theta_2$ directions.
Next, we formulate the governing equations, i.e., Helical-Bloch wave form of Schrodinger's  equation over the fundamental domain using helical coordinates. To this end, we first note that:
\begin{align}
\label{Eq:Schrodinger_Helical}
\begin{split}
-\half\Delta\psi_j + V\psi_j 
= -\half\bigg[&\big(\psi_j\big)_{rr}+\frac{1}{r}\big(\psi_j\big)_r+\frac{1}{\tau^2}\big(\psi_j\big)_{\theta_1\theta_1}-\frac{2\alpha}{\tau^2}\psi_{\theta_1\theta_2} \\ &+\frac{1}{4\pi^2}\left(\frac{1}{r^2}+\frac{4\pi^2\alpha^2}{\tau^2}\right)\big(\psi_j\big)_{\theta_2\theta_2}\bigg] + V\psi_j = \lambda_j \psi_j
\end{split}
\end{align}
Then, we recast eq.~\ref{eq:Bloch_theorem_1} to imply that the wavefunctions admit the following Helical Bloch ansatz  \citep{yu2022density}:
\begin{align}
\label{Eq:Helical_Bloch_modes}
\psi_j(\theta_1,\theta_2, r;\eta,\nu)=e^{-i2\pi(\eta\theta_1+\nu\theta_2)}\phi_j(\theta_1,\theta_2, r;\eta,\nu)\,.
\end{align}
Here, $\eta \in \left[-\half,\half \right )$, $\nu \in \{0,1,2,\ldots, \mathfrak{N}-1\}$, and the auxiliary functions $\phi_j(\theta_1,\theta_2, r;\eta,\nu)$ are group invariant. In particular, this implies that these functions obey the conditions:
\begin{align}
\label{Eq:Bloch_Ansatz_BC}
\begin{split}
\phi_j(\theta_1,\theta_2, r;\eta,\nu) &= \phi_j(\theta_1 + 1,\theta_2, r;\eta,\nu) \\
\phi_j(\theta_1,\theta_2, r;\eta,\nu) &= \phi_j(\theta_1 ,\theta_2 +\frac{1}{\mathfrak{N}}, r;\eta,\nu)\,.
\end{split}
\end{align}
Substituting eq.~\ref{Eq:Helical_Bloch_modes} into the Schr\"odinger equation above (eq.~\ref{Eq:Schrodinger_Helical}) and after some algebra (\ref{App:Governing_Equation}), we arrive at:
 \begin{align}
\label{Eq:Expanded_schrodinger}
\begin{split}
\Bigg[-\frac{1}{2}\Delta\phi_j-\left(\frac{2\pi^2}{\tau^2}\Big\{\nu\alpha\left(2\eta-\nu\alpha\right)-\eta^2\Big\}-\frac{\nu^2}{2r^2}\right)\phi_j-\frac{2i\pi}{\tau^2}\left(\nu\alpha-\eta\right)\big(\phi_j\big)_{\theta_1} \\-2i\pi\left[\frac{\alpha}{\tau^2}\left(\eta-\nu\alpha\right)-\frac{\nu}{4\pi^2 r^2}\right]\big(\phi_j\big)_{\theta_2}+V\phi_j\Bigg]=\lambda_j\phi_j\,.
\end{split}
\end{align}
This serves as the governing equation for the computational method in this work. It needs to be discretized and solved over the fundamental domain along with the enforcement of periodic boundary conditions in the $\theta_1$ and $\theta_2$ directions (eq.~\ref{Eq:Bloch_Ansatz_BC}), and the imposition of wavefunction decay in the radial direction, i.e.:
\begin{align}
\phi_j(\theta_1,\theta_2, r = R;\eta,\nu)  = 0\,.
\label{eq:Phi_j_Decay}
\end{align} 
Note that due to eq.~\ref{eq:Potential_Symmetry}, the effective potential in helical coordinates, $V(\theta_1,\theta_2, r)$, also obeys conditions of the form outlined in eq.~\ref{Eq:Bloch_Ansatz_BC}, although it is generically not expected to obey the decay conditions similar to eq.~\ref{eq:Phi_j_Decay}.
\subsection{Basis Set and Discretization}
\label{Sec:Methodology}
We now discuss discretization of the governing equations using helical waves. The derivation of these basis functions as symmetry adapted eigenfunctions of the Laplacian is presented in \ref{App:Basis_set_derivation}.  In what follows, we will usually suppress the dependence of $\phi_j(\theta_1,\theta_2, r;\eta,\nu)$ on the band index ($j$) for the sake of simplicity of notation. Denoting the basis functions as $F_{m,n,k}\left(\theta_1,\theta_2, r\right)$ in helical coordinates, we write:
\begin{align}
\label{Eq:Phi_expanded_in_basis_set}
\nonumber
\phi\left(\theta_1,\theta_2, r\right) &=\sum_{(m,n,k)\in\Gamma}\!\hat{\phi}_{m,n,k}\,F_{m,n,k}\left(\theta_1,\theta_2, r\right) \\
&= \sum_{(m,n,k)\in\Gamma}\!\hat{\phi}_{m,n,k}\,c_{m,n,k}\,e^{i2\pi(m\theta_1+n\mathfrak{N}\theta_2)}\,J_{n\mathfrak{N}}\left(\frac{b^{n\mathfrak{N}}_{k}r}{R}\right)\,.
\end{align}
Here, $\hat{\phi}_{m,n,k}$ are the expansion coefficients, $J_{n\mathfrak{N}}(\cdot)$ denotes Bessel functions of the first kind of order $n\mathfrak{N}$, while $b^{n\mathfrak{N}}_{k}$ denote the zeros of the Bessel functions. The basis function normalization constants $c_{m,n,k}$ are:
\begin{align}
c_{m,n,k}&=\sqrt{\frac{\mathfrak{N}}{\pi\tau}}\frac{1}{RJ_{n\mathfrak{N}+1}\left(b^{n\mathfrak{N}}_{k}\right)}\,.
\end{align}
The set $\Gamma$ denotes triplets of integers $(m,n,k)$ such that $m \in [-M_{\text{max}},M_{\text{max}}]$, $n \in [-N_{\text{max}},N_{\text{max}}]$ and $k \in [1,K_{\text{max}}]$. The basis set size is $\calL = (2M_{\text{max}}+1)*(2N_{\text{max}}+1)*K_{\text{max}}$, i.e., it grows as $\calO(M_{\text{max}}N_{\text{max}}K_{\text{max}})$ in terms of the discretization sizes along the $\theta_1,\theta_2,r$ directions. By design, the basis functions are orthonormal, i.e.:
\begin{align}
\innprod{F_{m,n,k}}{F_{m',n',k'}}{\Lpspc{2}{}{\calD}} = \delta_{m,m'}\delta_{n,n'}\delta_{k,k'}\,,
\end{align}
and they satisfy (see \ref{App:Basis_set_derivation}):
\begin{align}
\label{Eq:Laplacian_of_f_mnk}
-\Delta F_{m,n,k} &=\lambda^{0}_{m,n,k}F_{m,n,k}\,,\,\lambda^{0}_{m,n,k}=\left(\frac{b^{n\mathfrak{N}}_k}{R}\right)^2+\left|\frac{2\pi}{\tau}\left(m-\alpha n\mathfrak{N}\right)\right|^2\,.
\end{align}
The above condition implies that the kinetic energy part of the single particle Schr\"odinger operator is diagonalized in this basis. 

Consistent with the literature, we will refer to the representation of a function in terms of its expansion coefficients (i.e., $\hat{\phi}_{m,n,k}$ in the above) as its \textit{reciprocal space representation}. Furthermore, we will refer to the representation of the function in terms of its values on a discrete set of grid points as its \textit{real space representation}. If the basis functions are also available on these same grid points, the real and reciprocal space representations of the function can be connected via eq.~\ref{Eq:Phi_expanded_in_basis_set}. In \ref{App:Gradients} we demonstrate how the gradients of quantities expressed via eq.~\ref{Eq:Phi_expanded_in_basis_set}  may be evaluated.

For common systems of interest, the number of basis functions required for discretizing the governing equations can number in the tens or hundreds of thousands (See Section \ref{sec:results}). Thus, it can become infeasible to explicitly store the discretized Hamiltonian. This scenario is also encountered in the classical plane-wave method for bulk systems \citep{Hutter_abinitio_MD, Martin_ES}, and can be addressed by working with the discretized Hamiltonian implicitly, and using iterative, matrix-free diagonalization techniques to compute the eigenstates \citep{Saad_Chelikowsky_Shontz_review,  Saad_large_eigenvalue_book}. For adopting such strategies, we need to be able to compute  action of the Hamiltonian on an arbitrary vector, such as the wavefunction, as represented in our basis set. To this end, we consider a vector $\phi \in \text{Span}\big\{F_{m,n,k}: {(m,n,k) \in \Gamma}\big\}$, substitute eq.~\ref{Eq:Phi_expanded_in_basis_set} into eq.~\ref{Eq:Expanded_schrodinger}, and use eq.~\ref{Eq:Laplacian_of_f_mnk}, to arrive at:
\begin{align}
\label{eq:Simplify_1}
\nonumber
\left(-\frac{1}{2}\Delta+V\right)\phi=\frac{1}{2}\sum_{\Gamma}\hat{\phi}_{m,n,k}\lambda^0_{m,n,k}F_{m,n,k}-a\left(\alpha,\tau,\eta,\nu\right)\phi-b\left(\alpha,\tau,\eta,\nu\right)\phi_{\theta_1}\\-c\left(\alpha,\tau,\eta,\nu\right)\phi_{\theta_2}+\frac{\nu^2}{2r^2}\phi+\frac{i\nu}{2\pi r^2}\phi_{\theta_2}+V\phi=\lambda\phi\,,
\end{align}
which further simplifies to:
\begin{align}
\label{eq:Simplify_2}
\nonumber
&\frac{1}{2}\sum_{\Gamma}\hat{\phi}_{m,n,k}\lambda^0_{m,n,k}F_{m,n,k}-a\left(\alpha,\tau,\eta,\nu\right)\sum_{\Gamma}\hat{\phi}_{m,n,k}f_{m,n,k} \\\nonumber
&-b\left(\alpha,\tau,\eta,\nu\right)\sum_{\Gamma}\hat{\phi}_{m,n,k}\left(i2\pi m\right)F_{m,n,k}-c\left(\alpha,\tau,\eta,\nu\right)\sum_{\Gamma}\hat{\phi}_{m,n,k}\left(i2\pi n\mathfrak{N}\right)F_{m,n,k}  \\\nonumber
&+\frac{\nu^2}{2}\sum_{\Gamma}\hat{\phi}_{m,n,k}\frac{F_{m,n,k}}{r^2}+\frac{i\nu}{2\pi}\sum_{\Gamma}\hat{\phi}_{m,n,k}\left(\frac{i2\pi n\mathfrak{N}}{r^2}\right)F_{m,n,k}+V\left(r,\theta_1,\theta_2\right)\sum_{\Gamma}\hat{\phi}_{m,n,k}F_{m,n,k}  
\\ &=\lambda\sum_{\Gamma}\hat{\phi}_{m,n,k}F_{m,n,k}\,.
\end{align}
The constants $a, b, c$ in the above are as follows:
\begin{align}
\label{eq:constants_abc}
\begin{split}
a\left(\alpha,\tau,\eta,\nu\right)=\frac{2\pi^2}{\tau^2}\Big\{\nu\alpha\left(2\eta-\nu\alpha\right)-\eta^2\Big\}\,,\\
b\left(\alpha,\tau,\eta,\nu\right)=\frac{2i\pi}{\tau^2}\left(\nu\alpha-\eta\right)\,,
\,c\left(\alpha,\tau,\eta,\nu\right)=\frac{2i\pi\alpha}{\tau^2}\left(\eta-\nu\alpha\right)\,.
\end{split}
\end{align}
The action of the Hamiltonian on the vector $\phi$ is simply the left hand side of eq.~\ref{eq:Simplify_2} above. We observe that due to orthonormality of the basis set, the first four terms on the left hand side are easily handled in reciprocal space. Specifically, the second term is simply a scaling of the input vector $\phi$ with the factor $a\left(\alpha,\tau,\eta,\nu\right)$, while the other three terms can be evaluated as element-wise product operations (Matlab operation $.* $). Thus, these terms can all be evaluated at a cost that scales linearly with the basis set size. The last term on the left hand side is associated with action of the effective potential $V(\bfx)$ on the wavefunction vector.  If the expansion coefficients of the potential are available as:
\begin{align}
\label{eq:V_expansion}
V(\theta_1, \theta_2, r) = \sum_{(\tilde{m},\tilde{n},\tilde{k})\in\Gamma}\!\widehat{V}_{\tilde{m},\tilde{n},\tilde{k}}\,F_{\tilde{m},\tilde{n},\tilde{k}}(\theta_1, \theta_2, r)\,,
\end{align}
then the expansion coefficients of $V(\bfx)\phi(\bfx)$ can be computed as:
\begin{align}
\nonumber
&\Big\langle{V(\theta_1,\theta_2, r)\,\phi(\theta_1,\theta_2, r)\,,\,F_{m',n',k'}(\theta_1,\theta_2, r)\Big\rangle}_{\Lpspc{2}{}{\calD}}
\\\nonumber
=\,&\Bigg\langle{\bigg( \sum_{\Gamma}\!\widehat{V}_{\tilde{m},\tilde{n},\tilde{k}} F_{\tilde{m},\tilde{n},\tilde{k}}(\theta_1, \theta_2, r) \bigg)\,\bigg(\sum_{\Gamma}\hat{\phi}_{m,n,k} F_{m,n,k}(\theta_1, \theta_2, r) \bigg)\,,\,F_{m',n',k'}(\theta_1,\theta_2, r)\Bigg\rangle}_{\Lpspc{2}{}{\calD}}
\\\nonumber
=\,&\Bigg\langle{\bigg( \sum_{\Gamma}\sum_{\Gamma}\!\widehat{V}_{\tilde{m},\tilde{n},\tilde{k}}\,\hat{\phi}_{m,n,k}\,F_{\tilde{m},\tilde{n},\tilde{k}}(\theta_1, \theta_2, r) F_{m,n,k}(\theta_1, \theta_2, r)\bigg)\,,\,F_{m',n',k'}(\theta_1,\theta_2, r)\Bigg\rangle}_{\Lpspc{2}{}{\calD}}\\
=&\sum_{\Gamma}\sum_{\Gamma}\hat{V}_{\tilde{m},\tilde{n},\tilde{k}}\,\hat{\phi}_{m,n,k}\,\big\langle F_{\tilde{m},\tilde{n},\tilde{k}}\,F_{m,n,k}\,,\,F_{m',n',k'}\big\rangle_{\Lpspc{2}{}{\calD}}\,.
\label{Eq:Potential_simplification_wrong}
\end{align}
There are two problems with the above evaluation strategy. First, the time complexity of the procedure scales in a cubic manner with respect to the basis set size, i.e., $O(M_{\text{max}}^3N_{\text{max}}^3K_{\text{max}}^3)$. Moreover, if the coupling coefficients:
\begin{align}
\nonumber
\big\langle &F_{\tilde{m},\tilde{n},\tilde{k}}\,F_{m,n,k}\,,\,F_{m',n',k'}\big\rangle_{\Lpspc{2}{}{\calD}} = c_{\tilde{m},\tilde{n},\tilde{k}}c_{m,n,k}c^{*}_{m',n',k'}\int^{1}_{0}e^{i2\pi(m+\tilde{m}-m')\theta_1}\,d\theta_1 \\ & \times \int^{\frac{1}{\mathfrak{N}}}_{0}e^{i2\pi\mathfrak{N}(n+\tilde{n}-n')\theta_2}\,d\theta_2\int^{R}_{0}J_{n\mathfrak{N}}\left(\frac{b^{n\mathfrak{N}}_{k}r}{R}\right)J_{n'\mathfrak{N}}\left(\frac{b^{n'\mathfrak{N}}_{k'}r}{R}\right)J_{\tilde{n}\mathfrak{N}}\left(\frac{b^{\tilde{n}\mathfrak{N}}_{\tilde{k}}r}{R}\right)2\pi\tau r\,dr\,,
\end{align}
are to be calculated and stored ahead of time for easier evaluation of eq.~\ref{Eq:Potential_simplification_wrong},  the memory complexity of the procedure would also scale cubically with the basis set size. By making use of the fact that the coupling coefficients are non-zero only for $m+\tilde{m} = m'$ and $n+\tilde{n} = n'$, their evaluation, storage and application to eq.~\ref{Eq:Potential_simplification_wrong}, can be somewhat simplified. Despite this, the overall complexity still continues to be cubic in the basis set size in both memory and time. Second, the potential $V(\bfx)$ is generally not expected to be equal to zero at $r = R$ and may be slowly decaying due to long range electrostatics effects. Hence, it is inappropriate to express this quantity in terms of helical waves obeying Dirichlet boundary conditions.

Both of the above issues can be remedied by adopting a pseudospectral evaluation strategy \citep{orszag_1, orszag1980spectral, Spectral_Scheme_Book_Analysis, Spectral_Scheme_Book_Orszag, Banerjee2015spectral}, as we now describe. This is related to the observation that if $V(\bfx)$ and $\phi(\bfx)$ are available in real space, as functions sampled at a common set of grid points, the product $\chi(\bfx) = V(\bfx)\phi(\bfx)$ can be evaluated with a cost proportional to the size of the grid. Thereafter, the function $\chi(\bfx)$ can be directly expanded in terms of helical waves to yield:
\begin{align}
\label{eq:g_mnk}
\widehat{\chi}_{m',n',k'} = \Big\langle{V(\theta_1,\theta_2, r)\,\phi(\theta_1,\theta_2, r)\,,\,f_{m',n',k'}(\theta_1,\theta_2, r)\Big\rangle}_{\Lpspc{2}{}{\calD}}\,.
\end{align}
Since $\chi(\bfx)$ obeys Dirichlet boundary conditions and inherits all symmetries of the group $\calG$, its expansion using helical waves is appropriate. To put this strategy into practice however, we need access to fast basis transforms so that functions expressed in reciprocal space (i.e., as expansion coefficients) and real space (i.e., on the grid), may be readily interconverted. We describe the formulation and implementation of such transform routines in Sections \ref{Sec:Inv_FFT} and \ref{Sec:Fwd_FFT}. The  overall computational cost of this strategy is the sum total of the costs of the forward and inverse transforms, and the cost of carrying out the real space product. Theoretically, the transforms described here scale in a manner that is slightly worse than the basis set size. However, as we show later, in practice they scale  more favorably, in a sub-linear manner (see Fig.~\ref{Fig:Fast_vs_Naive_transforms}). Furthermore, the real space grid size is usually a constant multiple of the basis set size, leading to the overall cost of the pseudospectral method scaling in a manner that is close to the first power of this quantity ($=\calO(M_{\text{max}}N_{\text{max}}K_{\text{max}}^2)$). The memory complexity is also reduced and scales as the basis set size itself, i.e., $\calO(M_{\text{max}}N_{\text{max}}K_{\text{max}})$.

Finally, we discuss the evaluation of the fifth and the sixth terms on the left hand side of eq.~\ref{eq:Simplify_2}. The fifth term, i.e.,
\begin{align}
\ell(\theta_1,\theta_2, r) = \frac{\nu^2}{2}\sum_{\Gamma}\hat{\phi}_{m,n,k}\frac{F_{m,n,k}}{r^2}\,,
\end{align}
satisfies $\ell(\theta_1,\theta_2, r = R) = 0$ and invariance under $\calG$, since it is a finite linear combination of terms which individually obey these conditions. Thus, the expansion coefficients are:
\begin{align}
\nonumber
&\widehat{\ell}_{m',n',k'} = \Big\langle{\ell(\theta_1,\theta_2, r)\,,\,F_{m',n',k'}(\theta_1,\theta_2, r)\Big\rangle}_{\Lpspc{2}{}{\calD}}\\\nonumber
&=\frac{\nu^2}{2}\sum_{\Gamma}\hat{\phi}_{m,n,k}\int^{1}_{0}\int^{\frac{1}{\mathfrak{N}}}_{0}\int^{R}_{0}\frac{F_{m,n,k}F^{*}_{m',n',k'}}{r^2}\,2\pi\tau r\,dr\,d\theta_2\,d\theta_1 \\\nonumber
&= \nu^2\pi\tau\sum_{\Gamma}\hat{\phi}_{m,n,k}c_{m,n,k}c^{*}_{m',n',k'}\Bigg[\int^{1}_{0}e^{i2\pi(m-m')\theta_1}d\theta_1\times& \\\nonumber &\hspace{12em}\int^{\frac{1}{\mathfrak{N}}}_{0}e^{i2\pi\mathfrak{N}(n-n')\theta_2}d\theta_2\int^{R}_{0}\frac{J_{n\mathfrak{N}}\left(\frac{b^{n\mathfrak{N}}_{k}r}{R}\right)J_{n'\mathfrak{N}}\left(\frac{b^{n'\mathfrak{N}}_{k'}r}{R}\right)}{r^2}rdr \Bigg]\\
&= \frac{\nu^2}{R^2}\sum^{K_{\text{max}}}_{k=1}\!\hat{\phi}_{m',n',k}\frac{1}{J_{n'\mathfrak{N}+1}\left(b^{n'\mathfrak{N}}_{k}\right)J_{n'\mathfrak{N}+1}\left(b^{n'\mathfrak{N}}_{k'}\right)}\int^{R}_{0}\frac{J_{n'\mathfrak{N}}\left(\frac{b^{n'\mathfrak{N}}_{k}r}{R}\right)J_{n'\mathfrak{N}}\left(\frac{b^{n'\mathfrak{N}}_{k'}r}{R}\right)}{r^2}rdr\,.
\label{ell_coeff}
\end{align}
In the above, we have made use of the orthonormality of the complex exponentials in the $\theta_1$ and $\theta_2$ directions. We may rewrite eq.~\ref{ell_coeff} as:
\begin{align}
\widehat{\ell}_{m',n',k'} = \frac{\nu^2}{R^2}\sum^{K_{\text{max}}}_{k=1} \hat{\phi}_{m',n',k}\,\calI(n',k,k')\,,
\label{eq:ell_mnk}
\end{align}
with:
\begin{align}
\calI(n',k,k') = \frac{1}{J_{n'\mathfrak{N}+1}\left(b^{n'\mathfrak{N}}_{k}\right)J_{n'\mathfrak{N}+1}\left(b^{n'\mathfrak{N}}_{k'}\right)}\int^{1}_{0}\frac{J_{n'\mathfrak{N}}\left(b^{n'\mathfrak{N}}_{k}\;q\right)J_{n'\mathfrak{N}}\left(b^{n'\mathfrak{N}}_{k'}\;q\right)}{q^2}\,q\,dq\,.
\label{eq:I_nkk_prime}
\end{align}
Thus, if the quantities $\calI(n',k,k')$ are known ahead of time, the coefficients $\ell_{m',n',k'}$ can be readily evaluated at a cost of $\calO(M_{\text{max}}N_{\text{max}}K_{\text{max}}^2)$, i.e., quite close to the overall basis set size, and similar in computational complexity to the evaluation of the potential term. Since $\calI(n',k,k')$ is problem independent (e.g., it has no dependence on $R$, $\alpha$, $\tau$ or the potential $V(\bfx)$), we may evaluate and store it as a table for a large range of values of $n, k$ and $k'$. During program execution, this table of values may be loaded into memory, and each $\ell_{m',n',k'}$ can be evaluated as a vector dot product (eq.~\ref{eq:ell_mnk}), after accessing the necessary values of $\calI(n',k,k')$. As for the evaluation of the $\calI(n',k,k')$ values themselves, we may use the recurrence relation \citep{abramowitz1988handbook}: 
\begin{align}
\frac{2\kappa}{q} J_{\kappa} (q) = J_{\kappa - 1} (q) +  J_{\kappa + 1} (q)\,,
\end{align}
to rid the integrand in eq.~\ref{eq:I_nkk_prime} of it's denominator, and obtain a pair of oscillatory integrals. We may then compute these by using Gauss-Jacobi quadrature as outlined in eq.~\ref{App:Oscillatory_integrals}.

In an analogous manner, the sixth term on the left hand side of eq.~\ref{eq:Simplify_2}, i.e.,
\begin{align}
\frac{i\nu}{2\pi}\sum_{\Gamma}\hat{\phi}_{m,n,k}\left(i2\pi n\mathfrak{N}\right)\int^{1}_{0}\int^{\frac{1}{\mathfrak{N}}}_{0}\int^{R}_{0} \frac{F_{m,n,k}F^{*}_{m',n',k}}{r^2} 2\pi\tau r\,dr\,d\theta_2\,d\theta_1\,,
\end{align}
can be simplified to:
\begin{align}
=\frac{i\nu}{\pi R^2}\sum^{K_{\text{max}}}_{k=1}\left(\hat{\phi}_{m',n',k} * i2\pi n'\mathfrak{N}\right)\calI(n',k,k')\,.
\end{align}
With the quantities $\calI(n',k,k')$ available, the above can be evaluated in a manner similar to the evaluation of the fifth term, at a computational cost of $\calO(M_{\text{max}}N_{\text{max}}K_{\text{max}}^2)$. The key difference is that instead of the vector $\{\hat{\phi}_{m,n,k}\}_{(m,n,k) \in \Gamma}$, we need to consider an alternate one with entries $\{ i2\pi n'\mathfrak{N}\hat{\phi}_{m,n,k}\}_{(m,n,k) \in \Gamma}$. However, this modified vector is already available as part of evaluation of the fourth therm on the left hand side of eq.~\ref{eq:Simplify_2}, and therefore, it can be reused. 
\section{Numerical Implementation}
\label{sec:MATLAB}
We have implemented the above computational strategies into a MATLAB \citep{MATLAB:2019} package called HelicES (\textbf{Helic}al \textbf{E}lectronic \textbf{S}tructure). To ensure efficiency, our code heavily relies on vectorization features of MATLAB. Various details of our implementation are as follows. 
\subsection{Wave function storage: reciprocal and real space considerations}
\label{subsec:storage}
For any quantity in reciprocal space, there are three indices $m,n,k$ associated with each expansion coefficient, making the collection of coefficients a three-dimensional object of dimensions $(2M_{\text{max}}+1)\times (2N_{\text{max}}+1)\times K_{\text{max}}= \calL$ . However, it is easier for linear algebra operations to have these coefficients stacked up as vector in $\cz^{\calL}$. To achieve this, we adopt the following mapping between $(m,n,k) \in \Gamma$ and the linear index $\mathfrak{i} \in \{1,2,\ldots, \calL\}$:
\begin{align}
\mathfrak{i}(m,n,k) = (m + M_{\text{max}})*(2N_{\text{max}} + 1)*K_{\text{max}} + (n + N_{\text{max}})*K_{\text{max}} + k\,.
\label{eq:wavefun_storage}
\end{align}
With this, a collection of $N_{\text{s}}$ wavefunctions can be stored as a complex matrix of dimensions $\calL \times N_{\text{s}}$.

For real space representation, the number of grid points to be chosen along each helical coordinate $\theta_1,\theta_2$ is dictated by the accuracy of the basis transforms (see Section \ref{subsec:fast_transforms}). We choose to work with Fourier nodes along the $\theta_1$ and $\theta_2$ directions and denote the corresponding number of grid points as ${N_{\theta_1}}$ and ${N_{\theta_2}}$ respectively. 
Along the radial direction, we choose $N_{r}$ Gauss-Jacobi nodes \citep{Gauss_Jacobi_Quad} over the interval $[0,R]$. This has the  advantage that the coordinate singularity at the origin is automatically avoided. In order to accommodate non-linearities and to reduce aliasing errors \citep{bylaska2011large, Banerjee2015spectral}, we typically choose $N_{\theta_1} = 4*M_{\text{max}}+1$, $N_{\theta_2} = 4*N_{\text{max}}+1$ and $N_{r} = 4 * K_{\text{max}}$. These choices generally allow transforms to be evaluated accurately up to machine precision. With this setup, quantities such as the wavefunction are available in real space over a three-dimensional grid (i.e., the tensor product grid resulting from the one-dimensional grids along the individual coordinate directions), and each grid point is indexed via $\textsf{i} \in \{1,2,\ldots, N_{\theta_1}\}$, $\textsf{j} \in \{1,2,\ldots, N_{\theta_2}\}$ and $\textsf{k} \in \{1,2,\ldots, N_{r}\}$. For storage, we stack this three dimensional representation into a complex column vector of size $N_{\theta_1}*N_{\theta_2}*N_{r}$, for which we use the following ordering:
\begin{align}
\mathfrak{j}(\mathsf{i},\mathsf{j}, \mathsf{k})  = (\mathsf{i} - 1)*N_{r}*N_{\theta_2} + (\mathsf{j} - 1)*N_{r} + \mathsf{k}\,.
\label{eq:wavefun_storage_real}
\end{align}
Since the memory requirement for storage of each wavefunction in real space is much higher than storing it in reciprocal space, we typically avoid storing real space versions of all $N_{\text{s}}$ wave functions simultaneously.
\subsection{Imposition of kinetic energy cutoff}
\label{sec:Energy_cutoff}
In conventional plane-wave calculations, it is common to specify a \textit{kinetic energy cutoff}, i.e., a limit on the $H^1$ Sobolev norm of the plane-waves to be used for discretization \citep{Hutter_abinitio_MD, Cances_planewave_numerical_analysis}. Once a suitable periodic unit cell has been identified, this criterion automatically provides a recipe for calculating the number of planewaves along each of the Cartesian axes, and in turn, the dimensions of the underlying real space grid to be used for Fast Fourier Transforms (FFTs). In a similar vein, we may wish to retain only helical waves with kinetic energies below a pre-specified cutoff in our calculation, since this has the advantage that the  basis set limits $M_{\text{max}}$, $N_{\text{max}}$, and $K_{\text{max}}$ get specified automatically in proportion to the computational domain's geometry parameters. At the gamma point ($\eta = 0, \nu = 0$) for example, the kinetic energy cutoff criterion requires that all helical waves $f_{m,n,k}$, with $m,n,k$ values satisfying:
\begin{align}
\label{Eq:Energy_cutoff}
\frac{1}{2}\lambda^{0}_{m,n,k} = \half \bigg[ \left(\frac{b^{n\mathfrak{N}}_k}{R}\right)^2+\left|\frac{2\pi}{\tau}\left(m-\alpha n\mathfrak{N}\right)\right|^2 \bigg ] \leq E_{\text{cut}}\,,
\end{align}
be included in our calculations. In our implementation, we first determine the largest absolute values of integers $m,n$ and the largest natural number $k$ consistent with  with eq.~\ref{Eq:Energy_cutoff}. We set the basis set limits $M_{\text{max}}$, $N_{\text{max}}$, and $K_{\text{max}}$ accordingly. The real space grids used for carrying out fast transforms (described below) are chosen based on these quantities. Within these $(2M_{\text{max}} + 1)*(2N_{\text{max}} + 1)*K_{\text{max}}$ helical waves, however, not every combination of $m,n,k$ would satisfy the kinetic energy criterion. To remedy this, we create a \textit{masking vector} to exclusively retain helical waves which satisfy eq.~\ref{Eq:Energy_cutoff}, in various operations of interest (such as the Hamiltonian times wavefunction products). Based on the linear ordering for reciprocal space storage outlined in eq.~\ref{eq:wavefun_storage}, we may express the masking vector as:
\begin{align}
\nonumber
\calM_{\mathfrak{i}(m,n,k)} &= 1,\;\text{for}\;\frac{1}{2}\lambda^{0}_{m,n,k}  \leq E_{\text{cut}}\\
&= 0,\;\text{otherwise.}
\end{align}
Element-wise multiplication of a given vector with the masking vector results in only kinetic energy limited helical waves being retained in the calculation. We implement the above strategy at each $\eta,\nu$ point (with the expression for the kinetic energy modified appropriately) to impose the kinetic energy cutoff in HelicES.
\subsection{\texorpdfstring{${\eta}\text{-}$}{eta}space discretization and parallelization}
\label{subsec:reciprocal_space}
As described earlier, to obtain the helical Bloch states (eq.~\ref{eq:Bloch_theorem_1}), i.e., solutions to the single electron problem with a symmetry adapted potential (eq.~\ref{eq:Potential_Symmetry}), the single electron Hamiltonian has to be diagonalized for every $\eta \in [-\half,\half)$ and $\nu \in \{0, {1},2, \ldots,{\mathfrak{N}-1}\}$. To make this calculation feasible,  we sample $\eta$ over a discrete set $\{{\eta_b}\}_{b=1}^{N_{\eta}} \subset [-\half,\half)$. The specific choice of the values $\eta_b$ is based on the Monkhorst-Pack scheme \citep{monkhorst1976special}. This procedure akin to ``k-point sampling'' in conventional periodic codes \citep{Martin_ES}. With this choice, the Hamiltonian needs to be diagonalized at $N_{\calK} = N_{\eta} \times \mathfrak{N}$ points, and integrals  in $\eta$ can be calculated via quadrature:
\begin{align}
\int_{-\half}^{\half} p(\eta)\,d\eta \approx \sum_{{b=1}}^{N_{\eta}}\,w_b\,p(\eta_b)\,.
\end{align}
Here, $\{{w_b}\}_{b=1}^{N_{\eta}}$ are the Monkhorst-Pack quadrature weights and are uniformly equal to $1/N_{\eta}$. Integrals of the above kind appear, for example, while computing the electronic band energy, or the electron density from helical Bloch states \citep{banerjee2021ab, yu2022density}.

If the single electron Hamiltonian does not include magnetic fields --- as is the case here, time reversal symmetry allows further reduction in the number of $\eta,\nu$ points at which the Hamiltonian has to be diagonalized \citep{geru2018time, ghosh2019symmetry}. Specifically, it holds that for any $\eta \in [-\half,\half)$ the helical Bloch states and the associated electronic bands obey:
\begin{equation}
\left.\begin{aligned}
\psi_j(\bfx; \eta,\nu) &= \overline{\psi_j(\bfx;-\eta,\mathfrak{N}-\nu)}\\
\lambda_j(\eta,\nu) &= \lambda_j(-\eta,\mathfrak{N}-\nu) 
\end{aligned}
\,\right\}
\;\text{for}\,\nu \in \{0, {1},2, \ldots,{\mathfrak{N}-1}\}\,,
\end{equation}
and:
\begin{equation}
\left.\begin{aligned}
\psi_j(\bfx; \eta,0) &= \overline{\psi_j(\bfx;-\eta,0)}\\
\lambda_j(\eta,0) &= \lambda_j(-\eta,0)
\end{aligned}
\,\right\}
\;\text{for}\,\nu = 0\,.
\end{equation}
Overall, this reduces the number of discrete points in reciprocal space by a factor of $2$.

Since the diagonalization problem arising from distinct sets of $\eta,\nu$ values are independent of one another, they can be dealt with in an embarrassingly parallel manner.  In our implementation, we have used MATLAB's Parallel Computing Toolbox (specifically, the $\textsf{parfor}$ function) to carry out this parallelization.
\subsection{Fast basis transforms}
\label{subsec:fast_transforms}
Since our strategy for carrying out Hamiltonian matrix-vector products involves fast basis transforms, we now elaborate on various aspects of the implementation of such operations within the HelicES code. To arrive at fast transforms, we exploit the separability of the basis functions into radial and $\theta_1,\theta_2$ dependence. This allows us to make use of quadrature along the radial direction, and subsequently, efficient two-dimensional fast Fourier transforms (FFTs) along the $\theta_1,\theta_2$ directions for each radial grid point, or for each radial basis function. Since the radial part of the basis functions consists of Bessel functions, we have also investigated the use of Hankel and discrete Bessel transforms \citep{lemoine1994discrete, johansen1979fast, bisseling1985fast}. However, we found that the quadrature approach adopted here resulted in better performance for the basis set sizes considered, consistent with some earlier studies \citep{key2012fast}.

In what follows, $O_{M\times N}$ is used to denote a zero matrix of size $M\times N$. The typical real space grid point for sampling a function will be denoted as $(\theta_{1}^{\textsf{i}}, \theta_{2}^{\textsf{j}}, r^{\textsf{k}})$, with $\textsf{i} \in \{1,2,\ldots, N_{\theta_1}\}$, $\textsf{j} \in \{1,2,\ldots, N_{\theta_2}\}$ and $\textsf{k} \in \{1,2,\ldots, N_{r}\}$. We will use the MATLAB commands \textsf{ifft2} and \textsf{fft2} to denote two-dimensional fast inverse and forward Fourier Transforms respectively \citep{MATLAB:2019}. Additionally, we will use the MATLAB commands \textsf{ifftshift} and  \textsf{fftshift} to denote rearrangements of Fourier transform coefficients related to shifting of zero frequency components to matrix center \citep{MATLAB:2019}. Finally, we will use the MATLAB `:' (colon) operator notation to denote array/matrix indices with regular increments. Thus, $\mathtt{i}:\mathtt{k}:\mathtt{j}$ will denote indices starting at $\mathtt{i}$, incremented by $\mathtt{k}$ and ending at $\mathtt{j}$, $\mathtt{i}:\mathtt{j}$ will denote the indices $\mathtt{i}, \mathtt{i}+1, \mathtt{i}+2, \ldots, \mathtt{j}-1, \mathtt{j}$, and simply `:' will denote all indices along a particular matrix dimension.
\subsubsection{Fast Inverse Basis Transform}
\label{Sec:Inv_FFT}
Given the expansion coefficients $\{\hat{g}_{m,n,k}\}_{(m,n,k) \in \Gamma}$, a na{i}ve way of implementing the inverse basis transform would be to calculate each basis function $\{F_{m,n,k}\}_{(m,n,k) \in \Gamma}$ at every grid point $(\theta_{1}^{\textsf{i}}, \theta_{2}^{\textsf{j}}, r^{\textsf{k}})$, and to then evaluate the sum:
\begin{align}
\label{Eq:Naive_Inverse_Transform}
g(\theta_{1}^{\textsf{i}}, \theta_{2}^{\textsf{j}}, r^{\textsf{k}})=\sum_{\Gamma}\hat{g}_{m,n,k}F_{m,n,k}(\theta_{1}^{\textsf{i}}, \theta_{2}^{\textsf{j}}, r^{\textsf{k}})\,.
\end{align}
The computational complexity of this ``naive inverse transform'' is easily seen to be $\calO(M_{\text{max}}N_{\theta_1} N_{\text{max}}N_{\theta_2}K_{\text{max}}N_{r})$, which simplifies to  $\calO(M_{\text{max}}^2N_{\text{max}}^2K_{\text{max}}^2)$. The constant involved in the latter estimate can be seen to be quite large based on the discussion in Section \ref{subsec:storage}. To remedy this situation, we express the basis functions as in \ref{App:Basis_set_derivation}, i.e., $F_{m,n,k}(\theta_1,\theta_2, r)=e^{i2\pi(m\theta_1 + n\mathfrak{N}\theta_2)}\,\xi_{n,k} (r)$ and also rewrite the eq.~\ref{Eq:Naive_Inverse_Transform} as:
\begin{align}
\nonumber
g(\theta_{1}^{\textsf{i}}, \theta_{2}^{\textsf{j}}, r^{\textsf{k}})&=\sum_{m=-M_{\text{max}}}^{M_{\text{max}}}\;\sum_{n=-N_{\text{max}}}^{N_{\text{max}}}\;\sum_{k=1}^{K_{\text{max}}}\,\hat{g}_{m,n,k}\,e^{i2\pi(m\,\theta_1^{\textsf{i}} + n\mathfrak{N}\,\theta_2^{\textsf{j}})}\,\xi_{n,k} (r^{\textsf{k}})\\
&= \sum_{m=-M_{\text{max}}}^{M_{\text{max}}}\;\sum_{n=-N_{\text{max}}}^{N_{\text{max}}}\;e^{i2\pi(m\,\theta_1^{\textsf{i}} + n\mathfrak{N}\,\theta_2^{\textsf{j}})}\;\bigg(\sum_{k=1}^{K_{\text{max}}}\,\hat{g}_{m,n,k}\,\,\xi_{n,k} (r^{\textsf{k}})\bigg)\,.
\end{align}
Since the quantity in parentheses is independent of the basis function index $k$, we may rewrite the above as:
\begin{align}
\begin{split}
g(\theta_{1}^{\textsf{i}}, \theta_{2}^{\textsf{j}}, r^{\textsf{k}}) &=\sum_{m=-M_{\text{max}}}^{M_{\text{max}}}\;\sum_{n=-N_{\text{max}}}^{N_{\text{max}}}e^{i2\pi(m\,\theta_1^{\textsf{i}} + n\mathfrak{N}\,\theta_2^{\textsf{j}})}\;G_{m,n}(r^{\textsf{k}})\,\\
\text{with}:\quad G_{m,n}(r^{\textsf{k}}) &= \sum_{k=1}^{K_{\text{max}}}\,\hat{g}_{m,n,k}\,\,\xi_{n,k} (r^{\textsf{k}})\,.
\end{split}
\end{align}
Thus, if the quantities $G_{m,n}(r^{\textsf{k}})$ are known, calculation of the inverse basis transform amounts to computing an inverse two-dimensional fast Fourier transform at each radial grid point $r^{\textsf{k}}$. Additionally, we observe that at each radial grid point $r^{\textsf{k}}$, $G_{m,n}(r^{\textsf{k}})$ can be expressed as a vector dot product between two $K_{\text{max}}$ dimensional vectors, i.e., $\{\hat{g}_{m,n,k}\}_{k=1}^{K_{\text{Max}}}$  and $\{\xi_{n,k} (r^{\textsf{k}})\}_{k=1}^{K_{\text{Max}}}$. In fact, by grouping the evaluation of $G_{m,n}(r^{\textsf{k}})$ for different grid points together, the above operation may be expressed as the product of a $N_r\times K_{\text{max}}$ matrix with a $K_{\text{max}}$ dimensional vector, which allows for the use of Level-2 BLAS \citep{blackford2002updated} operations. If the radial part of the basis functions (i.e., $\xi_{n,k} (r^{\textsf{k}})$) are available ahead of time, the above steps provide a convenient recipe of computing the inverse basis transforms with computational complexity $\calO\left(M_{\text{max}}N_{\text{max}}K_{\text{max}}\left(K_{\text{max}}+\log\left(M_{\text{max}}\right)+\log\left(N_{\text{max}}\right)\right)\right)$, a significant improvement over the naive algorithm discussed earlier. We outline the overall steps of our implementation in Algorithms \ref{Algo:Fast_Inverse_Transform} and \ref{Algo:Fast_theta1_theta2_inv_transform} below, and also illustrate some key aspects through Fig.~\ref{Fig:Fast_theta1_theta2_inv_transform}.

\begin{figure}[ht]
    \centering
    \includegraphics[width=0.8\textwidth]{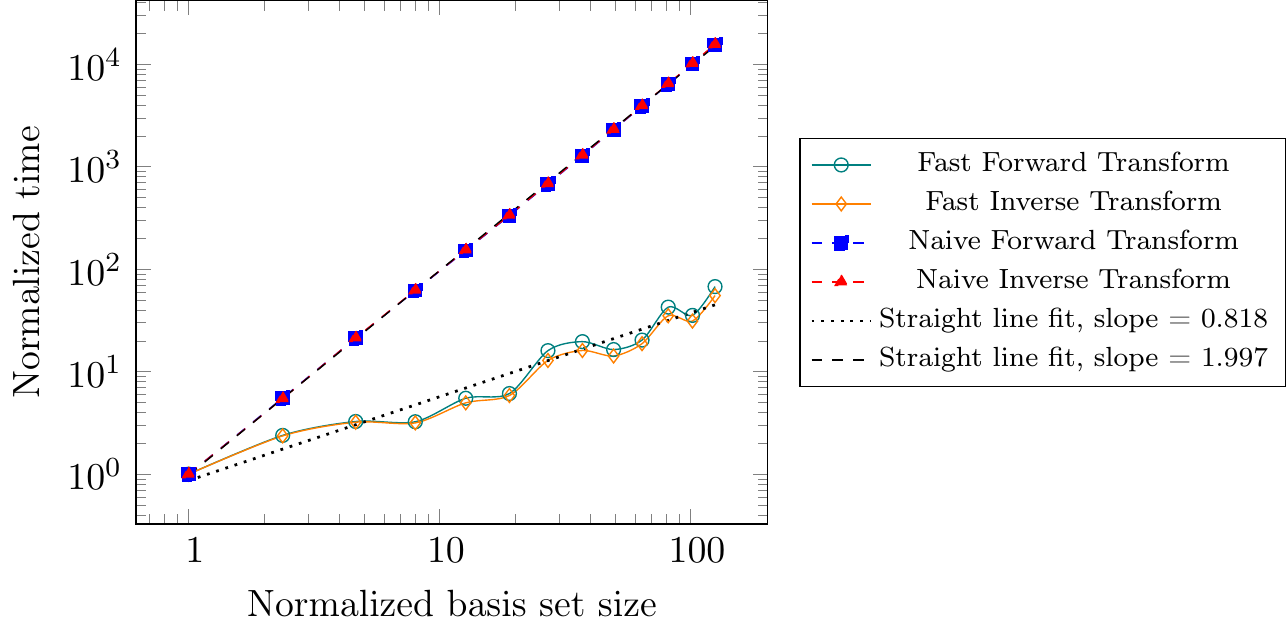}
    \caption{Variation of the normalized time for basis transforms plotted against the basis set size. Both axes are logarithmic. Straight lines were fit using the average of the forward and inverse transform times in each case.}
\label{Fig:Fast_vs_Naive_transforms}
\end{figure}

In Fig.~\ref{Fig:Fast_vs_Naive_transforms} we compare the naive and fast inverse transforms as implemented in HelicES. The starting vectors $\{\hat{g}_{m,n,k}\}_{(m,n,k)\in\Gamma}$ were randomly chosen for the tests. The results from these two methods always agreed with each other to machine precision, guaranteeing consistency of the implementations. However, consistent with the discussion above, the computational time for the naive transforms is found to scale in a quadratic manner with the basis set size, while for the fast transforms, it is close to being linear. The fact that the observed scaling of our fast transform implementation is actually \textit{sublinear}, is almost certainly related to our use of machine optimized linear algebra and Fourier transform routines as available within MATLAB.

\scalebox{0.8}{
\begin{algorithm}[H]
\label{Algo:Fast_Inverse_Transform}
\SetAlgoLined
\caption{Fast Inverse Basis Transform}
\textbf{Input}: The vector of expansion coefficients $\{\hat{g}_{m,n,k}\}\in\mathbb{C}^{\mathcal{L}}$ \\
\underline{Prerequisite:} The radial basis functions sampled on the grid $\{r^\textsf{k}\}_{k=1}^{N_r}$,  \\
 $\qquad\qquad\qquad$ i.e., for each integer $n \in [-N_{\text{max}},N_{\text{max}}]$, the matrix:\\
 $\qquad\qquad\qquad$ $\mathfrak{R}_n =\begin{pmatrix}\xi_{n,1}(r^{\mathsf{1}}) & \ldots & \xi_{n,K_{\text{max}}}(r^{\mathsf{1}}) \\
\vdots & \ddots & \vdots \\
\xi_{n,1}(r^{N_r}) & \ldots & \xi_{n,K_{\text{max}}}(r^{N_r}) \end{pmatrix}$ \\
$-------------------------------------------$\\
Initialize $g=O_{N_{r}N_{\theta_1}N_{\theta_2}\times 1}$, $G = O_{(2M_{\text{max}} +1)(2N_{\text{max}} +1)N_r \times 1}$ \\
Initialize $i,j, p=1$ \\
\For{$m,n\in\Gamma$}{
 Calculate $G(i:i+N_r-1) = \mathfrak{R}_n * \hat{g}_{m,n,k}(j:j+K_{\text{max}}-1)$\\
 $i = i+ N_r$\\
 $j = j+ K_{\text{max}}$
}
\For{$p\leq N_r$}{
 Set $\hat{v}=G(p:N_r:\text{end})$ \\
 $g(p:N_r:\text{end})=\;$AngularInverseTransform$(\hat{v})$
}
\KwResult{The inverse basis transform $g\left(\theta_1,\theta_2, r\right)$ of the vector $\{\hat{g}_{m,n,k}\}$}
\end{algorithm}
}

\scalebox{0.8}{
\begin{algorithm}[H]
\label{Algo:Fast_theta1_theta2_inv_transform}
\SetAlgoLined
\caption{AngularInverseTransform \\(Fast Angular 2D Inverse Fourier Transform)}
\textbf{Input}: Vector $\hat{v}\in\mathbb{C}^{(2M_{\text{max}}+1)*(2N_{\text{max}}+1)}$ \\
Initialize $\mathcal{W}=O_{N_{\theta_1}\times N_{\theta_2}}$ $\quad\slash\slash\,$\textsf{Note:} $N_{\theta_1} = 4M_{\text{max}}+1$, $N_{\theta_2} = 4N_{\text{max}}+1$\\
Reshape $\hat{v}$ into a matrix $\calV \in \cz^{(2N_{\text{max}}+1)\times (2M_{\text{max}}+1)}$  \\
$\mathcal{W}(N_{\text{max}}+1:3N_{\text{max}}+1,\;M_{\text{max}}+1:3M_{\text{max}}+1)=\calV$ \\
$\calU=\;$\textsf{ifft2}$($\textsf{ifftshift}$(\mathcal{W}))$) \\
Scale $\calU=\calU *(N_{\theta_1}*N_{\theta_2})$ \\
Reshape $\calU$ into a column vector $v \in \cz^{N_{\theta_1}*N_{\theta_2}}$ \\
\KwResult{The angular inverse Fourier transform $v(\theta_1,\theta_2)$ of the vector $\hat{v}$}
\end{algorithm}
}
\begin{figure}[ht]
    \centering
    \includegraphics[width=\textwidth]{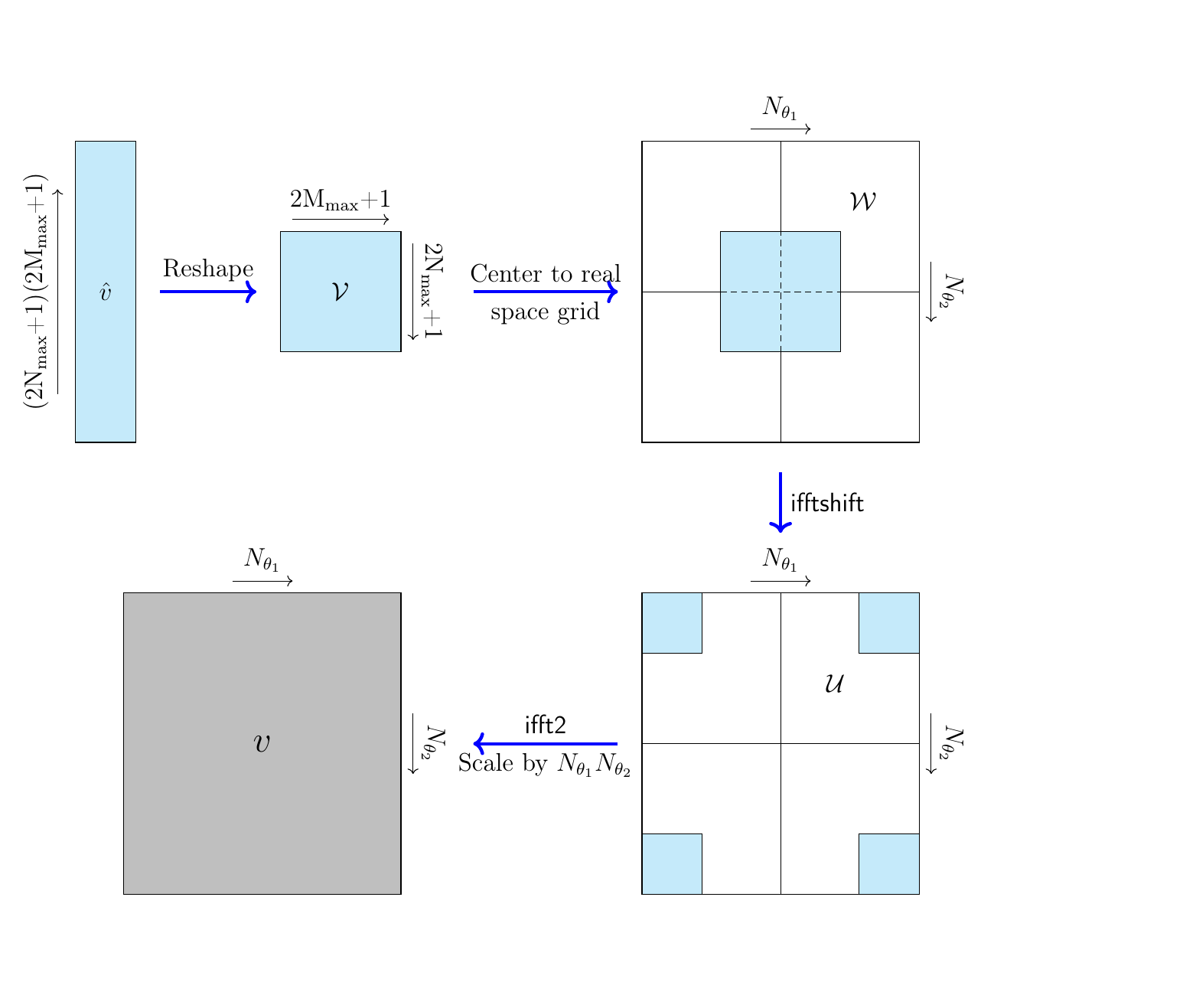}
    \caption{Pictorial representation of the workings of the fast angular 2D inverse Fourier transform (Algorithm \ref{Algo:Fast_theta1_theta2_inv_transform})}
\label{Fig:Fast_theta1_theta2_inv_transform}
\end{figure}

\subsubsection{Fast Forward Basis Transform}
\label{Sec:Fwd_FFT}
We now discuss the implementation of forward basis transforms within HelicES. Given a function $g(\theta_1,\theta_2,r)$, the forward basis transform is defined as:
\begin{align}
\hat{g}_{m,n,k}&= \innprod{g}{F_{m,n,k}}{\Lpspc{2}{}{\calD}} =  \int^{1}_{0}\int^{\frac{1}{\mathfrak{N}}}_{0}\int^{R}_{0}g(\theta_1,\theta_2, r)\,F^{*}_{m,n,k}(\theta_1,\theta_2, r)\,2\pi\tau r\,dr\,d\theta_{2}\,d\theta_1\,.
\end{align}
With $F_{m,n,k}$ and $g$ both sampled on the real space grid, this can be approximated via quadrature as:
\begin{align}
\label{Eq:Fast_fwd_transform_cont}
\hat{g}_{m,n,k}\approx 2\pi\tau*\bigg(\sum^{N_{\theta_1}}_{\mathsf{i}=1}\sum^{N_{\theta_2}}_{\mathsf{j}=1}\sum^{N_{r}}_{\mathsf{k}=1}g(\theta_{1}^{\textsf{i}}, \theta_{2}^{\textsf{j}}, r^{\textsf{k}})e^{-i2\pi(m\,\theta_1^{\textsf{i}} + n\mathfrak{N}\,\theta_2^{\textsf{j}})}\,\xi_{n,k}(r^{\textsf{k}})\,\omega^{\textsf{k}}_r\,\omega_{\theta_1}^{\mathsf{i}}\omega_{\theta_2}^{\mathsf{j}} \bigg)\,.
\end{align}
Here, the quadrature weights along the $\theta_1,\theta_2$ directions are constants, i.e., $\omega_{\theta_1}^{\mathsf{i}} = 1/N_{\theta_1}$ and $\omega_{\theta_2}^{\mathsf{j}} = 1/(\mathfrak{N}N_{\theta_2})$, due to the use of Fourier nodes (i.e., trapezoidal rule). The radial weights $\{\omega^{\mathsf{k}}_r\}_{\mathsf{k}=1}^{N_{r}}$ correspond to Gauss-Jacobi quadrature. We can see that like the case of the inverse transforms, a naive implementation of the above expression will lead to a computational complexity of $\calO(M_{\text{max}}^2N_{\text{max}}^2K_{\text{max}}^2)$. Instead, we deal with the evaluation of this expression along the $\theta_1,\theta_2$ directions simultaneously at each radial grid point using 2D FFTs, and then perform quadrature in the radial direction. Thus, we compute:
\begin{align}
\label{Eq:Fast_fwd_transform_cont_3}
H_{m,n}(r^{\textsf{k}})=\frac{1}{N_{\theta_1}N_{\theta_2}}*\bigg(\sum^{N_{\theta_1}}_{\mathsf{i}=1}\sum^{N_{\theta_2}}_{\mathsf{j}=1}g(\theta_{1}^{\textsf{i}}, \theta_{2}^{\textsf{j}}, r^{\textsf{k}})e^{-i2\pi(m\,\theta_1^{\textsf{i}} + n\mathfrak{N}\,\theta_2^{\textsf{j}})}\bigg)\,,
\end{align}
followed by:
\begin{align}
\label{Eq:Fast_fwd_transform_cont_2}
\hat{g}_{m,n,k}= \frac{2\pi\tau}{\mathfrak{N}}*\bigg(\sum^{N_{r}}_{\mathsf{k}=1}H_{m,n}(r^{\textsf{k}})\,\xi_{n,k}(r^{\textsf{k}})\,\omega^{\textsf{k}}_r\bigg)\,.
\end{align}
The radial quadratures in the above expression can be conveniently cast in terms of Level-2 BLAS \citep{blackford2002updated} operations if the radial basis functions scaled by the corresponding quadrature weights (i.e. $\{\omega^{\textsf{k}}_r\,\xi_{n,k} (r^{\textsf{k}})\}_{k=1}^{K_{\text{Max}}}$) are available as a matrix ahead of time. We outline the steps of our implementation in Algorithms \ref{Algo:Fast_Forward_Transform} and \ref{Algo:Fast_theta1_theta2_fwd_transform} below, and illustrate key aspects in Figure \ref{Fig:Fast_theta1_theta2_fwd_transform}.

\vspace{0.5cm}
\scalebox{0.8}{
\begin{algorithm}[H]
\SetAlgoLined
\caption{Fast Forward Basis Transform}
\textbf{Input}: Real space representation of function $\{g(\theta_{1}^{\textsf{i}}, \theta_{2}^{\textsf{j}}, r^{\textsf{k}})\} \in \cz^{N_{r}N_{\theta_1}N_{\theta_2}}$\\
\underline{Prerequisite:} The radial basis functions sampled on the grid $\{r^\textsf{k}\}_{k=1}^{N_r}$, scaled by\\
 the corresponding quadrature weights i.e., for each integer $n \in [-N_{\text{max}},N_{\text{max}}]$,\\
the following matrix:\\ 
 $\mathfrak{O}_n =\begin{pmatrix}\omega^{\mathsf{1}}_r\,\xi_{n,1}(r^{\mathsf{1}}) & \ldots & \omega^{{N_r}}_r\,\xi_{n,1}(r^{N_r}) \\
\vdots & \ddots & \vdots \\
\omega^{\mathsf{1}}_r\,\xi_{n,K_{\text{max}}}(r^{\mathsf{1}}) & \ldots & \omega^{{N_r}}_r\,\xi_{n,K_{\text{max}}}(r^{N_r}) \end{pmatrix}$ \\
$-------------------------------------------$\\
Initialize $\hat{g}=O_{\calL\times 1}$, $H = O_{N_r \times (2M_{\text{max}} +1)(2N_{\text{max}} +1)}$ \\
Initialize $i,j, p=1$ \\
\For{$p\leq N_r$}{
 Set $u=g\left(p:N_r:\text{end}\right)$ \\
 $H(p,:)=\;$AngularForwardTransform$(u)$
}
\For{$m,n\in\Gamma$}{
 Calculate $\hat{g}(i:i+K_{\text{max}}-1)=\mathfrak{O}_n *H(:,j)$ \\
 $i = i + K_{\text{max}}$ \\
 $j = j + 1$ \\
}
Scale $\hat{g} = ({2\pi\tau}/{\mathfrak{N}}) * \hat{g}$ \\
\KwResult{The forward basis transform $\{\hat{g}_{m,n,k}\}$ of the function  $g\left(r,\theta_1,\theta_2\right)$}
\label{Algo:Fast_Forward_Transform}
\end{algorithm}
}
\vspace{0.5cm}

\scalebox{0.8}{
\begin{algorithm}[H]
\label{Algo:Fast_theta1_theta2_fwd_transform}
\SetAlgoLined
\caption{AngularForwardTransform \\(Fast Angular 2D Forward Fourier Transform)}
\textbf{Input}: Vector containing real space representation of 2D angular function on\\
$\qquad\;\quad$ Fourier grid, i.e., $\{u(\theta_{1}^{\textsf{i}}, \theta_{2}^{\textsf{j}})\} \in \mathbb{C}^{N_{\theta_{1}}N_{\theta_2}}$\\
Reshape $u$ into a matrix $\mathcal{W} \in \cz^{{N_{\theta_2}\times N_{\theta_1}}}$\\
$\calV = \textsf{fft2}(\mathcal{W})$\\
Scale $\mathcal{V} = (\frac{1}{N_{\theta_1}N_{\theta_2}})*\mathcal{V}$\\
$\calU = \textsf{fftshift}$($\mathcal{V}$) \\
$\hat{u}=\mathcal{U}(N_{\text{max}}+1:3N_{\text{max}}+1,\;M_{\text{max}}+1:3M_{\text{max}}+1)$ \\
\KwResult{The 2D angular Fourier transform $\{\hat{u}_{m,n}\}$ of the function $u\left(\theta_1,\theta_2\right)$}
\end{algorithm}
}

\begin{figure}[!ht]
    \centering
    \includegraphics[width=\textwidth]{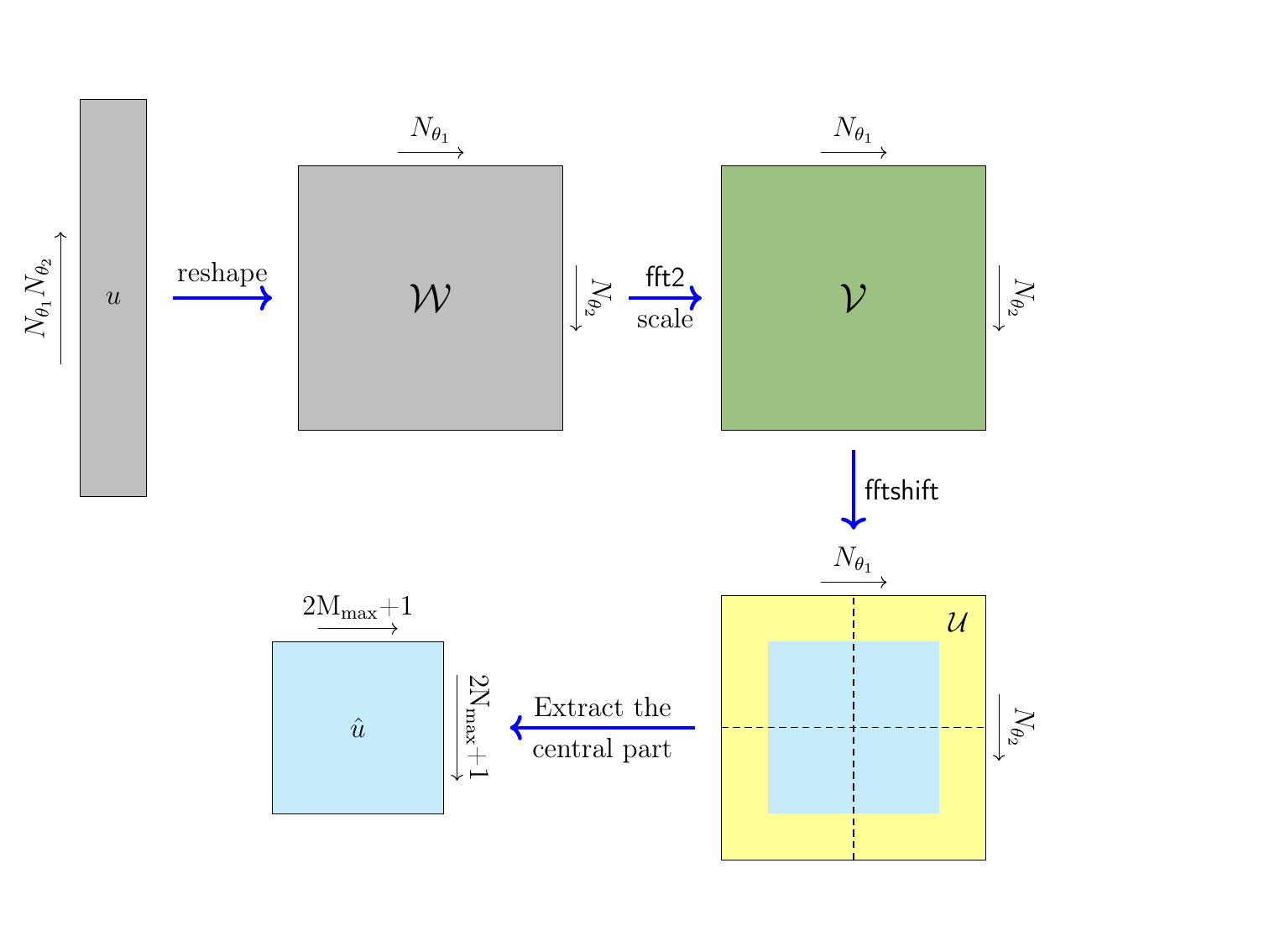}
\caption{Pictorial representation of the workings of the fast angular 2D forward Fourier transform (Algorithm \ref{Algo:Fast_theta1_theta2_fwd_transform}). `Scale' indicates dividing the result of the 2D FFT by $({N_{\theta_1}N_{\theta_2}})$}
\label{Fig:Fast_theta1_theta2_fwd_transform}
\end{figure}

Referring to Fig.~\ref{Fig:Fast_vs_Naive_transforms}, we see that like the case of the fast inverse basis transforms, our implementation of the fast forward basis transforms scale in a \emph{sublinear} manner with respect to basis set size increase, although a somewhat worse performance is expected theoretically. In contrast, a naive implementation of the forward transform scales in a quadratic manner with respect to basis set size, although both implementations of the transforms always agree with each other to machine precision.

In practice, the differences between the efficiencies of the fast and the naive transform implementations (both forward and inverse transforms) are not only apparent in terms of their respective scalings with respect to basis set size, but also the actual computational wall times. Indeed, we found that the fast transform implementations can be orders of magnitude faster as compared to the naive ones, even for relatively small basis set sizes. In Algorithm \ref{Algo:MATVEC}, we outline the steps of calculating the product of the Hamiltonian matrix with a wavefunction vector block by use of the fast transforms, as implemented in HelicES.

\vspace{0.2cm}
\scalebox{0.8}{
\begin{algorithm}[H]
\label{Algo:MATVEC}
\SetAlgoLined
\caption{Product of Hamiltonian Matrix with a block vector of wavefunctions}
\textbf{Input}: Block of $N_{\text{s}}$ wavefunctions expressed in reciprocal space, i.e., $\hat{X} \in \cz^{\calL \times N_{\text{s}}}$, \\
real space representation of local potential $V(\theta_1,\theta_2,r)$ as a vector $\mathscr{V} \in \cz^{N_{\theta_1}N_{\theta_2}N_r}$,\\
cyclic k-point $\nu$ and helical k-point $\eta$.\\
\underline{Prerequisites:} Indexing function $\mathfrak{i}:\Gamma \to \{1,2,\ldots, \calL\}$ (eq.~\ref{eq:wavefun_storage}), \\
for each $n\in [-N_{\text{max}},N_{\text{max}}]$, the matrix ${I}_n \in \rz^{K_{\text{max}}\times K_{\text{max}}}$ with entries\\ 
given by ${I}_n(k,k') = \calI(n,k,k')$ (eq.~\ref{eq:I_nkk_prime}),\\
vector $\Lambda \in \rz^{\calL}$ with entries corresponding to eq.~\ref{Eq:Laplacian_of_f_mnk}, i.e., $\Lambda(\mathfrak{i}(m,n,k)) = \lambda^{0}_{m,n,k}$, \\
vector $\mathscr{M} \in \cz^{\calL}$ with entries $\mathscr{M}(\mathfrak{i}(m,n,k)) = i2\pi m$,  \\
and vector $\mathscr{N} \in \cz^{\calL}$ with entries $\mathscr{N}(\mathfrak{i}(m,n,k)) = i2\pi\mathfrak{N}n$.\\
$-------------------------------------------$\\
Initialize $a=\frac{2\pi^2}{\tau^2}\Big\{\nu\alpha\left(2\eta-\nu\alpha\right)-\eta^2\Big\}$, $b=\frac{2i\pi}{\tau^2}\left(\nu\alpha-\eta\right)$, $c=\frac{2i\pi\alpha}{\tau^2}\left(\eta-\nu\alpha\right)$\\
Set $\hat{Y}=O_{\mathcal{L}\times N_{\text{s}}}$ $\quad\slash\slash\,$\textsf{Result to be stored in this}\\
\For{$j\leq N_{\text{s}}$}{
 $\hat{Z} = \hat{X}(:,j)$ $\quad\slash\slash\,$\textsf{Work on} $j^{\text{th}}$ \textsf{wavefunction}.\\
 $\hat{P} = \mathscr{M}\,.\!* \hat{Z}$\\
 $\hat{Q} = \mathscr{N} .\!* \hat{Z}$\\
 $\hat{Y}(:,j)=\half*(\Lambda\;.\!*\hat{Z})$ \\
 $\hat{Y}(:,j) = \hat{Y}(:,j) - a*\hat{Z} - b*\hat{P} - c*\hat{Q}$\\
 \If{$\nu\neq 0$}{
 Initialize $i=1$ \\
  \For{$m,n\in\Gamma$}{
  $T = \frac{\nu^{2}}{R^2} * \hat{Z}(i:i+K_{\text{max}}-1) + \frac{i\nu}{\pi R^2}*\hat{Q}(i:i+K_{\text{max}}-1)$\\
  $\hat{Y}(i:i+K_{\text{max}}-1,j) = \hat{Y}(i:i+K_{\text{max}}-1,j) + I_{n}*T$\\
  $i = i + K_{\text{max}}$ 
 }}
 $Z=\;$FastInverseBasisTransform$(\hat{Z})$ $\quad\slash\slash\,$\textsf{Use Algorithm} \ref{Algo:Fast_Inverse_Transform}.\\
 $\hat{Y}(:,j)= \hat{Y}(:,j)\,+\,$FastForwardBasisTransform$(Z\;.\!*\mathscr{V})$  $\quad\slash\slash\,$\textsf{Use Algorithm} \ref{Algo:Fast_Forward_Transform}.
}
\KwResult{$\hat{Y} \in \cz^{\calL \times N_\text{s}}$, i.e., the product of the Hamiltonian with $\hat{X}$ at the given values of $\eta, \nu$.}
\end{algorithm}
}
\subsection{Eigensolvers and Preconditioning}
\label{sec:eigensolvers_preconditioning}
\begin{figure}[H]
    \centering
    \includegraphics[width=0.7\textwidth]{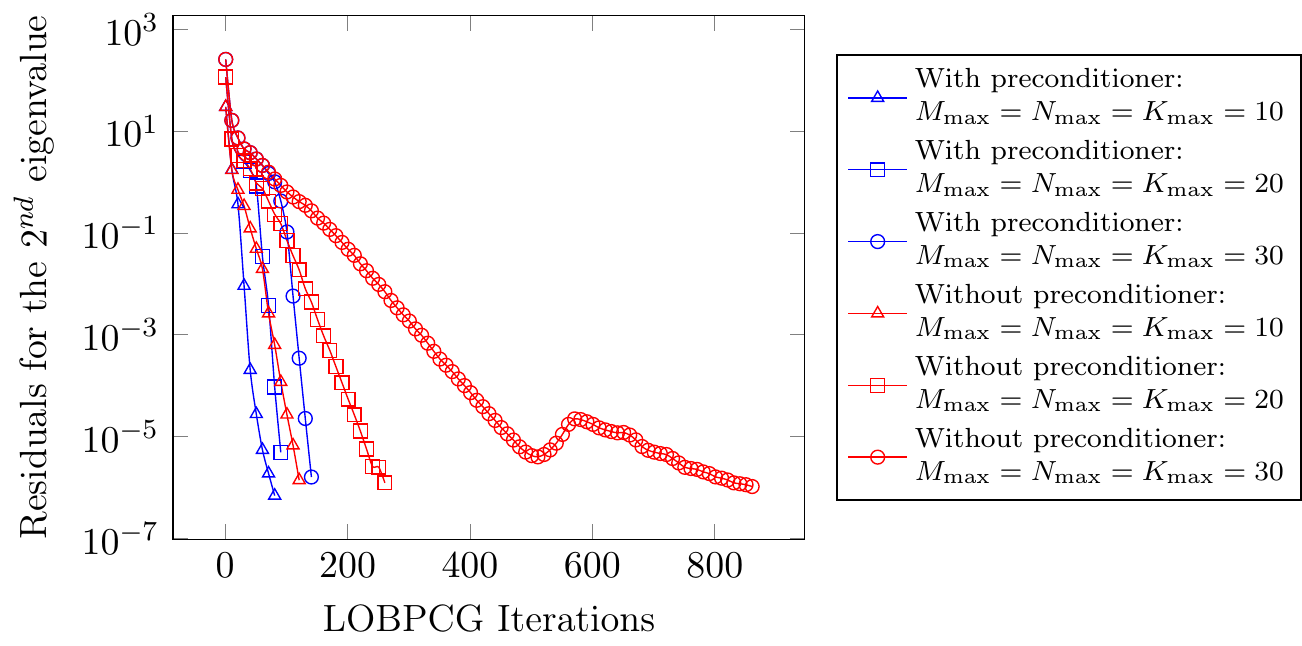}
\caption{Effect of the Teter-Payne-Allan preconditioner \citep{Teter_Payne_Allan_1} on LOBPCG \citep{LOBPCG_1} iterations for diagonalizing the discretized Hamiltonian in HelicES.  An untwisted $(6,6)$ armchair carbon nanotube (Mayer pseudopotentials \citep{mayer2004band}) has been used, and the residual associated with the $2^{\text{nd}}$ eigenvalue for $\eta = 0, \nu = 0$ as been monitored. For clarity, the residual for every $10^{\text{th}}$ iteration has been plotted. Without preconditioning, the number of iterations required to reach a given convergence threshold tends to dramatically increase as the basis set grows larger.}
\label{Fig:residuals_TPA}
\end{figure}

As mentioned earlier, we make use of matrix-free iterative eigenvalue solvers for diagonalization of the discretized Hamiltonian. Within HelicES, we have investigated two different diagonalization strategies for this purpose. The first is based on the Krylov-Schur method as implemented in the MATLAB \textsf{Eigs} function \citep{stewart2002krylov,Teter_Payne_Allan_2, zhou2015preconditioning}. The second is based on a MATLAB implementation \citep{LOBPCG_Matlab} of the Locally Optimal Block Preconditioned Conjugate Gradient (LOBPCG) scheme \citep{LOBPCG_1,LOBPCG_3,duersch2018robust}. LOBPCG requires the use of a preconditioner, for which we have adopted the Teter-Payne-Allan (TPA) recipe \citep{Teter_Payne_Allan_1, Teter_Payne_Allan_2, zhou2015preconditioning}. This preconditioner was originally developed in the context of plane-wave calculations of bulk systems, but has also been successfully applied to other spectral methods \citep{Banerjee2015spectral}. During LOBPCG iterations, use of the TPA preconditioner requires the calculation and application of a diagonal matrix $\mathscr{K} \in \rz^{\calL \times \calL}$ to the residual vector. The entries of the matrix are:
\begin{align}
\mathscr{K}_{i,j} = \frac{27 + 18\,g_i+12\,g_i^2+ 8\,g_i^3}{27 + 18\,g_i + 12\,g_i^2 +8\,g_i^3+16\,g_i^4}\;\delta_{i,j}\,,
\end{align}
with:
\begin{align}
g_i = \frac{\text{kinetic energy of basis function}\,i}{\text{kinetic energy of the residual vector}}\,.
\end{align}
As shown in  Fig. \ref{Fig:residuals_TPA}, the preconditioner can have quite a dramatic effect on the convergence of the diagonalization procedure, especially as the basis set size (and therefore, the size of the discretized Hamiltonian) is increased. Thus, the benefits of the preconditioner are likely to become more apparent when larger systems and/or harder pseudopotentials are considered.

We found that use of LOBPCG along with the TPA preconditioner generally tends to require longer diagonalization wall times as compared to \textsf{Eigs} along with an energy cutoff. Therefore, the latter strategy is adopted for most of the the examples considered in the next section. Implementation of more efficient eigensolvers in HelicES, particularly, ones that work well within self consistent field iterations \citep{My_DG_Cheby_paper,banerjee2018two,zhou_2014_chebyshev}, is the scope of future work.
\section{Results}
\label{sec:results}
We now present results obtained using HelicES and investigate the convergence and accuracy properties of our implementation. All of our calculations have been carried out using smooth empirical pseudopotentials \citep{mayer2004band, laturia2020generation}. We have used the planewave code PETRA \citep{van2019scalable}, as well as two separate MATLAB based finite  difference codes to generate reference data for comparison purposes. Specifically, we have employed the helical symmetry adapted finite difference code Helical DFT \citep{banerjee2021ab, yu2022density} and the Cartesian grid finite difference code RSDFT \citep{chelikowsky2019introductory}. The original versions of these finite difference codes were designed for self consistent field calculations, and were modified to work with the empirical pseudopotentials used in HelicES. We have also carried out comparisons of results obtained from HelicES against data obtained from the literature \citep{mayer2004band, laturia2020generation}. We have used the WebPlotDigitizer tool \citep{Rohatgi2020} for extracting data from published plots.
\subsection{Computational Platform}
\label{subsec:comp_platform}
All simulations involving HelicES were carried out using dedicated desktop workstations (Dell Precision 7920 Tower\REV{, iMac,} and iMac Pro) or on single nodes of the Hoffman2 cluster at UCLA's Institute for Digital Research and Education (IDRE). The Dell Precision workstation has an $18$-core Intel Xeon Gold 5220 processor ($24.75$ L3 MB cache, $2.2$ GHz clock speed), $256$ GB of RAM and $1$ TB of SATA Class 20 Solid State Drive (SSD) storage. \REV{The iMac has an $8$-core Apple M1 processor ($12$ MB L2 cache, $3.2$ GHz clock speed), $16$ GB of RAM and a $2$ TB Solid State Drive (SSD).} The iMac Pro has an $18$-core Intel Xeon W processor ($24.75$ MB L3 cache, $2.3$ GHz clock speed),  $256$ GB of RAM and a $2$ TB SSD. Every compute node of the Hoffman2 cluster has two $18$-core Intel Xeon Gold 6140 processors ($24.75$ MB L3 cache, clock speed of $2.3$ GHz), $192$ GB of RAM and local SSD storage. MATLAB version $9.7.0$ (R2019b) was used for the simulations. Parallelization was achieved by use of using MATLAB's Parallel Computing Toolbox. Reference results  generated using Helical DFT \citep{banerjee2021ab}, RSDFT \citep{chelikowsky2019introductory} and PETRA \citep{van2019scalable} employed the above platforms as well.
\subsection{Convergence Studies}
\label{sec:convergence}
\begin{figure}[H]
    \centering
    \includegraphics[width=0.75\textwidth]{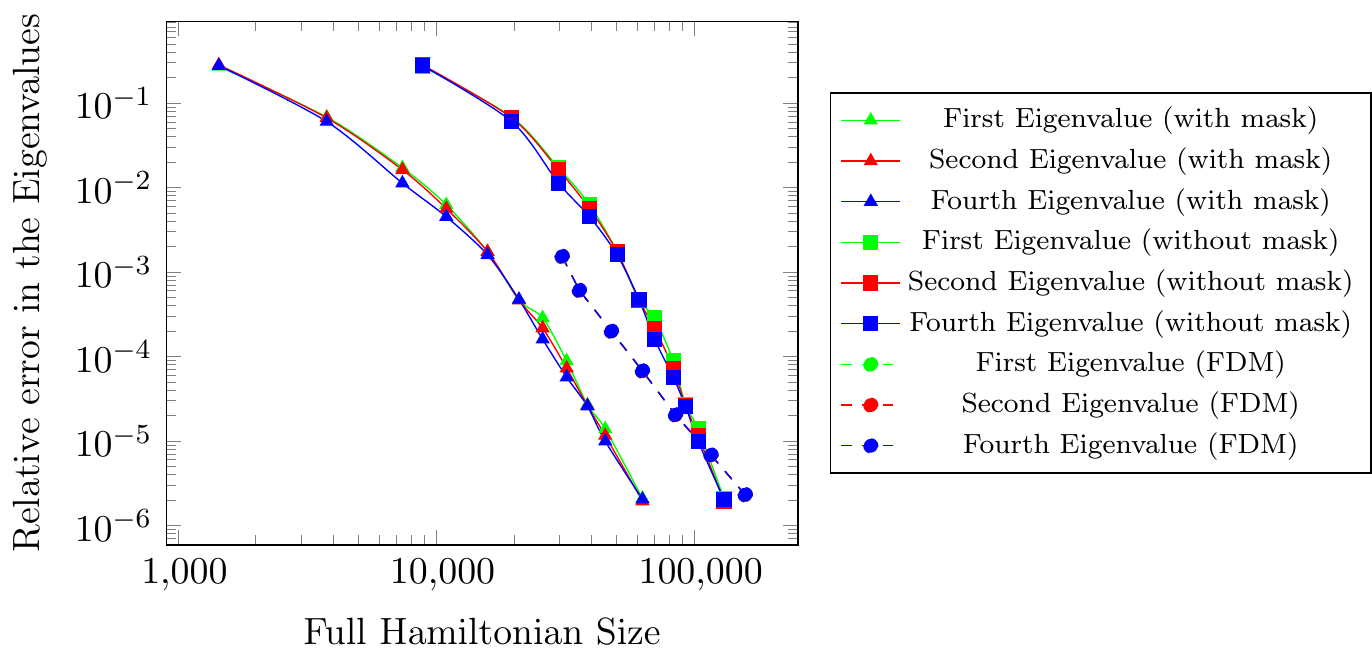}
\caption{Convergence of the first three non-degenerate eigenvalues of an armchair $(16,16)$ carbon nanotube with a twist parameter of $\alpha = 0.002$ using HelicES (both with and without the energy cutoff mask implemented) and a finite difference method (FDM), i.e., Helical DFT \citep{banerjee2021ab, yu2022density}. \REV{The mesh size of the FDM decreases from $0.7$ Bohrs to $0.4$ Bohrs (in steps of $0.05$ Bohrs) as the full Hamiltonian size varies from $30,744$ to $157,440$. The sparsity factor for the FDM Hamiltonian was $0.0055$.} The reference eigenvalues were taken to be the ones using an energy cutoff of $40$ Ha for HelicES and a mesh spacing of $0.10$ Bohr for Helical DFT. The $\eta = 0, \nu = 0$ case (``gamma point'') is considered here. \REV{Note that the errors in the eigenvalues, for different eigenvalues, differ by $\calO(10^{-4})$ or less in the FDM case, which makes them indistinguishable in the plot above.}}
\label{Fig:eigenvalue_convergence_with_ecut}
\end{figure}

Using a twisted armchair carbon nanotube as an example system (Mayer pseudopotentials \citep{mayer2004band}), we first investigate the convergence properties of HelicES. Considering first the case of eigenvalues of the Hamiltonian at $\eta = 0, \nu = 0$, we see in Fig.~\ref{Fig:eigenvalue_convergence_with_ecut} that as the number of basis functions in HelicES is increased, there is a rapid convergence to the reference values, regardless of which eigenvalue is considered. Consistent with earlier results for electronic structure calculations using spectral basis sets \citep{Cances_planewave_numerical_analysis, Banerjee2015spectral}, HelicES shows a curvature on a log-log scale, indicative of super-polynomial convergence. In contrast, the finite difference method, also shown on the same figure, shows slower, polynomial convergence. This is consistent with earlier findings for finite difference electronic structure calculations using curvilinear coordinates \citep{ghosh2019symmetry, yu2022density}. Furthermore, when the energy cutoff criterion is engaged, HelicES appears to employ noticeably fewer degrees of freedom than the finite difference method (Helical DFT) in reaching the same levels of convergence.

\begin{figure}[htb]
    \centering
    \includegraphics[width=0.6\textwidth]{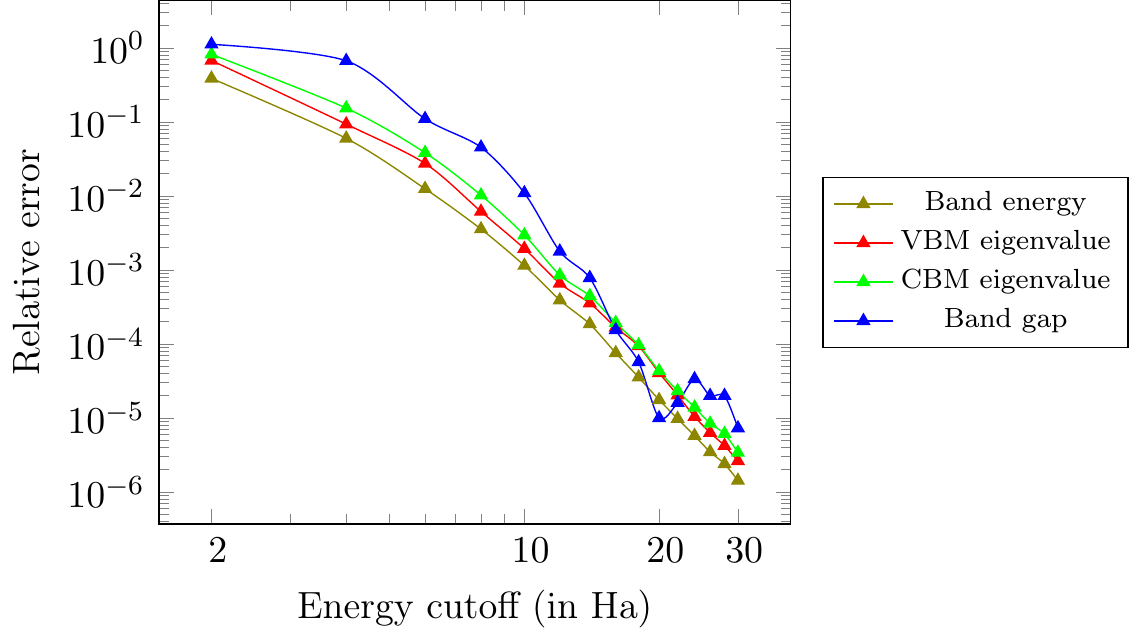}
\caption{Convergence of the electronic band energy, the Valence Band Maximum (VBM) eigenvalue, the Conduction Band Minimum (CBM) eigenvalue, and the band gap, with respect to the energy cutoff, in the HelicES code. An armchair $(16,16)$ carbon nanotube with a twist parameter of $\alpha = 0.002$ has been investigated. The reference values were generated using an energy cutoff of $40$ Ha.}
\label{Fig:band_energy_gap_homo_lumo_wrt_ecut}
\end{figure}

The electronic features of quasi-one-dimensional systems can be characterized by one-dimensional band diagrams \citep{banerjee2021ab, yu2022density}, and these can be readily calculated for systems of interest using HelicES. As the next step in our studies, we checked the convergence behavior of the code with regard to a few quantities that are associated with the overall features of the one-dimensional band diagram of the aforementioned armchair carbon nanotube system. These include the the electronic band energy --- which for an insulating system is simply twice the sum of all occupied state eigenvalues, the valence band maximum eigenvalue, the conduction band minimum eigenvalue and the band gap. As shown in Fig.~\ref{Fig:band_energy_gap_homo_lumo_wrt_ecut}, we see that all these quantities, except for the band gap, show monotonic convergence to reference values. We also note that convergence of the band gap is nearly monotonic until the curve enters regions of very high accuracy ($\calO(10^{-6})$ in the figure) and this behavior is likely related to the fact that the band gap is calculated as the difference of two quantities. 
\begin{figure}[htb]
    \centering
    \includegraphics[width=0.6\textwidth]{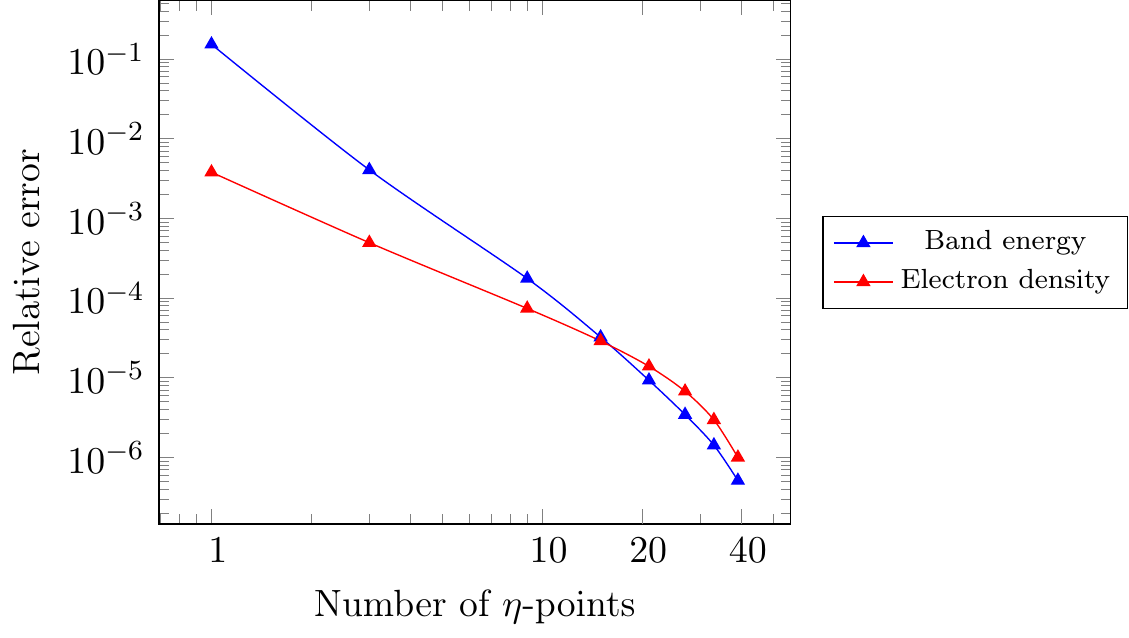}
\caption{Convergence of the band energy and electron density of a $(16,16)$ armchair carbon nanotube with a twist parameter of $\alpha = 0.002$. The reference value was taken to be from a calculation with $45$ $\eta$-points}
\label{Fig:band_energy_convergence_with_kpts}
\end{figure}

Within HelicES, the electronic properties of quasi-one-dimensional systems are also expected to exhibit convergence with respect to the number of points used to discretize the $\eta$-space (Section \ref{subsec:reciprocal_space}). In Fig.~\ref{Fig:band_energy_convergence_with_kpts}, we explore the convergence behavior of the electron density (in terms of the $L^{1}$ norm per electron) and the electronic band energy for the aforementioned carbon nanotube system, as the number of $\eta$ points in the calculation is increased. We see that both the above quantities show excellent convergence. We note that the electron density can be calculated from the wavefunctions $\phi_j(\theta_1,\theta_2,r; \eta, \nu)$, and the corresponding electronic occupation numbers $\varsigma_j(\eta,\nu)$ as:
\begin{align}
\rho(\theta_1,\theta_2,r) = \frac{2}{\mathfrak{N}}  \sum_{j=1}^{N_{\text{s}}} \sum_{b=1}^{N_{\eta}} \sum_{\nu=0}^{\mathfrak{N}-1} w_b\, \varsigma_j(\eta_b,\nu) \big\lvert{\phi_j(\theta_1,\theta_2,r; \eta_b, \nu)}\big\rvert^2\,.
\end{align}
This requires inverse basis transforms to be carried out on each wavefunction vector, at the end of the diagonalization procedure. We also note from Figs.~\ref{Fig:band_energy_gap_homo_lumo_wrt_ecut},\ref{Fig:band_energy_convergence_with_kpts} that for the Mayer pseudopotential employed in the above calculations, an energy cutoff of $16$ Ha and $15$ $\eta$-points are more than sufficient to reach chemical accuracy.

    
    
    

\subsection{Accuracy Studies}
\label{sec:accuracy_studies}
While the discussion in Section \ref{sec:convergence} serves to illustrate the systematic convergence properties of HelicES, it does not address the accuracy or correctness of the converged results produced by the code. Therefore, we now carry out a series of systematic tests and compare the results produced by HelicES against solutions produced by other methods, for a variety of systems.

Our first set of tests compares the results produced by HelicES against those computed through the Finite Difference Method (FDM). For these studies, the Mayer pseudopotential \citep{mayer2004band} was once again employed and the energy cutoff in HelicES was set at $16$ Ha. Reference results using the FDM codes were generated using a mesh spacing of $0.2$ Bohr, this being the finest mesh that could be uniformly employed for all systems of interest, within computational resource constraints. We first used the RSDFT code \citep{chelikowsky2019introductory} for calculating the electronic structure of a variety of finite (cluster-like) systems. The bound state eigenvalues for these same systems, as calculated by HelicES are compared against RSDFT results in Table \ref{Tab:FD_vs_spectral}. We see that for these discretization parameters, the agreement between the codes with respect to individual eigenvalues is about $1.3\times 10^{-4}$ Ha or better, \REV{while the band energies agree to within millihartree range}, suggesting excellent accuracy. 

\begin{figure}[H]
\centering
\scalebox{.85}{
\begin{subfigure}[b]{0.65\textwidth}
\centering
\includegraphics[width=1.2\textwidth]{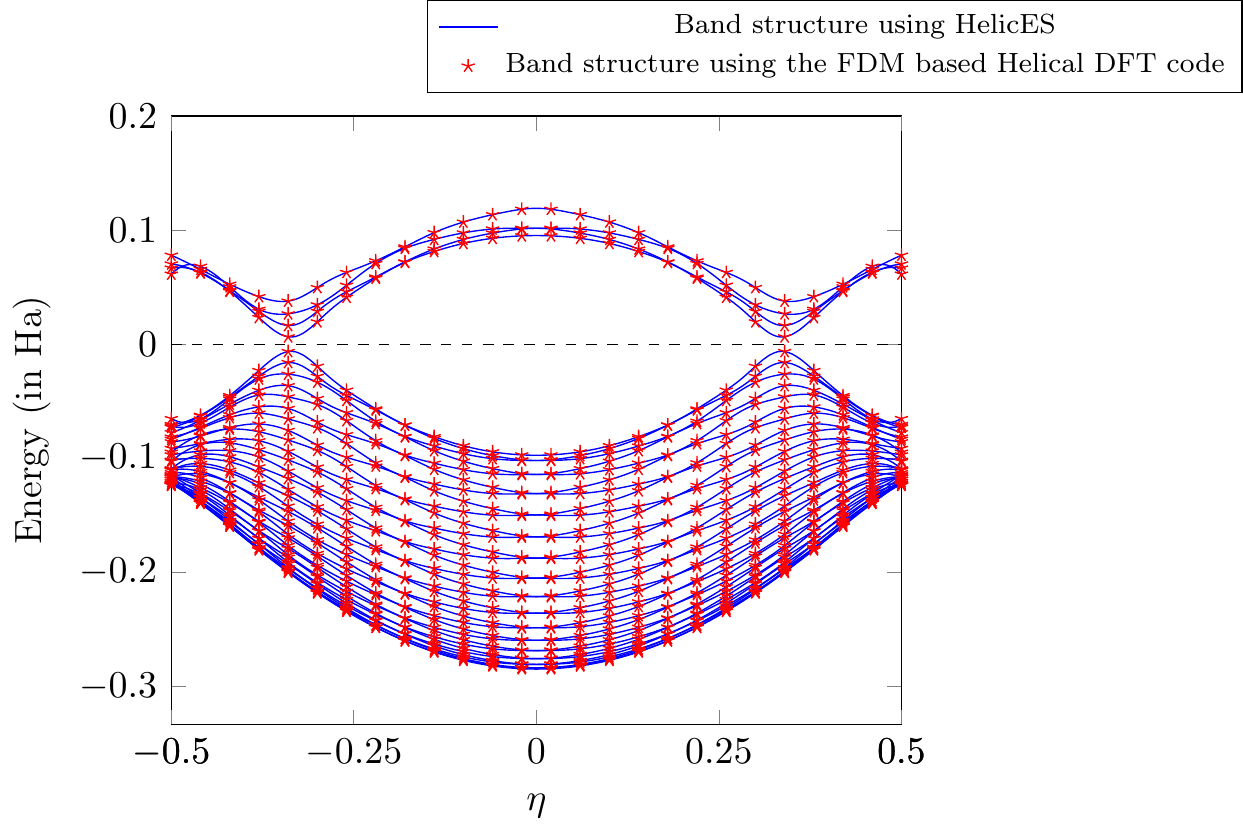}
\end{subfigure}
\hspace{-0.35cm}
\begin{subfigure}[b]{0.3\textwidth}
\centering
\raisebox{2.5cm}{
\includegraphics[width=2.2cm]{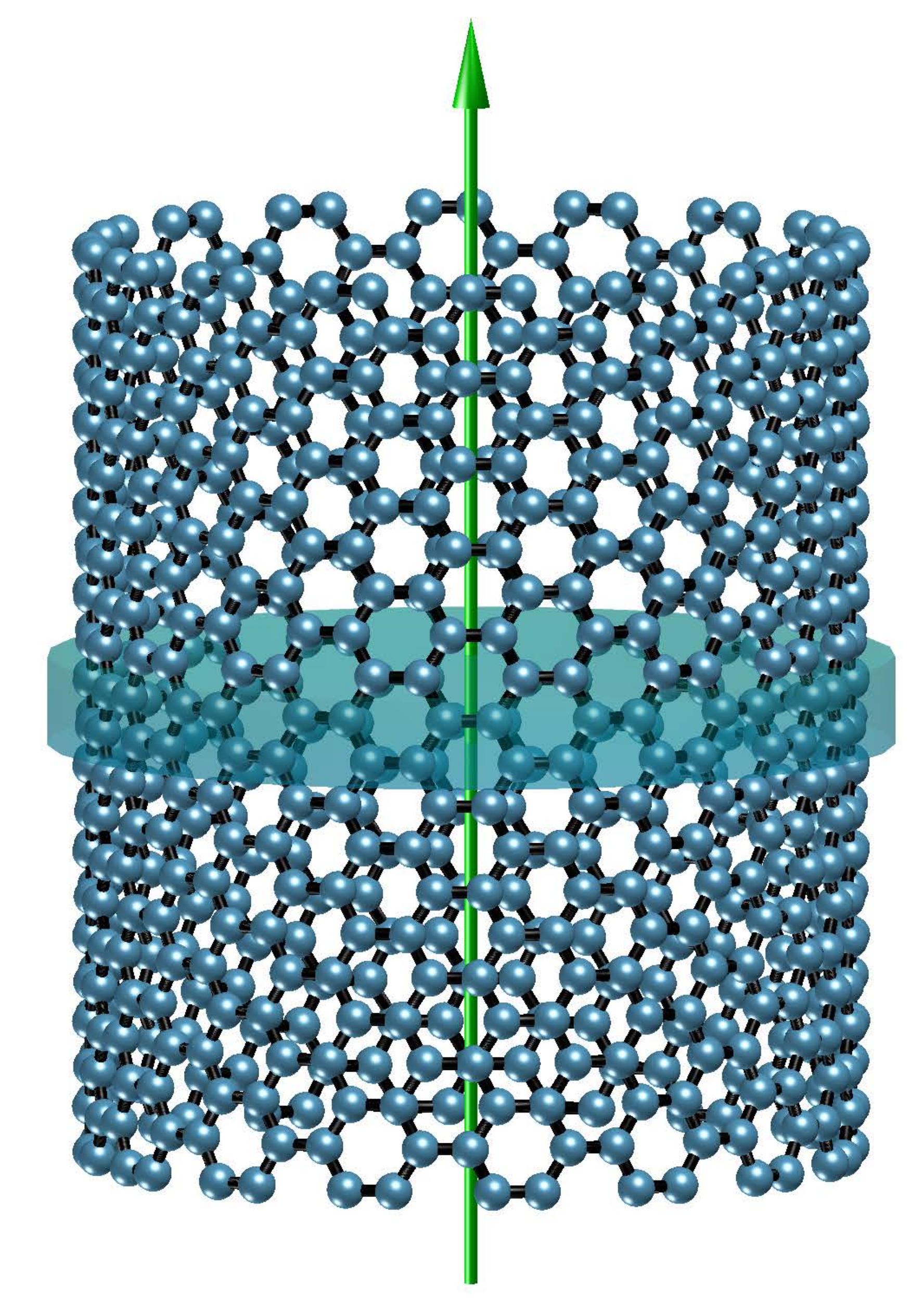}}
\end{subfigure}
}
\caption{Comparison of band diagrams for a twisted $(16,16)$ armchair carbon nanotube \REV{(diameter = 2.726 nm)} with twist parameter of $\alpha = 0.002$, generated using HelicES and the FDM based Helical DFT code \cite{yu2022density, banerjee2021ab}. The green shaded region in the structure on the right is the fundamental domain used in HelicES, while the green arrow denotes the $\textbf{e\textsubscript{Z}}$ axis.}
\label{Fig:band_diagram_DFT_vs_HelicES}
\end{figure}

Next, we generated the electronic band diagram associated with a deformed quasi-one-dimensional system, namely an armchair nanotube subjected to about $\beta = 2.95^{\circ}$ of twist per nanometer. Reference calculations were carried out using the Helical DFT code. Both Helical DFT and HelicES were made to use $21$ $\eta$-points and the \textsf{Eigs} eigensolver in MATLAB. As shown in Figure \ref{Fig:band_diagram_DFT_vs_HelicES}, the band diagrams produced by the two codes are virtually identical, once again suggesting the excellent accuracy of HelicES. Overall, these findings illustrate that HelicES adequately addresses many of the the computational bottlenecks in existing methods for the study of electronic properties of quasi-one-dimensional systems, commensurate with its design goals.
\begin{center}
\begin{table}[H]
\centering
\scalebox{0.63}{
\begin{tabularx}{1.3\textwidth}{|C|C|C|C|}
		\hline
		\textbf{System (\# atoms)} & \textbf{Hamiltonian size} & \textbf{Maximum difference in the eigenvalues (in Ha) between HelicES  and FDM} & \REV{\textbf{Difference in band energy (in Ha/atom) between HelicES  and FDM}}\\
		\hline 
		\vspace{0.1cm}
		Carbon dimer (2 atoms)
		\begin{minipage}{.3\textwidth}
		\centering
         \includegraphics[scale=0.15]{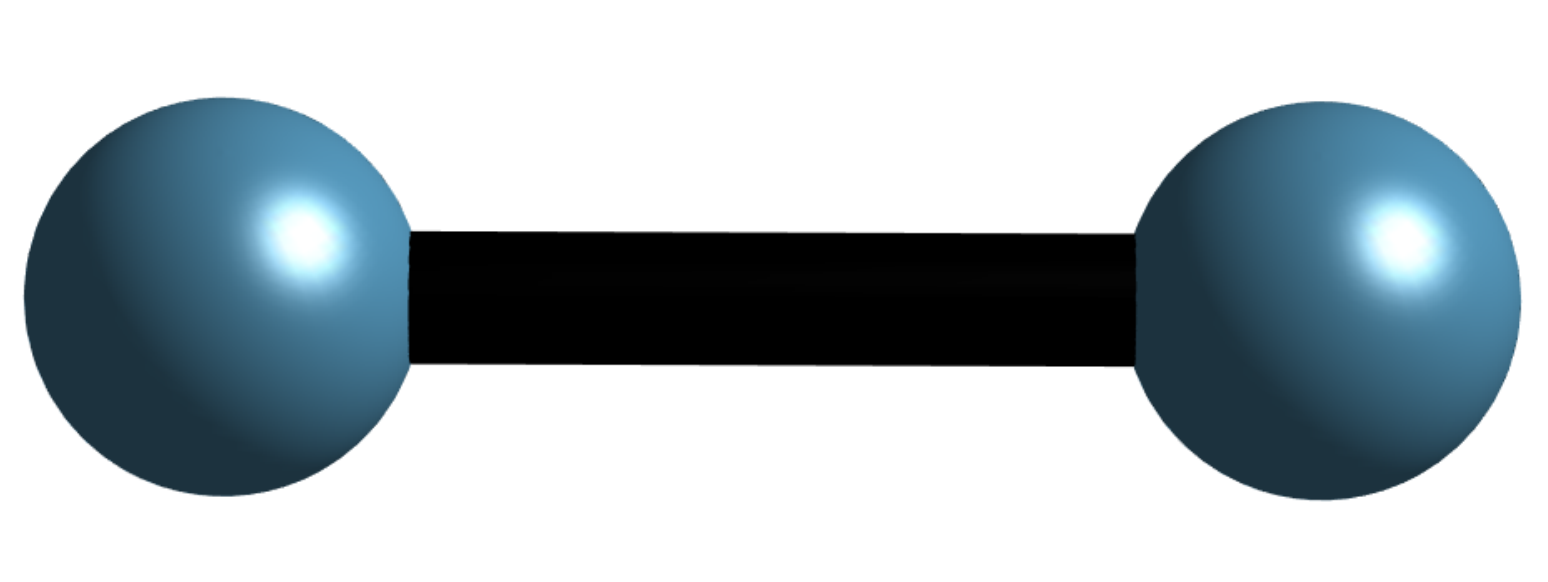}
         \end{minipage} & \vspace{0.15cm} \footnotesize{HelicES without mask: $148625$ \newline HelicES with mask: $96353$ \newline FDM: 1030301}  & \vspace{0.25cm} $6.0698\times 10^{-5}$ & \vspace{0.25cm} \REV{$5.7475\times 10^{-5}$}\\
		\hline 
		\vspace{0.1cm}
		Carbon ring (6 atoms)
		\begin{minipage}{.3\textwidth}
		\centering
         \includegraphics[scale=0.15]{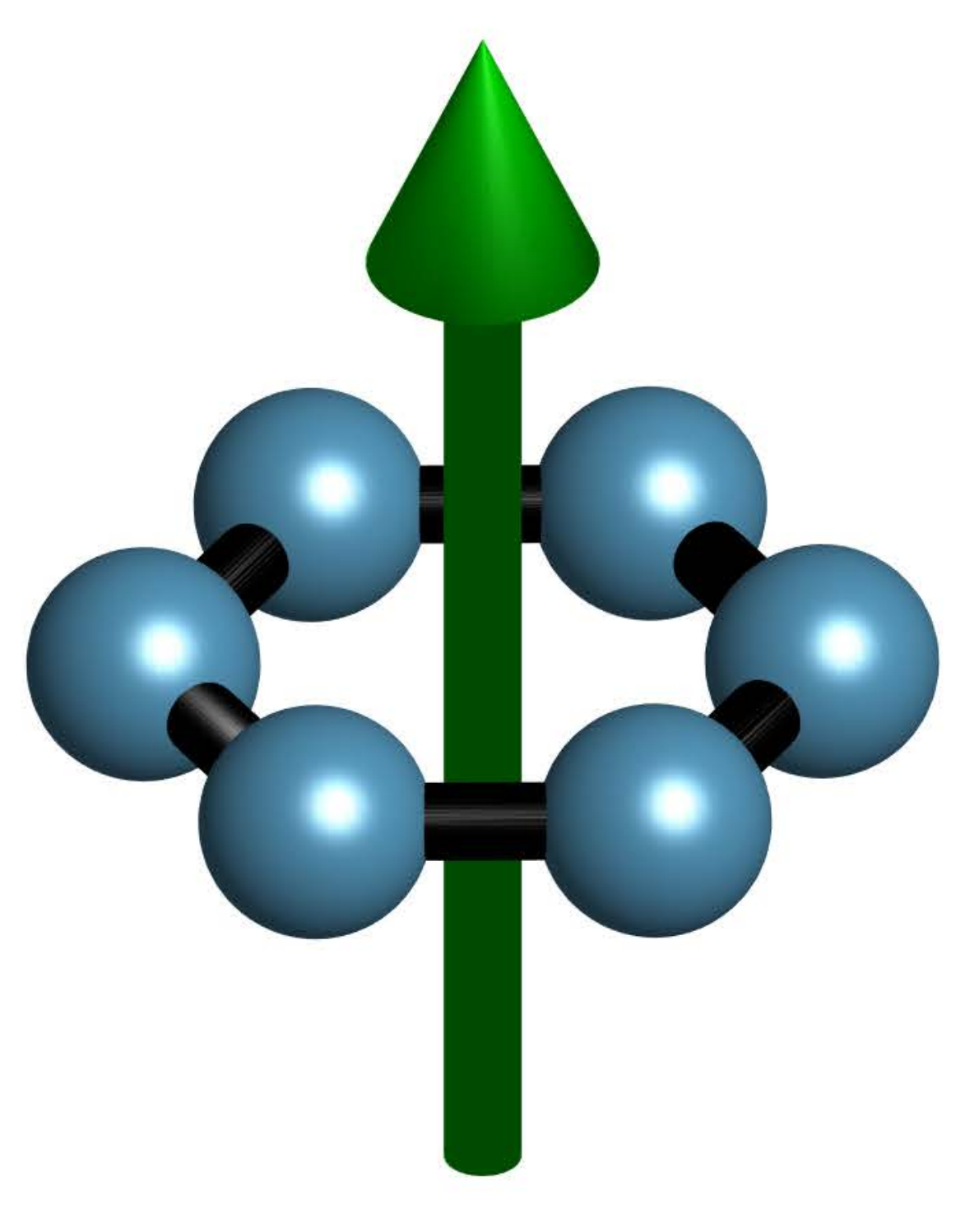}
         \end{minipage} & \vspace{0.8cm} \footnotesize{HelicES without mask: $148625$ \newline HelicES with mask: $96353$ \newline FDM: 8120601}  & \vspace{0.85cm} $3.8266\times 10^{-5}$ & \vspace{0.85cm} \REV{$3.7686\times 10^{-5}$}\\
		\hline 
		\vspace{0.1cm}
		Carbon disk (24 atoms)
		\begin{minipage}{.3\textwidth}
		\centering
         \includegraphics[scale=0.15]{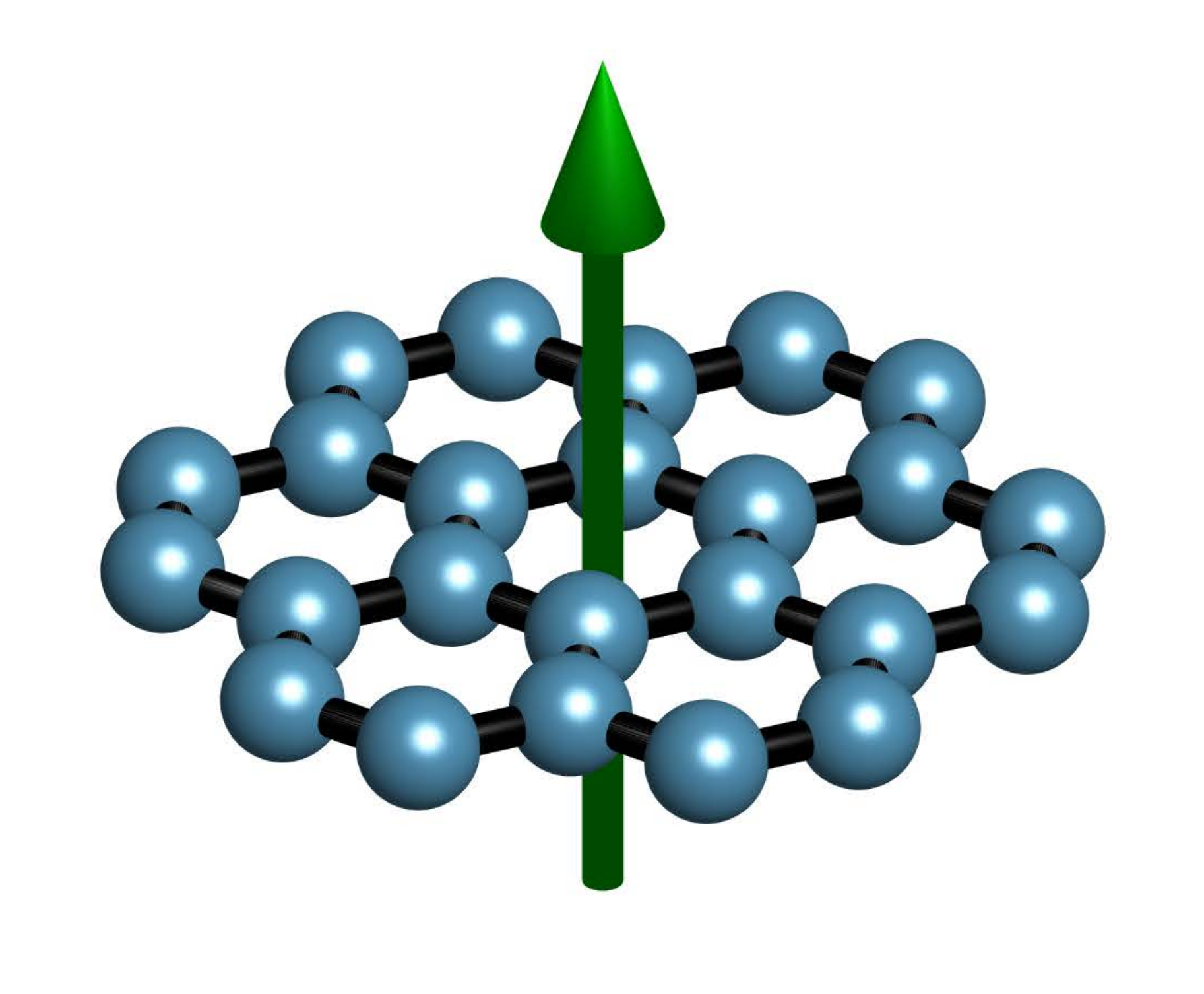}
         \end{minipage} & \vspace{0.9cm} \footnotesize{HelicES without mask: $240096$ \newline HelicES with mask: $152556$ \newline FDM: 4173281}  & \vspace{1cm} $7.7064\times 10^{-5}$ & \vspace{1cm} \REV{$4.7145\times 10^{-5}$}\\
		\hline 
		\vspace{0.1cm}
		Carbon pillar (120 atoms)
		\begin{minipage}{.3\textwidth}
		\centering
         \includegraphics[scale=0.25]{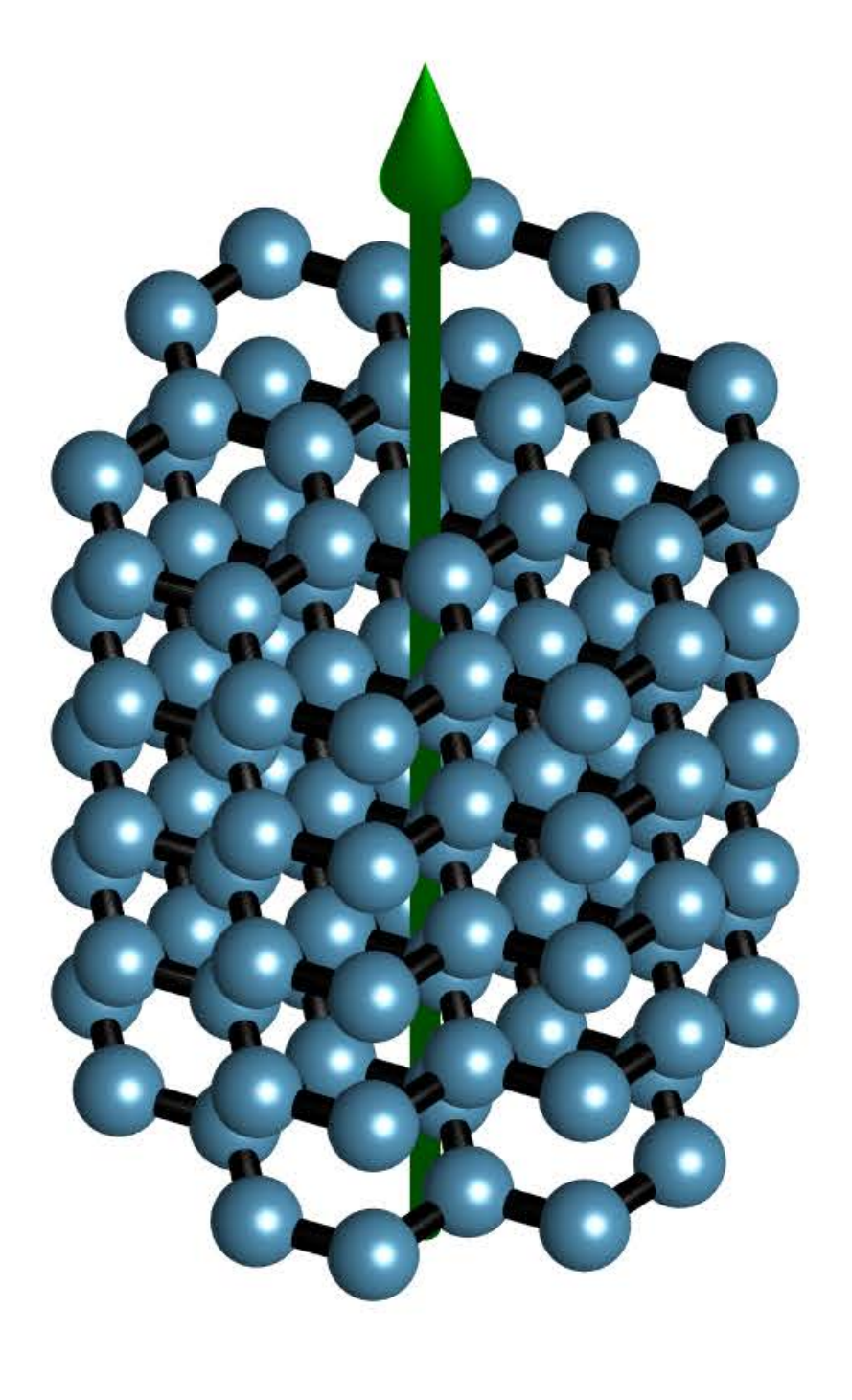}
         \end{minipage} & \vspace{1.4cm} \footnotesize{HelicES without mask: $129591$ \newline HelicES with mask: $85741$ \newline FDM: 8120601} &  \vspace{1.5cm} $1.2604\times 10^{-4}$ & \vspace{1.5cm} \REV{$3.7118\times 10^{-5}$}\\
		\hline 
\end{tabularx}
}
\caption{Accuracy of the HelicES code while studying finite systems (green arrow denotes the $\textbf{e\textsubscript{Z}}$ axis). Reference data was generated using RSDFT \citep{chelikowsky2019introductory}, a finite difference method (FDM) based MATLAB code. \REV{The last two columns show the maximum differences in the eigenvalues and the band energy per atom computed using the two methods.}}
\label{Tab:FD_vs_spectral}
\end{table}
\end{center}

\begin{figure}[H]
\centering
\scalebox{.85}{
\begin{subfigure}[b]{0.65\textwidth}
\centering
\includegraphics[width=1.5\textwidth]{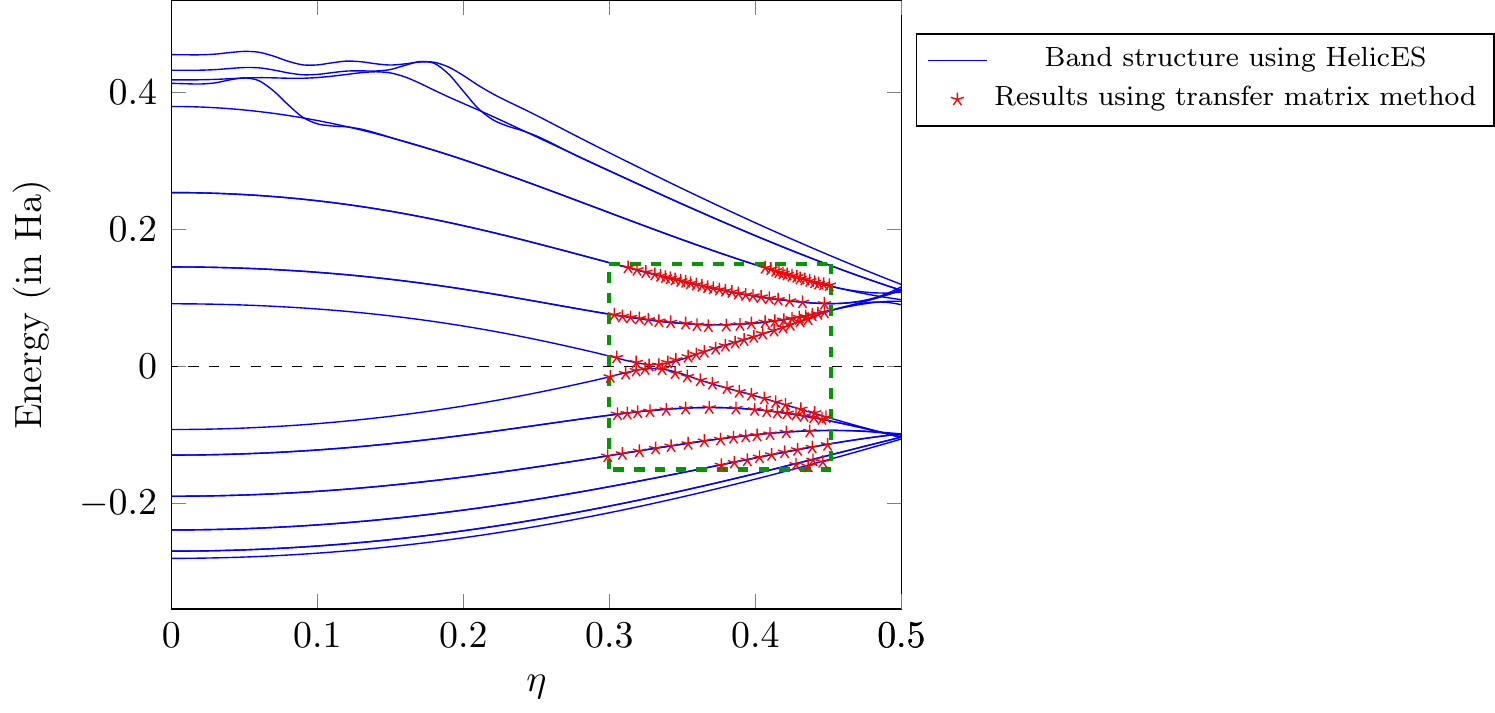}
\end{subfigure}
\begin{subfigure}[b]{0.3\textwidth}
\centering
\raisebox{0.75cm}{
\includegraphics[width=1.25cm]{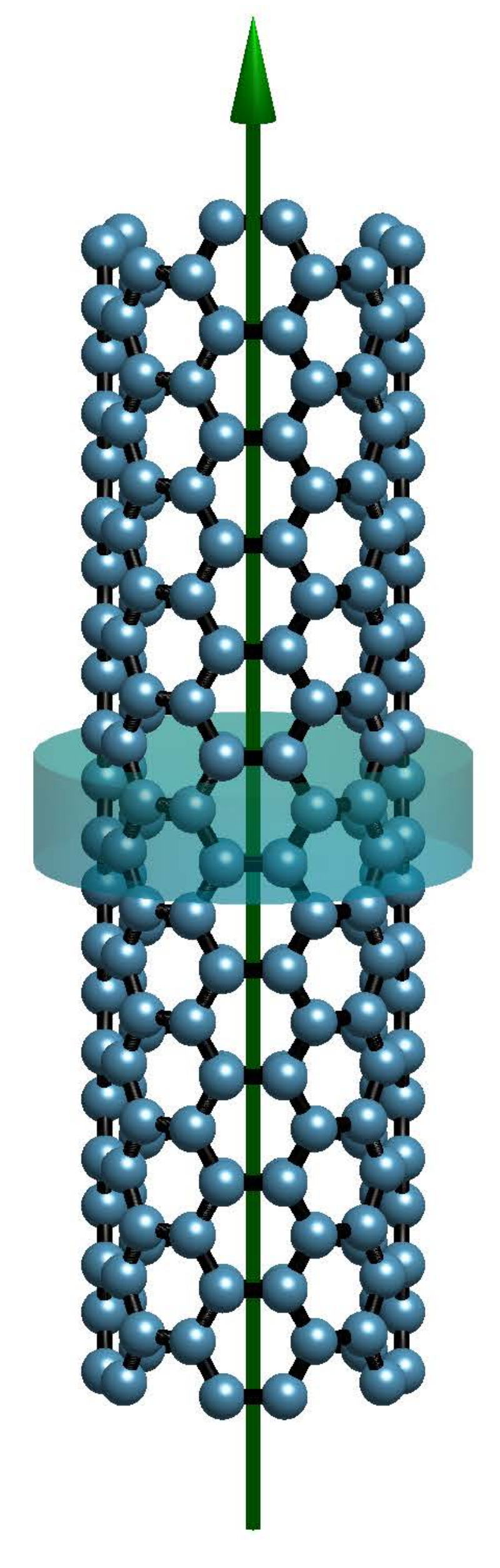}}
\end{subfigure}
}
\caption{Comparison of band diagrams for a $(5,5)$ armchair carbon nanotube \REV{(diameter = 0.851 nm)} generated using HelicES and a transfer-matrix technique \citep{mayer2004band}. The dashed green box in the plot represents the region of the band diagram over which the reference data was available for comparison. The green shaded region in the structure on the right is the fundamental domain used in HelicES and the green arrow denotes the $\textbf{e\textsubscript{Z}}$ axis.}
\label{Fig:band_diagram_5_5}
\end{figure}

\begin{figure}[H]
\centering
\scalebox{.85}{
\begin{subfigure}[b]{0.65\textwidth}
\centering
\includegraphics[width=1.5\textwidth]{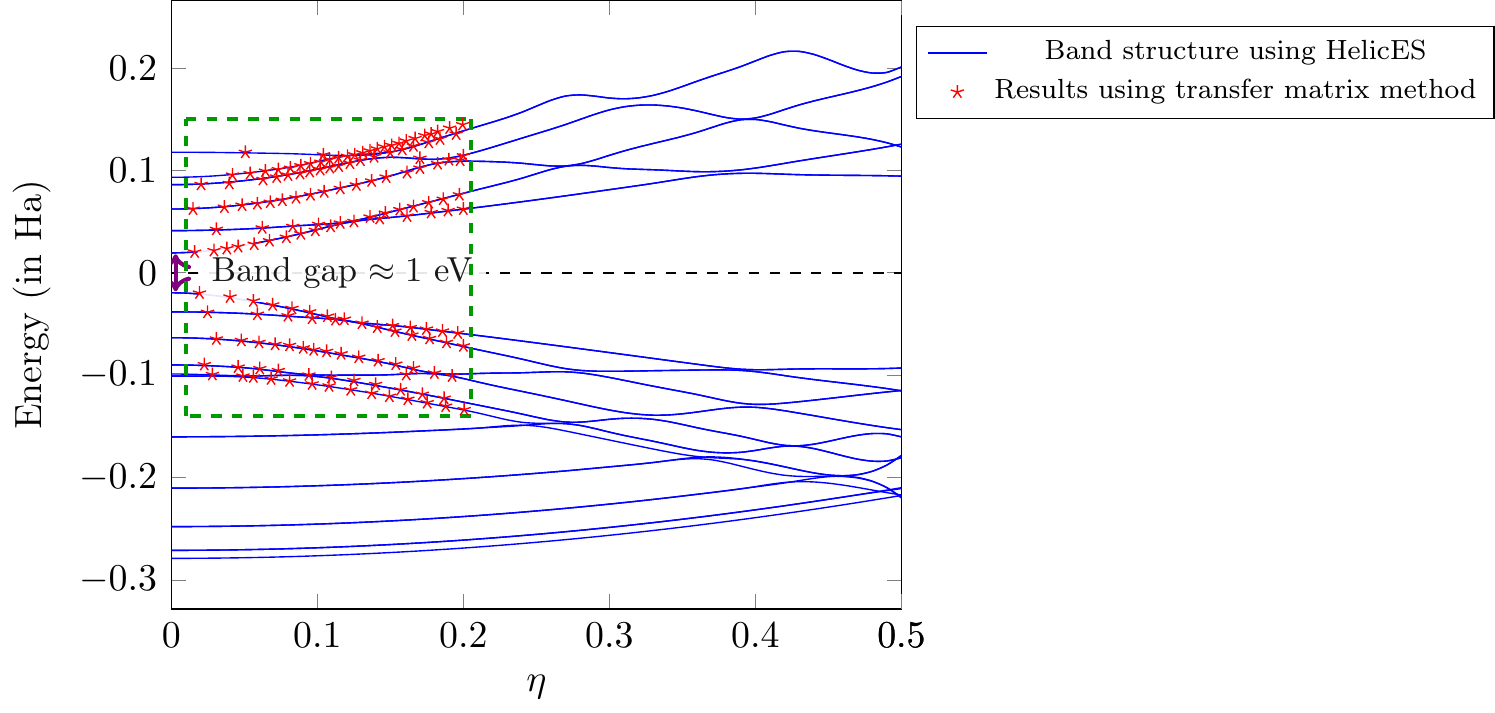}
\end{subfigure}
\begin{subfigure}[b]{0.3\textwidth}
\centering
\raisebox{0.75cm}{
\includegraphics[width=1.25cm]{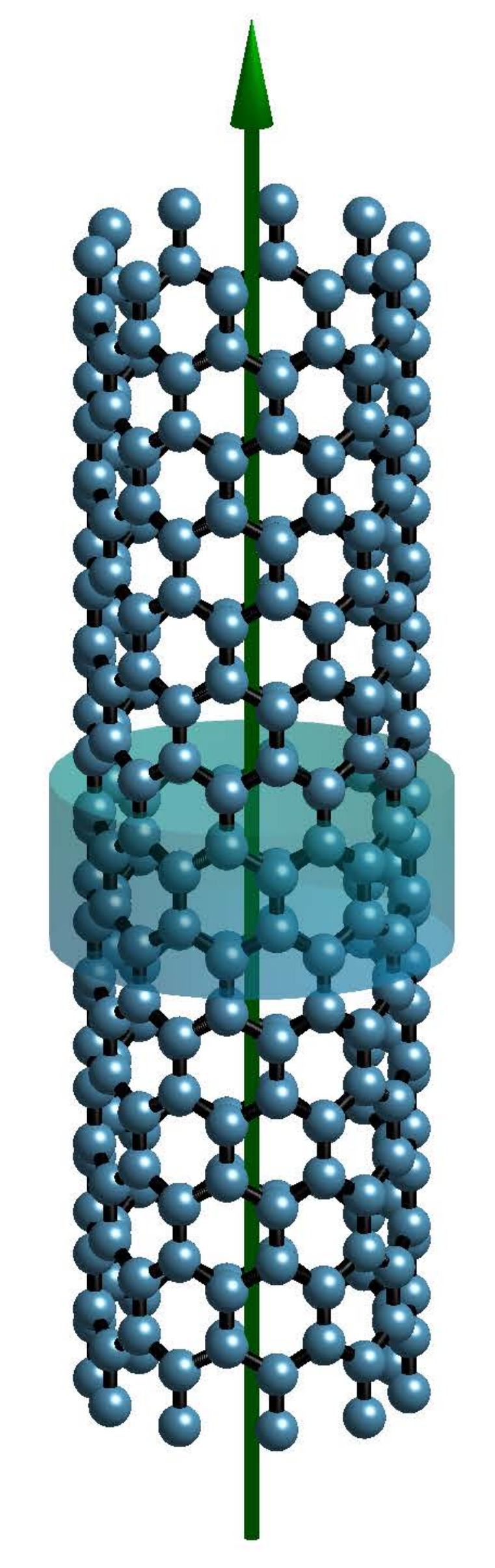}}
\end{subfigure}
}
\caption{Comparison of band diagrams for a $(10,0)$ zigzag carbon nanotube \REV{(diameter = 0.983 nm)} generated using HelicES and a transfer-matrix technique \citep{mayer2004band}. The dashed green box in the plot represents the region of the band diagram over which the reference data was available for comparison. The green shaded region in the structure on the right is the fundamental domain used in HelicES and the green arrow denotes the $\textbf{e\textsubscript{Z}}$ axis.}
\label{Fig:band_diagram_10_0}
\end{figure}

Due to inherent design limitations, the aforementioned FDM codes are unable to simulate quasi-one-dimensional nanostructures which have atoms situated near or along the system axis (e.g. nanoribbons, nanowires or small diameter nanotubes). However, these systems can be conveniently dealt with by HelicES. To carry out accuracy tests for such systems therefore, we compared the band structures calculated by HelicES against those generated through alternate electronic structure calculation techniques. The first of these is based on the transfer matrix method \citep{mayer1999accuracy, pendry1992calculation, tamura1991conductance}, often used in electromagnetics calculations. In Figs.~\ref{Fig:band_diagram_5_5} and \ref{Fig:band_diagram_10_0} we see that the band structure calculated by HelicES is in nearly perfect agreement with results calculated using this technique in \cite{mayer2004band}. The systems considered here are carbon nanotubes with radii about $0.3$ to $0.4$ nanometers. For the $(5,5)$ armchair nanotube, the position of the Dirac cone is correctly predicted to be at $\eta = \pm \frac{1}{3}$. Additionally, the $(10,0)$ zigzag nanotube, the band gap calculated by HelicES is $1.05$ eV which is very close to the value of $1.04$ eV obtained in \citep{mayer2004band}.

\begin{figure}[H]
\centering
\scalebox{.85}{
\begin{subfigure}[b]{0.65\textwidth}
\centering
\includegraphics[width=1.45\textwidth]{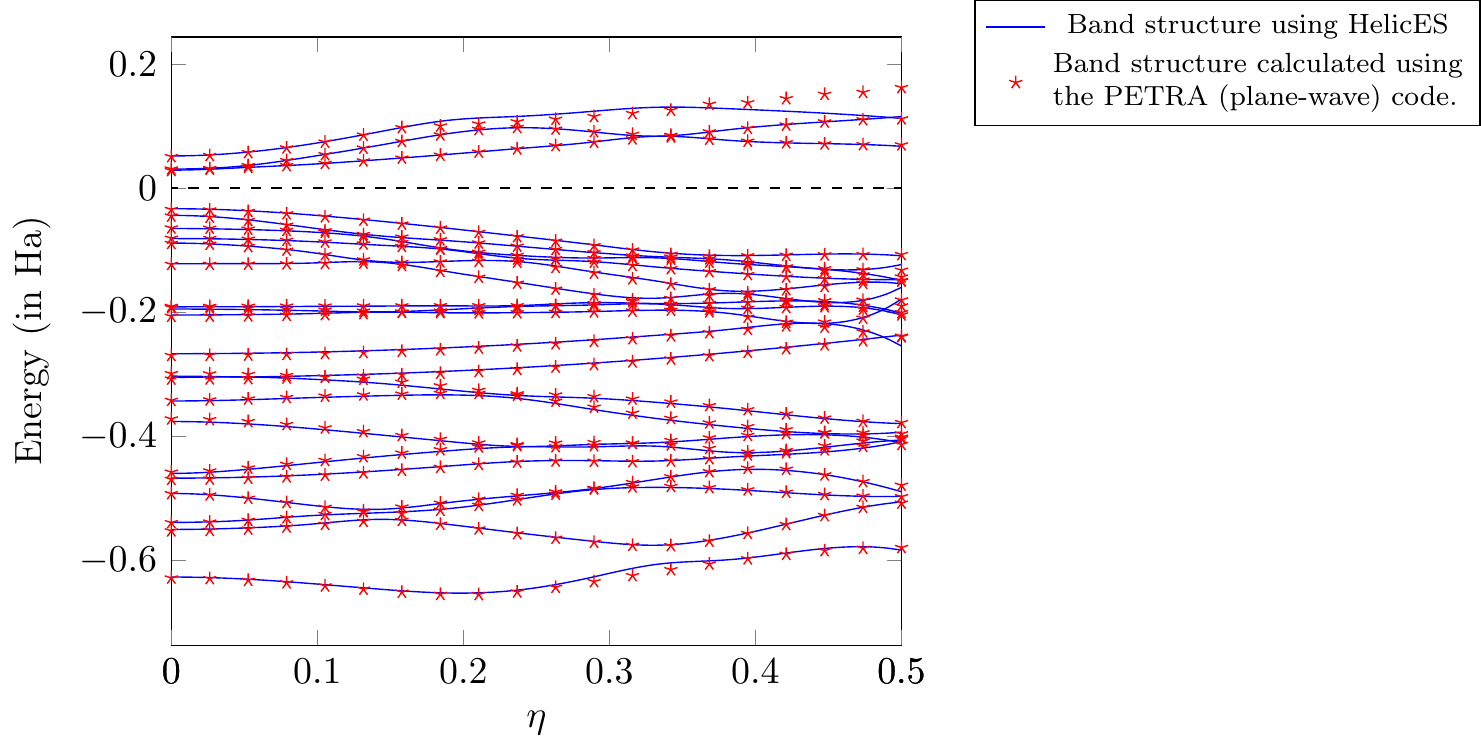}
\end{subfigure}
\hspace{-0.35cm}
\begin{subfigure}[b]{0.3\textwidth}
\centering
\raisebox{0.75cm}{
\includegraphics[width=1.5cm]{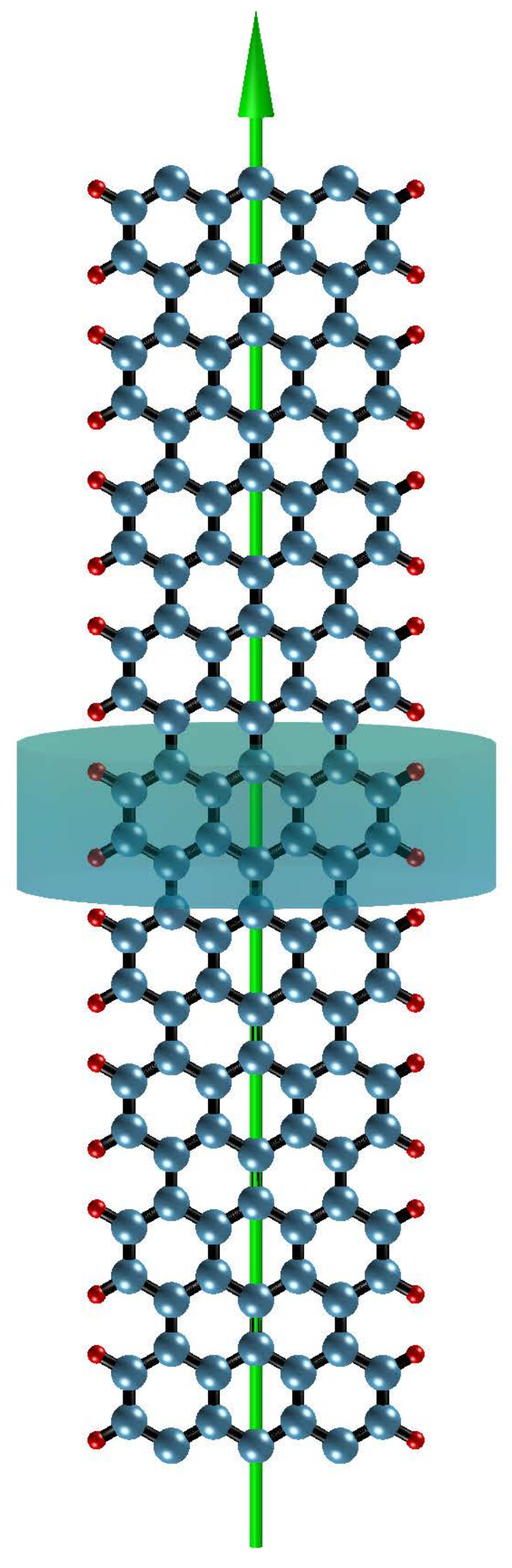}}
\end{subfigure}
}
\caption{Comparison of band diagrams for a hydrogen passivated armchair graphene nanoribbon generated using HelicES and a plane-wave technique \citep{laturia2020generation, van2019scalable}. The green shaded region in the structure on the right is the fundamental domain used in HelicES and the green arrow denotes the $\textbf{e\textsubscript{Z}}$ axis.}
\label{Fig:band_diagram_GNR}
\end{figure}

\begin{figure}[H]
\centering
\scalebox{.85}{
\begin{subfigure}[b]{0.65\textwidth}
\centering
\includegraphics[width=1.4\textwidth]{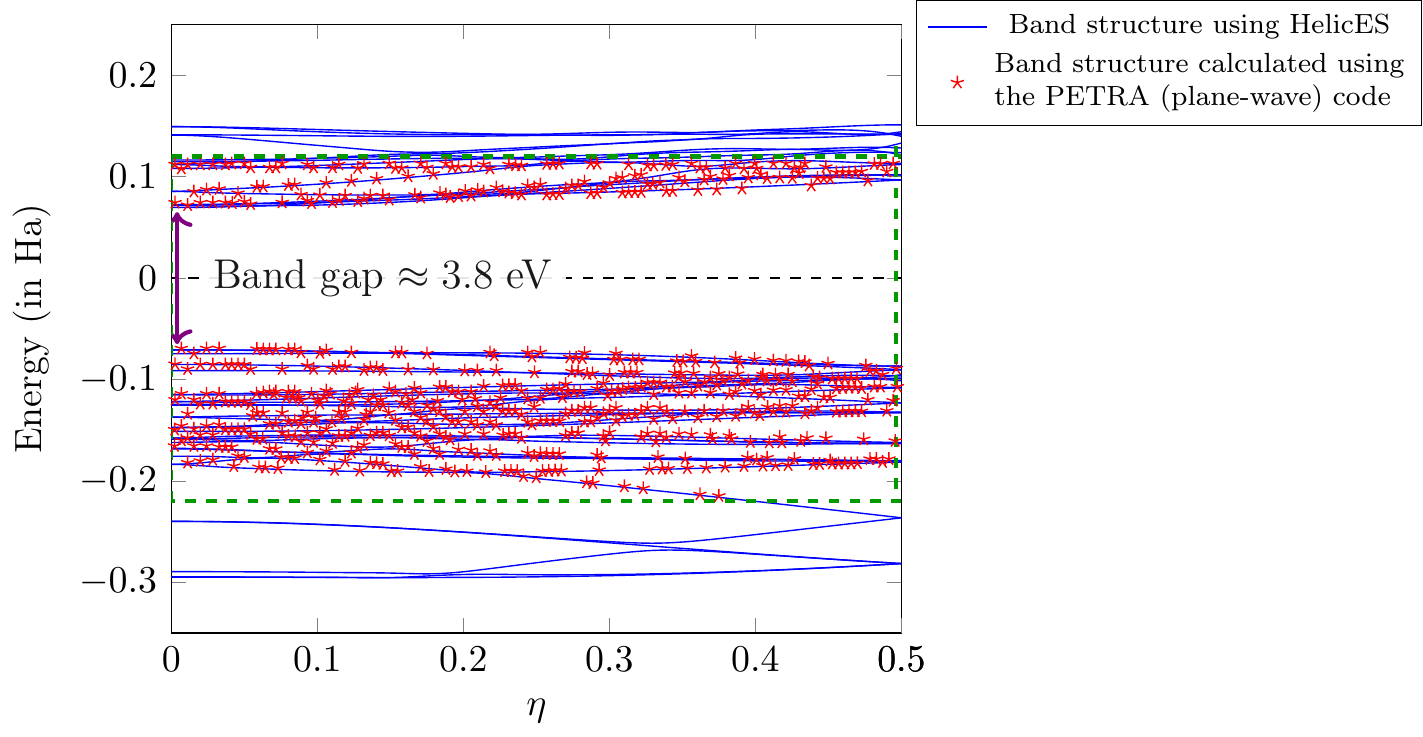}
\end{subfigure}
\hspace{-0.35cm}
\begin{subfigure}[b]{0.3\textwidth}
\centering
\raisebox{1cm}{
\includegraphics[width=1.25cm]{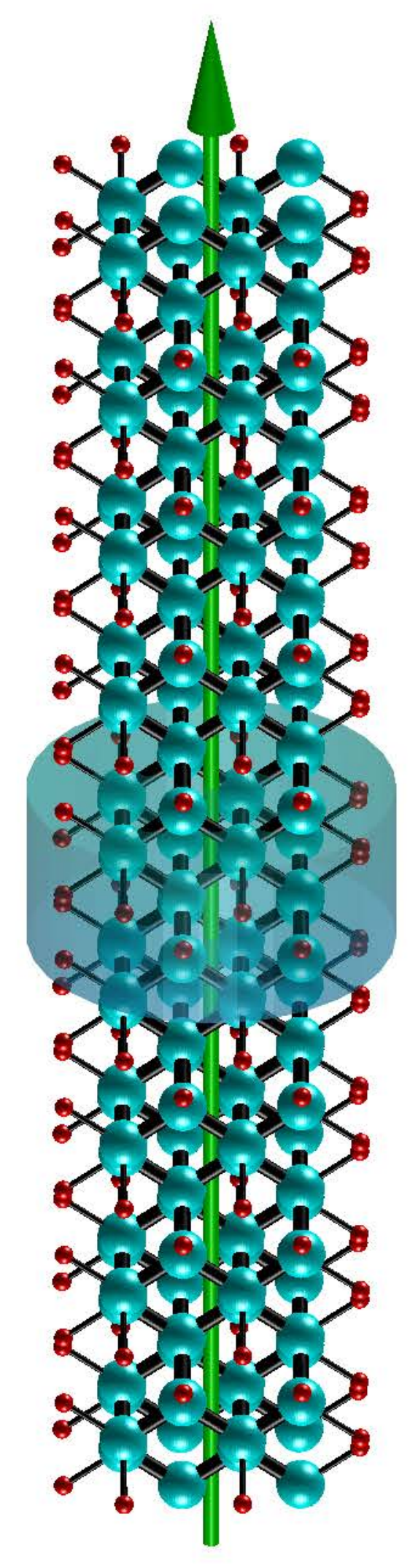}
\raisebox{1.5cm}{
\includegraphics[width=2cm]{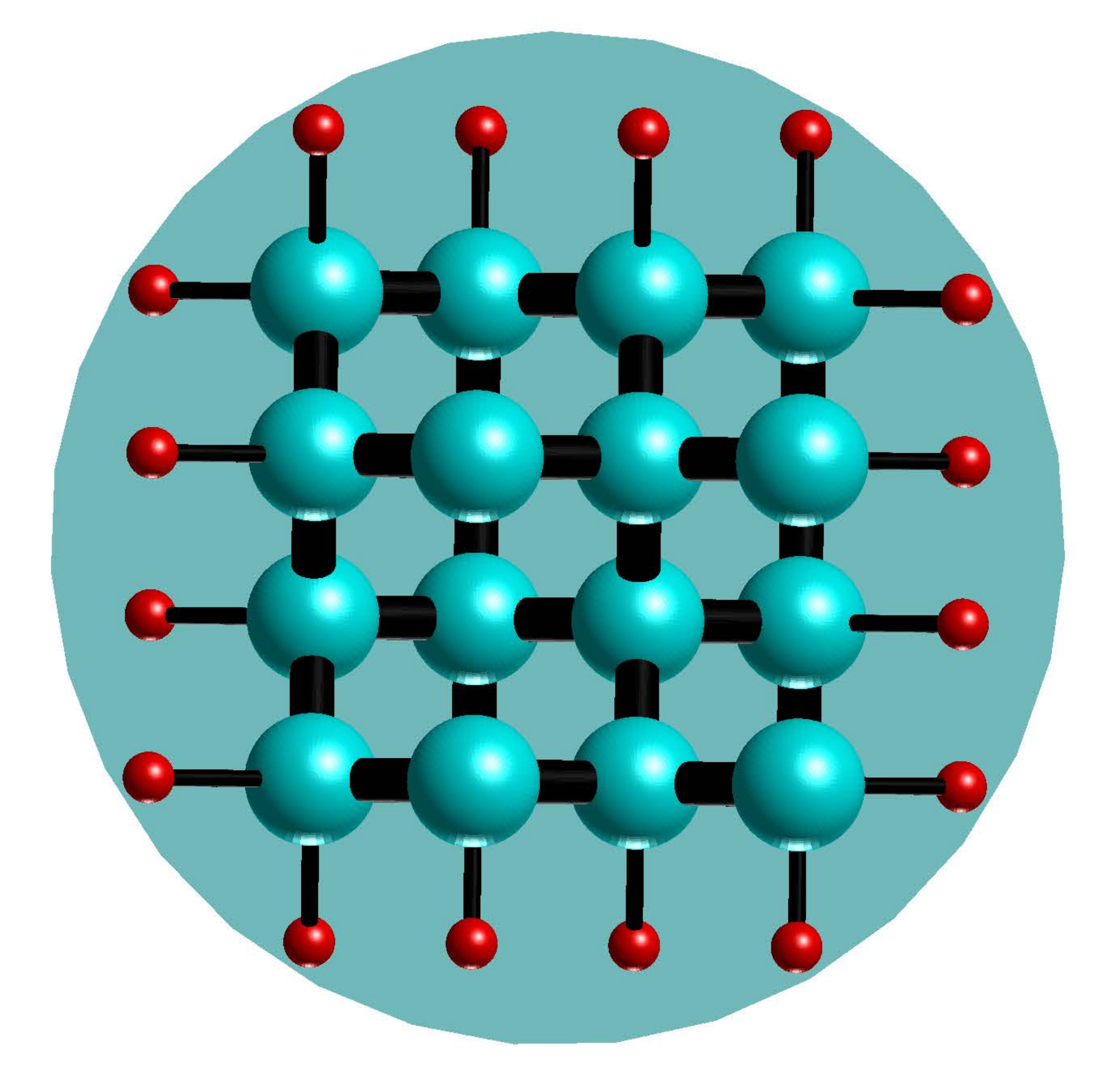}}
}
\end{subfigure}
}
\caption{Comparison of band diagrams for a hydrogen passivated, $\langle 100 \rangle$ oriented silicon nanowire generated using HelicES and a plane-wave technique \citep{laturia2020generation, van2019scalable}. The dashed green box in the plot represents the region of the band diagram over which the reference data was available for comparison. The green shaded region in the structure in the middle is the fundamental domain used in HelicES, with the green arrow denoting the $\textbf{e\textsubscript{Z}}$ axis. The right image shows a top view of the structure (i.e., looking down along $\textbf{e\textsubscript{Z}}$).}
\label{Fig:band_diagram_Si_100}
\end{figure}

Next, we used the PETRA code for studying an armchair graphene nanoribbon, as well as a silicon nanowire oriented along the $\langle 100 \rangle$ direction. Both these systems were treated using the empirical pseudopotentials developed in \citep{laturia2020generation} and feature hydrogen passivation. Figs.~\ref{Fig:band_diagram_GNR} and ~\ref{Fig:band_diagram_Si_100} show that the overall agreement between HelicES and PETRA is excellent, although some minor variations at the edge of the highest energy band for the nanoribbon case may be observed. This is possibly due to the different boundary conditions being employed by PETRA and HelicES in the directions orthogonal to the ribbon axis. We also note that the band gap for the silicon nanowire calculated by HelicES is $3.82$ eV, which is very close to the value of $3.84$ eV reported in \citep{laturia2020generation}. 

\subsection{\REV{Comments on computational efficiency and timing studies}}
\label{sec:timing}
\REV{We now discuss issues connected to the computational efficiency of HelicES. By design, the code is meant to overcome the computational limitations of prior approaches in modeling quasi-one-dimensional systems. We highlight this aspect of the code by providing timing comparisons between HelicES and other existing methods, for a few systems of interest. We have focused on the Plane-wave Electronic TRAnsport (PETRA) code \cite{van2019scalable, Laturia_Proceedings, laturia2020generation} which can model periodic systems, and Helical DFT \citep{banerjee2021ab, yu2022density} which models quasi-one-dimensional structures within a finite difference framework. For comparisons with PETRA, we chose a twisted hydrogen-passivated graphene nanoribbon. Note that while realistic values of $\alpha$ range from $0.0005$ to $0.0025$ (i.e., less than about $2.1^{\circ}$ per nanometer), it is not feasible to use these values in PETRA. This is because to simulate such a system in a typical plane-wave code like PETRA, we would require $1/\alpha$ times the number of atoms needed for untwisted geometries (for rational values of $\alpha$). However, the number of atoms required in the fundamental domain in HelicES is independent of the amount of twist. Thus, while a realistic twisted nanoribbon can be studied using only 20 atoms in HelicES, PETRA would require at least $10,000$ atoms in the fundamental domain for the same system. Keeping this in mind, we use larger values of $\alpha=0.25,0.2,0.1$, and $0.05$, so that the simulation and timing data from PETRA could be obtained within reasonable wall times. For both codes, we used the same diagonalization technique. The simulations were carried out on dedicated workstations, or on a single node of the Hoffman2 cluster when larger memory was needed. 
In our studies, we noted factors of $1.26, 1.84, 3.83$, and $12.25$ improvement in the total diagonalization wall time of HelicES over PETRA, for $\alpha=0.25,0.2,0.1$, and $0.05$ respectively. Based on the above discussion, we anticipate that the performance gap between HelicES and PETRA, as well as the memory requirements of the latter, will only increase when more realistic values of $\alpha$ or more complicated unit cells are considered.

Due to the fundamental limitations of plane-wave codes to efficiently represent helical symmetries, it also makes sense to compare HelicES to Helical DFT, since the right symmetries are incorporated into both these codes, although the latter uses finite differences. For this purpose, we studied a twisted $(16,16)$ armchair carbon nanotube with a diameter of $2.726$ nm and a twist parameter of $\alpha=0.002$, and we used 21 $\eta-$points. As we showed earlier (Section \ref{sec:accuracy_studies}), while the two codes produce nearly identical results, the diagonalization wall time for HelicES was about a factor of $27$ lower, and the memory footprint was also significantly less. These observations continue to be true when larger values of the energy cutoff are used in HelicES, with the diagonalization wall time of the code being about a factor of $8$ lower than Helical DFT, even when an energy cutoff of $40$ Ha is employed.}

\begin{figure}[H]
\centering
\scalebox{.85}{
\begin{subfigure}[b]{0.65\textwidth}
\centering
\includegraphics[width=0.95\textwidth]{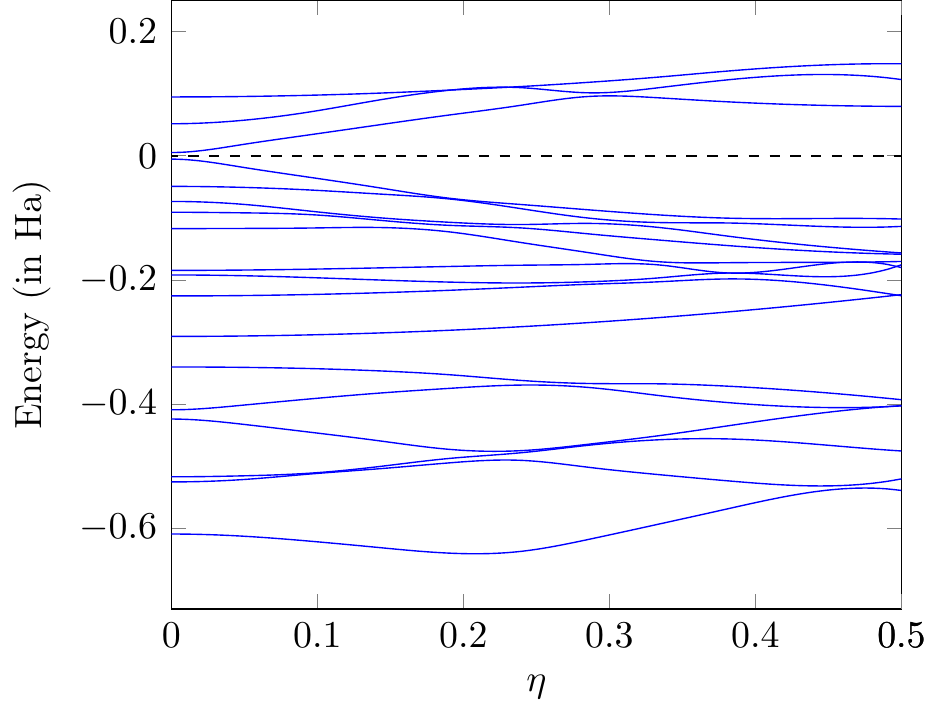}
\end{subfigure}
\hspace{-0.35cm}
\begin{subfigure}[b]{0.3\textwidth}
\centering
\raisebox{0.6cm}{
\includegraphics[width=2.5cm]{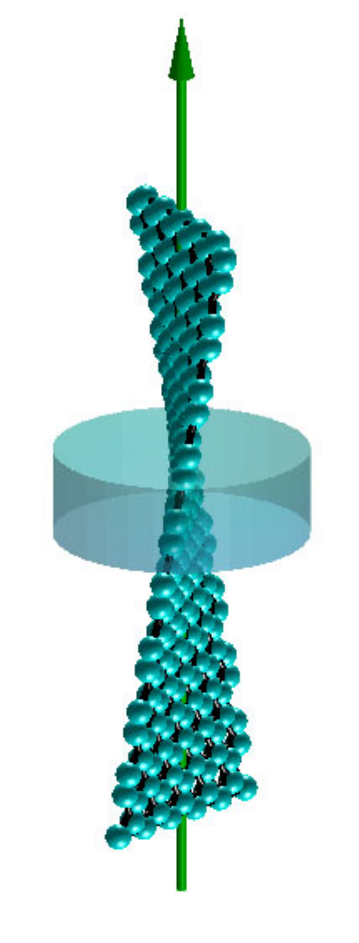}}
\end{subfigure}
}
\caption{\REV{Band diagram for an armchair graphene nanoribbon with a twist parameter of $\alpha=0.02$ (corresponding to a rate of twist $\beta = 16.9^{\circ}$ per nanometer) generated using HelicES. The green shaded region in the structure on the right is the fundamental domain used in HelicES. The green arrow denotes the $\textbf{e\textsubscript{Z}}$ axis.}}
\label{Fig:band_diagram_GNR_twisted_unpassivated}
\end{figure}

\REV{To finish this discussion on computational advantages of HelicES, we now present a system that cannot be simulated in Helical DFT, and one that will require extensive computational resources in typical periodic finite difference or plane-wave codes --- an armchair graphene nanoribbon with a twist of $\alpha=0.02$. Note that this is still a relatively high value of $\alpha$, but was chosen here for a better visual representation of the system. The band diagram of this system is presented in Fig.~\ref{Fig:band_diagram_GNR_twisted_unpassivated}. Noticeably, in contrast to the untwisted, passivated nanoribbon presented in Fig.~\ref{Fig:band_diagram_GNR}, this system appears to have a vanishingly small band-gap, indicative of metallic behavior.}

\subsection{Application to the study of the electromechanical response of a nanotube}
\label{sec:twist_comparison}
\begin{figure}[H]
\centering
\scalebox{0.85}{
\includegraphics[width=0.8\textwidth]{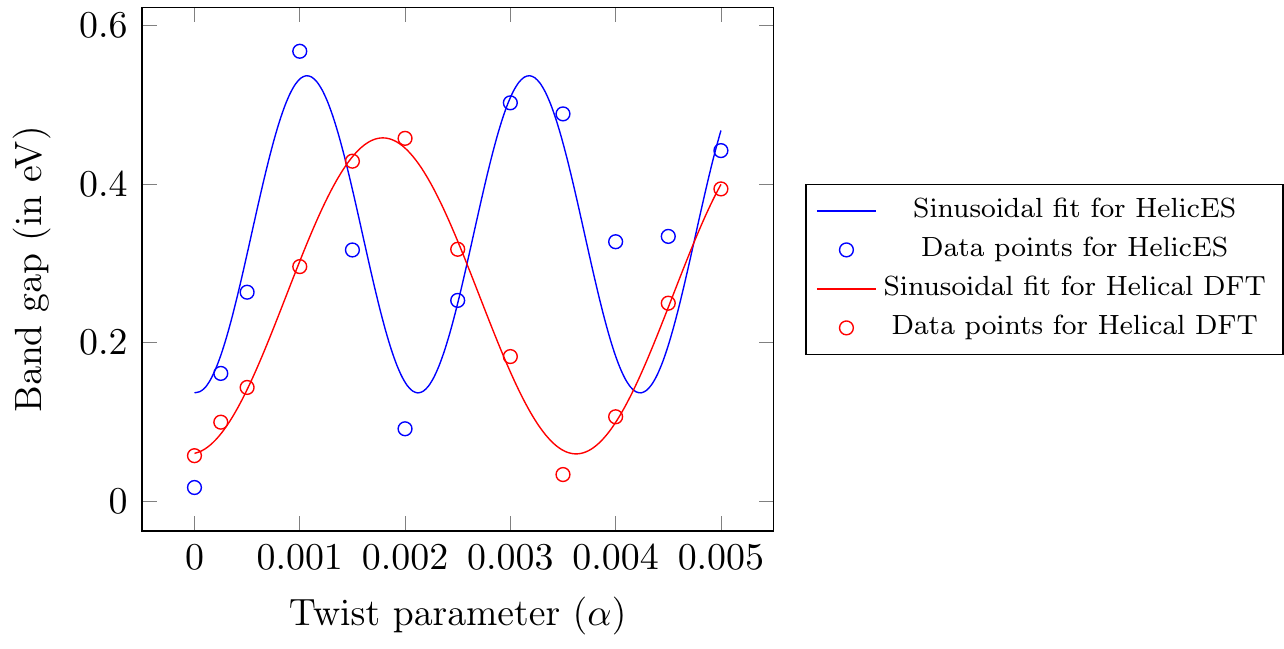}
}
\caption{Band gap trend as the twist parameter $\alpha$ is varied for a $(16,16)$ armchair carbon nanotube. Results from HelicES (empirical pseudopotentials) and the Helical DFT code (self consistent calculations with ab initio pseudopotentials and atomic relaxation effects included \cite{yu2022density, banerjee2021ab}) are both shown.}
\label{Fig:twist_DFT_vs_ES}
\end{figure}

Finally, as a demonstration of the utility of the computational method developed here, we study the the electromechanical response of a quasi-one-dimensional nanomaterial as it undergoes deformations. Specifically, we consider a carbon nanotube with a radius of about $1.0$ nanometer (an armchair $(16,16)$ tube), and subject it to twisting. We start from the untwisted structure and increase the rate of applied twist, considering up to about $\beta = 7.4^{\circ}$ , in our simulations. Fig.~\ref{Fig:twist_DFT_vs_ES} shows the variation of the band gap of the material with applied twist. For comparison purposes, results from full self consistent Kohn-Sham DFT calculations using ab initio Troullier Martins pseudopotentials \citep{Troullier_Martins_pseudo} and Local Density Approximation based exchange correlation  \citep{Kohn1965self, Perdew_Wang}, are also shown (obtained from \citep{yu2022density}). It is well known that upon twisting, armchair nanotubes --- which are generally metallic in untwisted form --- show metal-to-semiconductor transitions, and that these changes manifest themselves as oscillatory behavior in the band gap \citep{yu2022density, Dumitrica_Tight_Binding1, yang1999band, yang2000electronic}. We see from Fig.~\ref{Fig:twist_DFT_vs_ES} that the results from HelicES do reproduce this qualitative behavior correctly, but the actual response curve is quantitatively different from the first principles data. This is very likely due to the lack of inclusion of atomic relaxation effects in HelicES, as well as the general failure of the Mayer pseudopotential to model scenarios where the carbon atoms do not form a perfect honeycomb lattice --- a consequence of the shearing distortions that arise from the applied twist in this case. Therefore, these results strongly suggest the need for building in ab initio pseudopotentials and self consistent iterations into HelicES, which constitutes ongoing work \citep{Poisson_Shivang}.
\section{Conclusions}
\label{sec:Conclusions}
In summary, we have presented a novel spectral method for efficiently solving the Schr\"odinger equation for quasi-one-dimensional materials and structures. The basis functions in our method --- helical waves --- are natural analogs of plane-waves, and allow systematically convergent electronic structure calculations of materials such as nanowires, nanoribbons and nanotubes to be carried out. We have discussed various mathematical, algorithmic and implementation oriented issues of our technique. We have also used our method to carry out a variety of demonstrative calculations and studied its accuracy, \REV{computational efficiency,} and convergence behaviors. 

We anticipate that the method presented here will find utility in the discovery and characterization of new forms of low dimensional matter. It is particularly well suited for coupling with specialized machine learning techniques \citep{pathrudkar2022machine} and for the multiscale modeling of low dimensional systems \citep{hakobyan2012objective}. Building self-consistency into the method, so as to enable \textit{ab initio} calculations (e.g. using Hartree-Fock or Kohn-Sham Density Functional Theory \citep{LeBris_ReviewBook}) remains the scope of ongoing and future work. An important first step in this direction is efficient solution of the associated electrostatics problem \citep{Dumitrical_Ewald}, towards which we have been making recent progress  \citep{liebman2022helical, Poisson_Shivang}. \REV{Finally, the full power of some of the techniques described here can be brought to bear upon complex materials problems, once a parallel, efficient, hardware optimized version of HelicES is available. Development of such a code constitutes yet another avenue of ongoing and future work.}
\begin{center}
---    
\end{center}
\appendix
\section{Derivation of the governing equation in helical coordinates}
\label{App:Governing_Equation}
We are interested in solutions of the Schr\"{o}dinger equation, i.e., $\big(-\half \Delta + V(\bfx)\big)\psi = \lambda\,\psi$, as it applies to a quasi-one-dimensional structure. For a function $\psi(\theta_1,\theta_2, r)$ expressed in helical coordinates, the Laplacian is given by \citep{My_PhD_Thesis, banerjee2021ab}:
\begin{align}
\label{Eq:Laplacian_helical}
\Delta\psi=\psi_{rr}+\frac{1}{r}\psi_r+\frac{1}{\tau^2}\psi_{\theta_1\theta_1}-\frac{2\alpha}{\tau^2}\psi_{\theta_1\theta_2}+\frac{1}{4\pi^2}\left(\frac{1}{r^2}+\frac{4\pi^2\alpha^2}{\tau^2}\right)\psi_{\theta_2\theta_2}\,.
\end{align}
Considering the helical and cyclic symmetry adapted Bloch ansatz, $\psi(\theta_1,\theta_2, r;\eta,\nu)=e^{-i2\pi(\eta\theta_1+\nu\theta_2)}\phi(\theta_1,\theta_2, r;\eta,\nu)$,  we first note:
\begin{align}
\begin{split}
\psi_r &= e^{-i2\pi(\eta\theta_1+\nu\theta_2)}\phi_r\,,\\
\psi_{rr} &=e^{-i2\pi(\eta\theta_1+\nu\theta_2)}\phi_{rr}\,,\\
\psi_{\theta_1}&=e^{-i2\pi(\eta\theta_1+\nu\theta_2)}\left[-i2\pi\eta\phi+\phi_{\theta_1}\right]\,,\\
\psi_{\theta_2} &= e^{-i2\pi(\eta\theta_1+\nu\theta_2)}\left[-i2\pi\nu\phi+\phi_{\theta_2}\right]\,,\\
\psi_{\theta_1\theta_2}&=e^{-i2\pi(\eta\theta_1+\nu\theta_2)}\left[-4\pi^2\eta\nu\phi-i2\pi\nu\phi_{\theta_1}-i2\pi\eta\phi_{\theta_2}+\phi_{\theta_1\theta_2}\right]\,, \\
\psi_{\theta_1\theta_1}&=e^{-i2\pi(\eta\theta_1+\nu\theta_2)}\left[-i2\pi\eta\left(-i2\pi\eta\phi+2\phi_{\theta_1}\right)+\phi_{\theta_1\theta_1}\right]\,, \\ 
\psi_{\theta_2\theta_2}&=e^{-i2\pi(\eta\theta_1+\nu\theta_2)}\left[-4\pi^2\nu^2\phi-4i\pi\nu\phi_{\theta_2}+\phi_{\theta_2\theta_2}\right] \,.
\end{split}
\end{align}
Thus, we get:
\begin{align}
\begin{split}
\Delta\psi=\left[\phi_{rr}+\frac{1}{r}\phi_r+\frac{1}{\tau^2}\phi_{\theta_1\theta_1}-\frac{4\pi^2\eta^2}{\tau^2}\phi-\frac{i4\pi\eta}{\tau^2}\phi_{\theta_1}-\frac{2\alpha}{\tau^2}\phi_{\theta_1\theta_2}+\frac{8\alpha\pi^2\eta\nu}{\tau^2}\phi \right. \\ \left. +\frac{4i\pi\nu\alpha}{\tau^2}\phi_{\theta_1}+\frac{4i\pi\eta\alpha}{\tau^2}\phi_{\theta_2}+\frac{1}{4\pi^2}\left(\frac{1}{r^2}+\frac{4\pi^2\alpha^2}{\tau^2}\right)\phi_{\theta_2\theta_2} \right. \\ \left. -\nu^2\left(\frac{1}{r^2}+\frac{4\pi^2\alpha^2}{\tau^2}\right)\phi-\frac{i\nu}{\pi}\left(\frac{1}{r^2}+\frac{4\pi^2\alpha^2}{\tau^2}\right)\phi_{\theta_2}\right]e^{-i2\pi\left(\eta\theta_1+\nu\theta_2\right)}\,,
\end{split}
\end{align}
which simplifies to:
\begin{align}
\label{Eq:Expanded_laplacian}
\begin{split}
\Delta\psi=\left[\Delta\phi+\left(\frac{4\pi^2}{\tau^2}\left[\nu\alpha\left(2\eta-\nu\alpha\right)-\eta^2\right]-\frac{\nu^2}{r^2}\right)\phi+\frac{4i\pi}{\tau^2}\left(\nu\alpha-\eta\right)\phi_{\theta_1} \right. \\ \left. +i\left[\frac{4\pi\alpha}{\tau^2}\left(\eta-\nu\alpha\right)-\frac{\nu}{\pi r^2}\right]\phi_{\theta_2}\right]e^{-i2\pi\left(\eta\theta_1+\nu\theta_2\right)}\,.
\end{split}
\end{align}
Hence the action of the Schr\"odinger operator on $\psi$ can be expressed as:
\begin{align}
\label{Eq:Expanded_schrodinger_1}
\begin{split}
\left(-\frac{1}{2}\Delta+V\right)\psi=\Bigg[-\frac{1}{2}\Delta\phi-\left(\frac{2\pi^2}{\tau^2}\Big\{\nu\alpha\left(2\eta-\nu\alpha\right)-\eta^2\Big\}-\frac{\nu^2}{2r^2}\right)\phi \\ -\frac{2i\pi}{\tau^2}\left(\nu\alpha-\eta\right)\phi_{\theta_1}-2i\pi\left[\frac{\alpha}{\tau^2}\left(\eta-\nu\alpha\right)-\frac{\nu}{4\pi^2 r^2}\right]\phi_{\theta_2}+V\phi\Bigg]e^{-i2\pi(\eta\theta_1+\nu\theta_2)}\,.
\end{split}
\end{align}
Since the phase $e^{-i2\pi(\eta\theta_1+\nu\theta_2)} \neq 0$, canceling it from both sides of the Schrodinger equation in $\psi$ leaves us with the following eigenvalue problem in $\phi$:
\begin{align}
\label{Eq:Expanded_schrodinger_2}
\begin{split}
\Bigg[-\frac{1}{2}\Delta\phi-\left(\frac{2\pi^2}{\tau^2}\Big\{\nu\alpha\left(2\eta-\nu\alpha\right)-\eta^2\Big\}-\frac{\nu^2}{2r^2}\right)\phi-\frac{2i\pi}{\tau^2}\left(\nu\alpha-\eta\right)\phi_{\theta_1}-\\2i\pi\left[\frac{\alpha}{\tau^2}\left(\eta-\nu\alpha\right)-\frac{\nu}{4\pi^2 r^2}\right]\phi_{\theta_2}+V\phi\Bigg]=\lambda\phi
\end{split}
\end{align}
This equation needs to be discretized and solved over the fundamental domain, along with suitable boundary conditions in $\phi$.
\section{Derivation of the Basis Set}
\label{App:Basis_set_derivation}
In analogy to the classical plane-wave method \citep{Hutter_abinitio_MD, Martin_ES}, the basis functions in our scheme are eigenfunctions of the Laplacian. However, instead of periodic boundary conditions obeyed by planewaves, we consider boundary conditions resulting from invariance under helical and cyclic symmetries. The calculation presented below is based on similar results in \citep{My_PhD_Thesis}, while a vector version of this calculation appears in \citep{friesecke2016twisted, justel2016bragg} in the context of x-ray diffraction patterns of twisted nanomaterials.

Let  $F(\theta_1,\theta_2, r)$ be a basis function expressed in helical coordinates. Then, invariance under helical and cyclic symmetries implies that this function must be periodic in $\theta_1$ with a period of $1$, and also periodic in $\theta_2$ with a period of $\frac{1}{\mathfrak{N}}$. Assuming $F(\theta_1,\theta_2, r)$ is separable, we characterize the dependence of the function on $\theta_1$ and  $\theta_2$ through Fourier modes (i.e., complex exponentials), and write:
\begin{align}
F_{m,n,k}(\theta_1,\theta_2, r)=e^{i2\pi(m\theta_1 + n\mathfrak{N}\theta_2)}\,\xi (r)\,.
\label{eq:basis_radial_theta1_theta2}
\end{align}
Here $\xi (r)$ is a purely radial function that possibly depends on $m,n,k$, and incorporates normalization constants. The Laplacian of the above function in the helical coordinates is:
\begin{align}
\label{Eq:Laplace_particular_separable_ansatz_1}
\begin{split}
\Delta F_{m,n,k}=\xi_{r}\,e^{i2\pi(m\theta_1+n\mathfrak{N}\theta_2)}+\frac{1}{r}\xi_{rr}\,e^{i2\pi(m\theta_1+n\mathfrak{N}\theta_2)}-\frac{4\pi^{2}m^{2}}{\tau^2}F_{m,n,k}\\ +\frac{8\alpha\pi^{2}nm\mathfrak{N}}{\tau^2} F_{m,n,k} -\left(\frac{1}{r^2}+\frac{4\pi^2\alpha^2}{\tau^2}\right)n^2\mathfrak{N}^2 F_{m,n,k}\,,
\end{split}
\end{align}
which can be rewritten as:
\begin{align}
\label{Eq:Laplace_particular_separable_ansatz_2}
\Delta f_{m,n,k}&=e^{i2\pi(m\theta_1+n\mathfrak{N}\theta_2)}\left[\xi_{rr}+\frac{1}{r}\xi_{r}\right]-F_{m,n,k}\left[\frac{n^{2}\mathfrak{N}^2}{r^2}+\frac{4\pi^2}{\tau^2}\left(m-\alpha n\mathfrak{N}\right)^2\right]\,.
\end{align}
Now, imposing the condition that $f_{m,n,k}$ is an eigenfunction of the Laplacian, i.e.,
\begin{align}
\label{Eq:Derive_basis_set}
-\Delta F_{m,n,k}=\lambda_{m,n,k}^{0}\,F_{m,n,k}\,,
\end{align}
we get:
\begin{align}
\label{Eq:Derive_basis_set_2}
\begin{split}
-e^{i2\pi(m\theta_1+n\mathfrak{N}\theta_2)}\left[\xi_{rr}+\frac{1}{r}\xi_{r}\right] &+ \,e^{i2\pi(m\theta_1+n\mathfrak{N}\theta_2)}\,\left[\frac{n^{2}\mathfrak{N}^2}{r^2}+\frac{4\pi^2}{\tau^2}\left(m-\alpha n\mathfrak{N}\right)^2\right] \xi \\
=&\,\lambda_{m,n,k}^{0}\,e^{i2\pi(m\theta_1+n\mathfrak{N}\theta_2)} \xi\,,
\end{split}
\end{align}
which simplifies to:
\begin{align}
\xi_{rr}+\frac{1}{r}\xi_{r}-\xi\left[\frac{n^{2}\mathfrak{N}^2}{r^2}-\lambda_{m,n,k}^{0}+\frac{4\pi^2}{\tau^2}\left(m-\alpha n\mathfrak{N}\right)^2\right]&=0\,.
\end{align}
Denoting $\xi_{m,n}^2=\frac{4\pi^2}{\tau^2}\left(m-\alpha n\mathfrak{N}\right)^2$ and performing the change of variables:
\begin{align}
\tilde{r}=r\sqrt{\lambda_{m,n,k}^{0}-\gamma_{m,n}^2}\,,\,\xi(r) = \widetilde{\xi}(\tilde{r})\,,
\end{align}
we see that the above equation reduces to:
\begin{align}
\tilde{r}^2\,\widetilde{\xi}_{\tilde{r}\tilde{r}} + \tilde{r}\,\widetilde{\xi}_{\tilde{r}} + (\tilde{r}^2 - n^2\mathfrak{N}^2)\widetilde{\xi} = 0\,.
\end{align}
This is simply Bessel's equation \citep{abramowitz1988handbook, NIST_DLMF} in $ \widetilde{\xi}(\tilde{r})$. Since $n\mathfrak{N}$ is real, the general solution of this equation can be expressed in terms of ordinary Bessel functions of the first and second kind as:
\begin{align}
\label{Eq:General_sol_Gamma_ode}
\widetilde{\xi}(\tilde{r})&=A\,J_{n\mathfrak{N}}(\tilde{r})+B\,Y_{n\mathfrak{N}}(\tilde{r})\,.
\end{align}
To evaluate the constants $A$ and $B$, we need to invoke boundary and normalization conditions. Since the wavefunctions are expected to be finite valued at the origin ($r=0$), and Bessel functions of the second kind approach infinity near $0$, we conclude that $B = 0$. Furthermore, since the wavefunctions obey Dirichlet boundary conditions on the lateral surface of the computational domain ($r=R$), so should the basis functions used to discretize them. Hence, we obtain:
\begin{align}
\xi\bigg(R\sqrt{\lambda_{m,n,k}^{0}-\gamma_{m,n}^2}\bigg) = A\,J_{n\mathfrak{N}}\bigg(R\sqrt{\lambda_{m,n,k}^{0}-\gamma_{m,n}^2}\bigg) = 0\,.
\end{align}
This implies that $R\sqrt{\lambda_{m,n,k}^{0}-\gamma_{m,n}^2}$ must be a root of the the ordinary Bessel function of the first kind. Denoting the $k^{th}$ root ($k=1,2,\ldots$) of the Bessel function of order $p$, as $b^{p}_{k}$, we see that:
\begin{align}
b^{n\mathfrak{N}}_{k} = R\sqrt{\lambda_{m,n,k}^{0}-\gamma_{m,n}^2}\,,
\end{align}
from which, it follows that:
\begin{align}
\lambda_{m,n,k}^{0} = \left(\frac{b^{n\mathfrak{N}}_{k}}{R}\right)^2+\left[\frac{2\pi}{\tau}\left(m-\alpha n\mathfrak{N}\right)\right]^2\,.
\end{align}
Thus, we have:
\begin{align}
\xi(r) = A\,J_{n\mathfrak{N}}\bigg(\frac{b^{n\mathfrak{N}}_{k}}{R}r\bigg)\,.
\end{align}
Finally, to determine the constant $A$, we apply the orthonormality condition between two distinct basis functions $F_{m,n,k}$ and $F_{m',n',k'}$:
\begin{align}
\innprod{F_{m,n,k}}{F_{m',n',k'}}{\Lpspc{2}{}{\calD}} = \delta_{m,m'}\,\delta_{n,n'}\,\delta_{k,k'}\,.
\end{align}
This requires that:
\begin{align}
\begin{split}
A^{2}\int^{1}_{0}e^{i2\pi(m-m')\theta_1}d\theta_1\times &\int^{\frac{1}{\mathfrak{N}}}_{0}e^{i2\pi\mathfrak{N}(n-n')\theta_2}d\theta_2 \\
\times &\int^{R}_{0}J_{n\mathfrak{N}}\left(\frac{b^{n\mathfrak{N}}_{k}r}{R}\right)J_{n'\mathfrak{N}}\left(\frac{b^{n'\mathfrak{N}}_{k'}r}{R}\right)2\pi\tau rdr=  \delta_{m,m'}\,\delta_{n,n'}\,\delta_{k,k'}\,.
\end{split}
\end{align}
Due to the properties of complex exponentials and Bessel functions, we note that this condition is readily satisfied for distinct basis functions (i.e., when any of the conditions $m\neq m'$, $n\neq n'$, $k \neq k'$ hold). For the case $m = m', n=n', k=k'$, we arrive at:
\begin{align}
\frac{2\pi\tau A^2}{\mathfrak{N}}\int^{R}_{0}J^{2}_{n\mathfrak{N}}\left(\frac{b^{n\mathfrak{N}}_{k}r}{R}\right)rdr =1\,,
\end{align}
i.e., 
\begin{align}
\frac{2\pi\tau A^2}{\mathfrak{N}}\frac{R^2}{2}J^{2}_{n\mathfrak{N}+1}\left(b^{n\mathfrak{N}}_{k}\right)&=1\,.
\end{align}
Thus it follows that the normalization constant:
\begin{align}
A&=\sqrt{\frac{\mathfrak{N}}{\pi\tau}}\frac{1}{RJ_{n\mathfrak{N}+1}\left(b^{n\mathfrak{N}}_{k}\right)}\,,
\end{align}
and that:
\begin{align}
\xi(r) \equiv \xi_{n,k}(r) = \sqrt{\frac{\mathfrak{N}}{\pi\tau}}\frac{1}{RJ_{n\mathfrak{N}+1}\left(b^{n\mathfrak{N}}_{k}\right)}\,J_{n\mathfrak{N}}\bigg(\frac{b^{n\mathfrak{N}}_{k}}{R}r\bigg)\,.
\end{align}
Hence, the basis functions in our method have the form:
\begin{align}
\label{Eq:Basis_fun_derived}
F_{m,n,k}\left(\theta_1,\theta_2, r\right)=\sqrt{\frac{\mathfrak{N}}{\pi\tau}}\frac{1}{RJ_{n\mathfrak{N}+1}\left(b^{n\mathfrak{N}}_{k}\right)}\,e^{i2\pi(m\theta_1+n\mathfrak{N}\theta_2)}\,J_{n\mathfrak{N}}\left(\frac{b^{n\mathfrak{N}}_{k}r}{R}\right)\,.
\end{align}

Note that if the computational domain were an annular cylinder (as employed in \citep{banerjee2021ab, yu2022density}),  instead of the solid cylinder considered here, the boundary conditions on the radial part of the wavefunction would be expected to change. For Dirichlet boundary conditions applied to the inner and outer walls of such an annular cylinder --- often employed in simulations of large diameter nanotubes --- Bessel functions of both kinds would be involved (i.e., $A,B \neq 0$ in eq.~\ref{Eq:General_sol_Gamma_ode}) and the zeros of the cross products of Bessel functions \citep{laslett1962evaluation} would be required.
\section{Calculation of Gradients}
\label{App:Gradients}
In electronic structure calculations, it can sometimes become necessary to compute the derivative of a quantity expressed using a chosen basis set, or over a grid. For instance, evaluation of the Hellmann-Feynman forces \cite{hellmann2015hans,feynman1939forces} on the atoms of a system involves calculation of Cartesian gradients, if atomic pseudopotentials and pseudocharges are used to compute total energies \citep{banerjee2021ab, banerjee2016cyclic,ghosh2019symmetry}. In this section, we describe how such gradients may be computed for quantities expressed using helical waves.

Let  $E(\theta_1,\theta_2, r)$ be a function expressed in helical coordinates over the fundamental domain, and let its expansion using helical waves be:
\begin{align}
\label{Eq:E_expanded_in_basis_set}
\nonumber
E\left(\theta_1,\theta_2, r\right)&=\sum_{\Gamma}\hat{E}_{m,n,k}\,F_{m,n,k}(\theta_1,\theta_2, r)\\ 
&= \sum_{\Gamma}\hat{E}_{m,n,k}\,c_{m,n,k}\,e^{i2\pi(m\theta_1+n\mathfrak{N}\theta_2)}\,J_{n\mathfrak{N}}\left(\frac{b^{n\mathfrak{N}}_{k}r}{R}\right)\,.
\end{align}
The Cartesian gradient of this quantity,
\begin{align}
\nabla E&=\frac{\partial E}{\partial x}\mathbf{e_\textsubscript{X}}+\frac{\partial E}{\partial y}\mathbf{e_\textsubscript{Y}}+\frac{\partial E}{\partial z}\mathbf{e_\textsubscript{Z}}\,,
\end{align}
may be evaluated by using the chain rule, i.e.,:
\begin{align}
\nonumber
\frac{\partial E}{\partial x}&=\frac{\partial E}{\partial \theta_1}\frac{\partial \theta_1}{\partial x}+\frac{\partial E}{\partial \theta_2}\frac{\partial \theta_2}{\partial x}+\frac{\partial E}{\partial r}\frac{\partial r}{\partial x} = -\frac{\sin\left(2\pi\left(\alpha\theta_1+\theta_2\right)\right)}{2\pi r} E_{\theta_2} + \cos\left(2\pi\left(\alpha\theta_1+\theta_2\right)\right) E_{r}\,,\\\nonumber
\frac{\partial E}{\partial y}&= \frac{\partial E}{\partial \theta_1}\frac{\partial \theta_1}{\partial y}+\frac{\partial E}{\partial \theta_2}\frac{\partial \theta_2}{\partial y}+\frac{\partial E}{\partial r}\frac{\partial r}{\partial y} = \frac{\cos\left(2\pi\left(\alpha\theta_1+\theta_2\right)\right)}{2\pi r} E_{\theta_2}+ \sin\left(2\pi\left(\alpha\theta_1+\theta_2\right)\right) E_{r}\,,\\
\frac{\partial E}{\partial z}&= \frac{\partial E}{\partial \theta_1}\frac{\partial \theta_1}{\partial z}+\frac{\partial E}{\partial \theta_2}\frac{\partial \theta_2}{\partial z}+\frac{\partial E}{\partial r}\frac{\partial r}{\partial z} = \frac{1}{\tau}(E_{\theta_1} - \alpha E_{\theta_2})\,.
\label{eq:Cartesian_Derivatives}
\end{align}
Based on eq.~\ref{Eq:E_expanded_in_basis_set}, we note immediately that:
\begin{align}
\label{Eq:Derivative_of_energy_wrt_theta}
\begin{split}
E_{\theta_1}(\theta_1,\theta_2, r)&=\sum_{\Gamma}\hat{E}_{m,n,k}\left(i2\pi m\right)F_{m,n,k}(\theta_1,\theta_2, r)\,, \\
E_{\theta_2}(\theta_1,\theta_2, r)&=\sum_{\Gamma}\hat{E}_{m,n,k}\left(i2\pi n\mathfrak{N}\right)F_{m,n,k}(\theta_1,\theta_2, r)\,.
\end{split}
\end{align}
These expressions correspond to the inverse basis transforms of vectors with entries $\displaystyle \{\left(i2\pi m\right)\hat{E}_{m,n,k}\}_{(m,n,k) \in \Gamma}$ and $\displaystyle \{\left(i2\pi n\mathfrak{N}\right)\hat{E}_{m,n,k}\}_{(m,n,k) \in \Gamma}$ respectively, and so they may be readily computed. To calculate the radial derivative $E_r$, we first note the following identity \citep{abramowitz1988handbook}:
\begin{align}
\label{Eq:Derivative_of_bessel_1}
\frac{\partial J_{\kappa}\left(q\right)}{\partial q}&=J_{\kappa-1}\left(q\right)-\frac{\kappa}{q}J_{\kappa}\left(q\right)\,.
\end{align}
This expression may be used for computing the radial derivative of all helical waves within the basis set. However, as $n$ varies from $-N_{max}$ to  $N_{max}$, the order of the Bessel functions involved range from $\kappa = -\mathfrak{N}N_{max}$ to $\kappa = \mathfrak{N}N_{max}$, and the above expression results in a Bessel function that lies beyond the range of the basis set. To remedy this, we may use the following alternate expression \citep{abramowitz1988handbook} for the $\kappa = -\mathfrak{N}N_{max}$ case:
\begin{align}
\label{Eq:Derivative_of_bessel_2}
\frac{\partial J_{\kappa}\left(q\right)}{\partial q}&=\frac{\kappa}{q}J_{\kappa}\left(q\right)-J_{\kappa+1}\left(q\right)\,.
\end{align}
Combining eqs.~\ref{Eq:Derivative_of_bessel_1} and \ref{Eq:Derivative_of_bessel_2} with eq.~\ref{Eq:E_expanded_in_basis_set}, we se that the radial derivative $E_r$ may be expressed as:
\begin{align}
E_r(\theta_1,\theta_2, r) &=\sum_{\Gamma}\hat{E}_{m,n,k}\,c_{m,n,k}\,e^{i2\pi(m\theta_1+n\mathfrak{N}\theta_2)}\,\calB_n(r)\,,
\label{eq:E_r}
\end{align}
where the radial functions $\calB_n(r)$ are:
\begin{align}
\nonumber
\calB_n(r) &= \frac{b^{n\mathfrak{N}}_{k}}{R}\left[J_{n\mathfrak{N}-1}\left(\frac{b^{n\mathfrak{N}}_{k}r}{R}\right)-\frac{n\mathfrak{N}R}{b^{n\mathfrak{N}}_{k}r}J_{n\mathfrak{N}}\left(\frac{b^{n\mathfrak{N}}_{k}r}{R}\right)\right],\, \text{for}\,n \neq N_{\text{max}}\,,\\
&= \frac{b^{n\mathfrak{N}}_{k}}{R}\left[\frac{n\mathfrak{N}R}{b^{n\mathfrak{N}}_{k}r}J_{n\mathfrak{N}}\left(\frac{b^{n\mathfrak{N}}_{k}r}{R}\right)-J_{n\mathfrak{N}+1}\left(\frac{b^{n\mathfrak{N}}_{k}r}{R}\right)\right],\,\text{for}\,n = N_{\text{max}}\,.
\label{eq:B_r}
\end{align}
With this, the radial derivative may be considered an inverse basis transform of the vector with entries $\displaystyle \{\hat{E}_{m,n,k}\}_{(m,n,k) \in \Gamma}$, provided we use the functions $\calB_n(r)$ along the radial direction. These functions may be computed ahead of time and stored, and Algorithm \ref{Algo:Fast_Inverse_Transform} (Section \ref{Sec:Inv_FFT}) may then be used for computing $E_r(\theta_1,\theta_2, r)$. With the derivatives $E_r(\theta_1,\theta_2, r), E_{\theta_1}(\theta_1,\theta_2, r)$ and $E_{\theta_2}(\theta_1,\theta_2, r)$ available on hand as values on a real space grid, we may use eq.~\ref{eq:Cartesian_Derivatives} to evaluate the Cartesian derivatives at each point on the same grid.

Instead of obtaining the derivatives as real space quantities as described above, it is also possible to directly obtain them in reciprocal space. The expansion coefficients of $E_{\theta_1}$ and $E_{\theta_2}$ are immediately seen to be:
\begin{align}
\nonumber
(\widehat{E_{\theta_1}})_{m',n',k'} = \left(i2\pi m'\right)\hat{E}_{m',n',k'}\,,\\
(\widehat{E_{\theta_2}})_{m',n',k'} = \left(i2\pi n'\mathfrak{N}'\right)\hat{E}_{m',n',k'}\,,
\end{align}
and they may be evaluated at a cost that is proportional to the basis set size. By considering the inner product of eq.~\ref{eq:E_r} with the basis functions, the expansion coefficients of the radial derivative may be expressed as :
\begin{align}
(\widehat{E_{r}})_{m',n',k'}  = 
\big\langle E_{r},\;F_{m',n',k'}\big\rangle_{\Lpspc{2}{}{\calD}}=\frac{2}{R}\sum_{\Gamma}\hat{E}_{m',n',k}\;\mathcal{A}(n',k,k')\,.
\label{Eq:Inner_product_cartesian_derivatives_r_final}
\end{align}
The numbers $\mathcal{A}(n',k,k')$ can be expressed in terms of oscillatory integrals:
\begin{align}
\nonumber
&\mathcal{A}\left(n',k,k'\right)=\frac{1}{J_{n'\mathfrak{N}+1}\left(b^{n'\mathfrak{N}}_{k}\right)J_{n'\mathfrak{N}+1}\left(b^{n'\mathfrak{N}}_{k'}\right)} \Bigg[ b^{n'\mathfrak{N}}_{k}\int_{0}^{1}J_{n'\mathfrak{N}-1}\left(b^{n'\mathfrak{N}}_{k}q\right) \\ \label{eq:A1}
&\times J_{n'\mathfrak{N}}\left(b^{n'\mathfrak{N}}_{k'}q\right)q\,dq-n'\mathfrak{N}\int_{0}^{1}\frac{J_{n'\mathfrak{N}}\left(b^{n'\mathfrak{N}}_{k}q\right)J_{n'\mathfrak{N}}\left(b^{n'\mathfrak{N}}_{k'}q\right)}{q}q\,dq \Bigg],\,\text{for}\,n'\neq -\mathfrak{N}N_{max}\\\nonumber 
&= \frac{1}{J_{n'\mathfrak{N}+1}\left(b^{n'\mathfrak{N}}_{k}\right)J_{n'\mathfrak{N}+1}\left(b^{n'\mathfrak{N}}_{k'}\right)} \Bigg[ n'\mathfrak{N}\int_{0}^{1}\frac{J_{n'\mathfrak{N}}\left(b^{n'\mathfrak{N}}_{k}q\right)J_{n'\mathfrak{N}}\left(b^{n'\mathfrak{N}}_{k'}q\right)}{q}q\,dq\\ \label{eq:A2}
&\qquad\qquad\qquad-b^{n'\mathfrak{N}}_{k}\int_{0}^{1}J_{n'\mathfrak{N}+1}\left(b^{n'\mathfrak{N}}_{k}q\right)J_{n'\mathfrak{N}}\left(b^{n'\mathfrak{N}}_{k'}q\right)q\,dq \Bigg],\,\text{for}\,n'= -\mathfrak{N}N_{max}\,.
\end{align}
We may precompute them using the techniques described in \ref{App:Oscillatory_integrals} and store them for use later. Note that such an expansion of the radial derivatives using the basis set implicitly requires these quantities to obey Dirichlet boundary conditions at $r=R$. However, this may not be satisfied in general. The real space expression outlined earlier (eqs.~\ref{eq:E_r},\ref{eq:B_r}) skirts  this issue.
\section{Evaluation of Oscillatory Radial Integrals}
\label{App:Oscillatory_integrals}
\begin{figure}[!ht]
\centering
\includegraphics[width=0.95\textwidth]{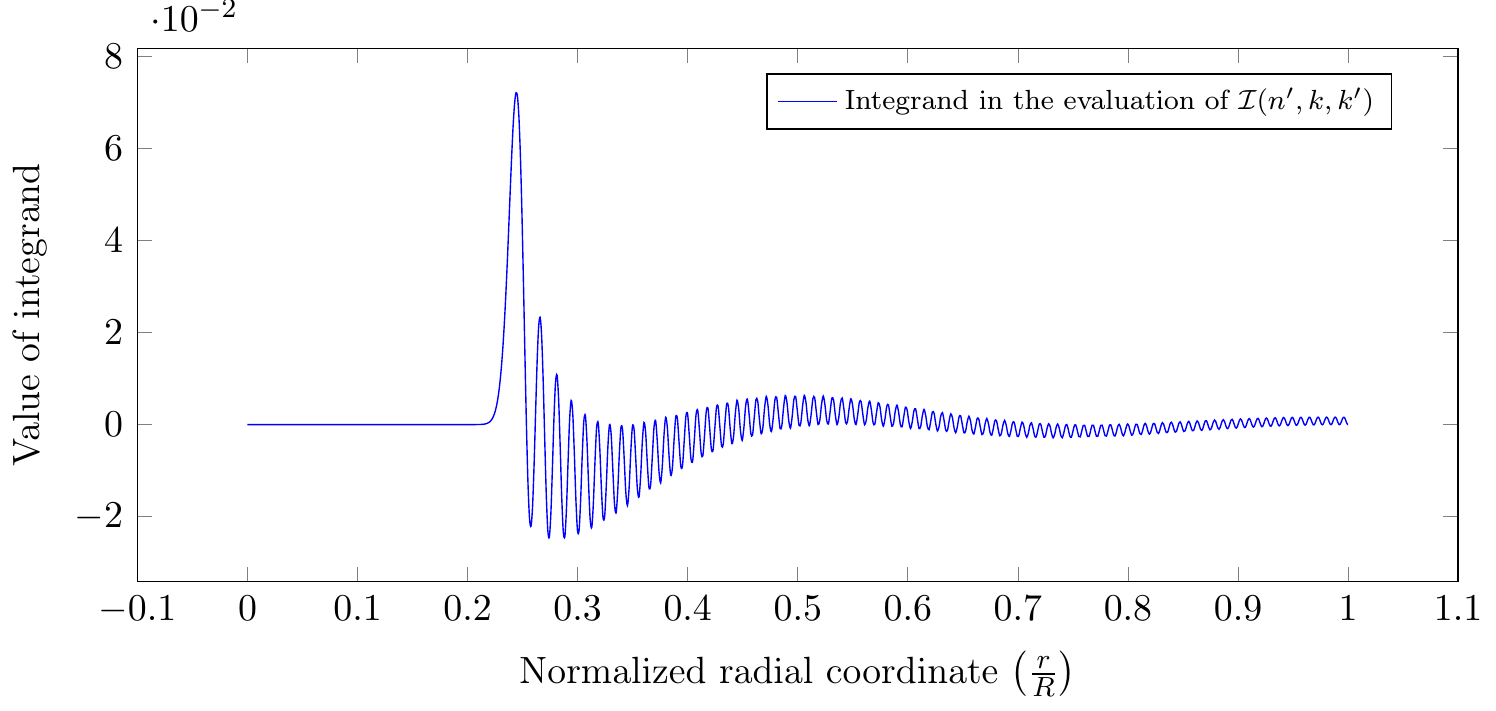}
\caption{The highly oscillatory behavior of the integrand involved in the evaluation of the quantity $\calI(n',k,k')$ in eq.~\ref{eq:I_nkk_prime}. The behavior increases as the values of $n,k,k'$ increase. For the above plot, we chose $n=-98, k=89,k'=85$}
\label{Fig:Oscillatory_integrals}
\end{figure}

This work has multiple instances in which integrals with oscillatory integrands along the radial direction make an appearance (e.g.,~eq.~\ref{eq:I_nkk_prime} and eqs.~\ref{eq:A1},\ref{eq:A2}). A typical scenario is depicted in Fig.~\ref{Fig:Oscillatory_integrals}. Techniques for the evaluation of such integrals have been extensively studied in the literature \citep{deano2017computing, ma2018computing, iserles2005efficient, iserles2004quadrature} and specialized methods for integrands involving Bessel functions are also available \citep{chen2012numerical}. Instead of adopting these more elaborate methods, we choose to evaluate the oscillatory integrals in this work by using the simpler procedure of employing a large number of Gauss-Jacobi quadrature \citep{ralston2001first} nodes and weights. Thus, denoting $q = r/R$, we write:
\begin{align}
\int_{0}^{1}\!f(q)\,q^{\sigma}\,dq=\sum_{i=1}^{N_q}f(q_i)w_i\,.
\end{align}
The values of the weights $w_i$ and the nodes $q_i$ dependent on the the exponent $\sigma$, as well as the quadrature order $N_q$. The  weights and nodes can be computed inexpensively \citep{hale2013fast} even when $N_q$ is of the order of a few thousand.

For the case of the integrals involved in the evaluation of $\calI(n',k,k')$ via eq.~\ref{eq:I_nkk_prime}, the number of oscillations of the integrand is approximately equal to $k+k'$. Thus, within a given basis set, the maximum number of oscillations is $2K_{\text{max}}$. For all the examples considered in this work, $K_{\text{max}}$ does not generally exceed $200$, and we have found that choosing $N_q$ to be a few thousand for such cases allows the integrals to be converged to $\calO(10^{-14})$. To verify our calculations, we have also used Gauss-Kronrod quadrature \citep{laurie1997calculation, shampine2008vectorized} as employed within Matlab (\textsf{quadgk} function). This allows for automatic adaptive placement of the integration nodes and monitoring of the quadrature error, and we verified that the latter was always $\calO(10^{-13})$ or lower, even for the cases involving the most oscillatory integrands.
\section*{Acknowledgements}
\label{sec:acknowledgements}
\REV{ASB and SA would like to acknowledge support through grant DE-SC0023432 funded by the U.S. Department of Energy, Office of Science. ASB acknowledges startup support from the Samueli School Of Engineering at UCLA, as well as funding from a Faculty Research Grant through UCLA's Council on Research (COR). ASB also acknowledges support through a Faculty Career Development Award from UCLA's Office of Equity, Diversity and Inclusion. ASB and SA would like to thank Alexandre Mayer (University of Namur, Belgium) and William Vandenberghe (University of Texas at Dallas) for insightful discussions (over email) on the use of the pseudopotentials developed by them. ASB and SA also acknowledge computational resource support from UCLA's Institute for Digital Research and Education (IDRE), and the National Energy Research Scientific Computing Center (NERSC awards BES-ERCAP0025205 and BES-ERCAP0025168), a DOE Office of Science User Facility supported by the Office of Science of the U.S. Department of Energy under Contract No. DE-AC02-05CH11231.}
\begin{center}
---
\end{center}

\end{document}